\documentclass[11pt]{article}

\usepackage{natbib}
 \bibpunct[, ]{(}{)}{,}{a}{}{,}%
\usepackage[ruled,linesnumbered]{algorithm2e}
\usepackage[margin=1in]{geometry}
\usepackage{subcaption}
\usepackage{multirow}
\usepackage{rotating}
\usepackage{comment}
\usepackage[normalem]{ulem}
\usepackage{amsmath}
\usepackage{amssymb}
\usepackage{amsthm}
\usepackage{natbib}
\usepackage{authblk}
\usepackage{setspace}
\bibpunct[, ]{(}{)}{,}{a}{}{,}%
\newtheorem{assumption}{Assumption}
\newtheorem{claim}{Claim}
\newtheorem{definition}{Definition}
\newtheorem{theorem}{Theorem}
\newtheorem{lemma}{Lemma}

 \def\E{\mathbb{E}}

 \def\A{\mathcal{A}}
 
 \def\Agood{\mathcal{A}_{\textrm{good}}}
 \def\sumT{\sum_{i=1}^{N^t}}
 \def\sumC{\sum_{j=1}^{N^c}}
 
 \def\ind{\mathbb{I}}
 \def\a{\mathbf{a}}
 
 \def\Hatt{\mathbb{H}_0^{SATT}}
 \def\Hsharp{\mathbb{H}_0^{sharp}}

 \def\chichi{(\chi^+, \chi^-)}
 \def\zz{(z^+, z^-)}
 
\newcommand{\indep}{\mathrel{\text{\scalebox{1.07}{$\perp\mkern-10mu\perp$}}}}
\DeclareMathOperator*{\argmax}{arg\,max}
\DeclareMathOperator*{\argmin}{arg\,min}

\newcommand{\Psia}{\Psi_{\mathcal{D}}(\a)}
\newcommand{\psia}{\psi_{\mathcal{D}}(\a)}

\newcommand{\Psimax}{\Psi^+_{\mathcal{D}}}
\newcommand{\Psimin}{\Psi^-_{\mathcal{D}}}

\newcommand{\Hzero}{\mathbb{H}_0}
\newcommand{\Hone}{\mathbb{H}_1}

\newcommand{\psimax}{\psi^+}
\newcommand{\psimin}{\psi^-}
\newcommand{\D}{\mathcal{D}}
\newcommand{\bX}{\mathbf{X}}

\newcommand{\bA}{\mathbf{A}}

\begin{document}
\title{
A robust approach to quantifying uncertainty in matching problems of causal inference
\footnote{No data ethics considerations are foreseen related to this paper. Code and data to reproduce all the results in this paper is available at https://github.com/marcomorucci/robust-tests}
}


\author[1]{Marco Morucci \thanks{marco.morucci@duke.edu}}
\affil[1]{Department of Political Science, Duke University}
\author[2]{Md. Noor-E-Alam \thanks{md.alam@northeastern.edu}}
\affil[2]{Department of Mechanical and Industrial Engineering, Northeastern University}
\author[3]{Cynthia Rudin \thanks{cynthia@cs.duke.edu}}
\affil[1,3]{Department of Computer Science, Duke University}

\date{}
\maketitle

\begin{abstract}%

Unquantified sources of uncertainty in observational causal analyses can break the integrity of the results. One would never want another analyst to repeat a calculation with the same dataset, using a seemingly identical procedure, only to find a different conclusion. However, as we show in this work, there is a typical source of uncertainty that is essentially never considered in observational causal studies: the choice of match assignment for matched groups, that is, which unit is matched to which other unit before a hypothesis test is conducted. The choice of match assignment is anything but innocuous, and can have a surprisingly large influence on the causal conclusions. 
Given that a vast number of causal inference studies test hypotheses on treatment effects after treatment cases are matched with similar control cases, we should find a way to quantify how much this extra source of uncertainty impacts results. What we would really like to be able to report is that \emph{no matter} which match assignment is made, as long as the match is sufficiently good, then the hypothesis test result still holds. In this paper, we provide methodology based on discrete optimization to create robust tests that explicitly account for this possibility. We formulate robust tests for binary and continuous data based on common test statistics as integer linear programs solvable with common methodologies. We study the finite-sample behavior of our test statistic in the discrete-data case. We apply our methods to simulated and real-world datasets and show that they can produce useful results in practical applied settings.
\end{abstract}%

\section{Introduction}

We have a reproducibility crisis in science. Part of the reason for the crisis is sources of uncertainty in the analysis pipeline that are not accounted for. Observational causal studies have serious problems with reproducibility, despite the fact that these studies underlie important policy decisions. In our paper, we focus on a source of uncertainty that can have a large impact on the result, but is almost never mentioned: match assignments of treatment units to control units.

Classically, assignments of treatment and control units to matches are constructed using a single method aimed at constructing pairs that achieve a good level of some measure of quality predefined by the analyst (e.g., Optimal Matching, \citealt{ros1989}, or Genetic Matching, \citealt{diamond2013}). Choosing a single fixed match ignores a major source of uncertainty, which is the design itself, or in other words, the uncertainty related to the choice of experimenter. What if there were two possible equally good match assignments, one where the treatment effect estimate is very strong and one where it is nonexistent? When we report a result on a particular matching assignment, we thus ignore the possibility of the opposite result occurring on an equally good assignment. It is entirely possible that two separate researchers studying the same effect on the same data, using two different equally good sets of matched groups, would get results that disagree.
When researchers follow the classic pipeline of match-then-estimate, their hypothesis tests are conditioned on the match assignment; in other words, only uncertainty \textit{after} matching is considered, and not the uncertainty stemming from the match assignment itself.


Our goal is to create robust matched pairs hypothesis tests for observational data. These tests implicitly consider \textit{all possible reasonably good assignments} and consider the \textit{range of possible outcomes} for tests on these data. This is a more computationally demanding approach to hypothesis testing than the standard approach where one considers just a single assignment, but the result is then more robust to the choice of experimenter who chooses the match assignment. It is computationally infeasible (and perhaps not very enlightening) to explicitly compute all possible assignments, but it is possible to look at the range of outcomes associated with them. In particular, \textit{our algorithms compute the maximum and minimum of quantities like the test statistic values and their associated $p$-values}, in accordance with the principles of robust optimization.


After motivation and formalization for our framework in Sections \ref{Sec:Sensitivity} and \ref{sec:approach}, we offer a robust formulation for McNemar's statistic for binary outcomes in Section \ref{SectionMcNemar}. 
In Section \ref{SectionZtest} we formulate an integer linear program (ILP) to compute the robust version of the canonical $z$-test in a general case. 
We then study the finite-sample distribution of our robust version of McNemar's test that includes the uncertainty stemming from the match assignment. Finally, we present evidence of the performance of our methods in Section \ref{Sec:Simulations} using simulated data where the ground truth is known, and two real-data case studies are discussed in Section \ref{SectionCaseStudies}. We know of no previous approaches to matching in observational causal inference that use robust optimization to handle uncertainty in the match assignment.


\subsection{Matching Analyses and Their Sensitivity to Arbitrary Choices}\label{Sec:Sensitivity}
Existing studies that use matching for hypothesis testing all roughly follow the following template:  (Step 1) Choose a test statistic, (Step 2) Define criteria that acceptable matches should satisfy (e.g., covariate balance), (Step 3) Use an algorithm to find matched groups (and corresponding subsamples) that satisfy the criteria, (Step 4) Implicitly choose the subsample of interest to be defined by the matched data, (Step 5) Compute the test statistic and its p-value on the matched data \citep[see, e.g.,][]{Rosenbaum2010}.
This procedure \textit{does not explicitly incorporate uncertainty arising from Steps 3 and 4: values of the test statistic computed under different but equally good matched subsamples could differ}. 
These matches are indistinguishable in terms of quality.
Which of the matched sets should be taken as representative of the whole sample? There is no clear answer to this question.

Even a trivial source of randomness, such as the order in which the data appear in the dataset, could produce two equivalently good match assignments under a variety of popular matching methods, such as nearest-neighbors. If we apply the same test statistic to each of the two matched samples, we would be implicitly testing two \textit{different} hypotheses: the first one is defined on the subset of the data selected by the first match assignment, and the second hypothesis on the second. 
Under the assumption that both matches are equally as good, common statistical tests will yield valid results for each of these two hypotheses, yet the hypothesis test conducted on the first matched sample could reject while the test conducted on the second could not. 
\textit{Since both assignments achieve similar quality, there is no clear way to choose one result over the other}, and arbitrarily reporting one of the two would hide the fact that the other exists.
Such uncertainty would likely be of interest to policy makers and 
researchers who would want their results to be reproducible under all arbitrary choices of analyst. 

There is no easy fix. Assigning multiple control units to each treatment unit, or using deterministic matching methods that yield a unique result for each dataset, ignore the possibility of a slightly different -- but still good -- match assignment giving opposite results. Asymptotics \citep[e.g.,][]{abadie2011} provide no remedy, since asymptotic results are usually intended to apply to all good matching algorithms asymptotically; the problem we encounter in practice occurs because we always work with finite samples. The problem extends not only to treatment effects but also to variance estimates and beyond.

\subsection{Empirical Evidence of the Problem}
The problem we identified is serious in practice.
\cite{Prelim} replicate several social science studies -- all published in top journals after 2010 -- that use matching. They perform the same hypothesis tests after applying several popular matching algorithms. 
Table \ref{Tab:Prelim} reports agreement between any two of six different matching methods on the same hypothesis test, i.e., the percentage of the tests replicated that have the same result (reject/fail to reject) under each pair of methods. If our hypothesis were false, and matching at the same level of quality always produced the same results, then we would see high agreement in the table. This is clearly not the case. Additionally, Table \ref{Tab:Prelim} also reports a measure of similarity of the balance between the datasets produced by each pair of matching methods. We measure balance as the proportion of covariates whose difference in means between the matched treated and control group is statistically insignificant at the 5\% level. The numbers in $b$ of each $a/b$ entry of Table \ref{Tab:Prelim} represent the correlation between indicator vectors that keep track of which covariates were balance by each method according to the criterion just presented. Clearly, while imperfectly, methods tend to ``agree'' on balance more than they do on rejection: agreement on balance is always greater than agreement on rejection, sometimes by a large margin. This implies that differences in balance produced are not, at least in full, to blame for the problem of different matching methods leading to different conclusions: even when methods produce similar balance, they still disagree. 

\begin{table}[!htbp]
  \centering
  \begin{tabular}{ll|r|r|r|r|r|r}
     \multicolumn{8}{c}{\textbf{Rejection agreement / Balance agreement}}\\
      \hline
      && CEM & Genetic & Optimal & Nearest & Subclass & Optimal \\ 
      \hline
      &CEM & -- & 0.15 / 0.82 & 0.40 / 0.68 & 0.60  / 0.71 & 0.41 / 0.74 & 0.23 / 0.62\\ 
      & Genetic &  & -- & 0.27 / 0.63 & 0.51 / 0.68 & 0.68 / 0.63  & 0.21 / 0.83 \\ 
     &Optimal &  &  & -- & 0.42 /  0.99 & 0.44 / 0.94 & 0.40 / 0.75 \\ 
      &Nearest & & &  & -- & 0.64 / 0.95 & 0.31 / 0.78 \\ 
      &Subclass &  & &  & & -- & 0.62 / 0.83 \\ 
      &Optimal &  & &  &  &  & -- \\ 
       \hline
     \multicolumn{8}{l}{Number of hypotheses tested: \textbf{68} times \textbf{6} methods to test each hypothesis}\\
     \multicolumn{8}{l}{\% of tests with at least one rejection and one non-rejection: \textbf{64.7}}\\
     \multicolumn{8}{l}{\% of tests with at least one positive rejection and one negative rejection: \textbf{5.9}}\\
     \multicolumn{8}{l}{\% agreement between two methods, on average: \textbf{40.5}}\\
     \hline\hline
  \end{tabular}\caption{Agreement between rejection decisions on the same data between different matching methods. In each $a$/$b$ entry, the $a$ is the agreement on rejecting the null between methods, and $b$ measures how often covariate balance agrees between the two methods. The bottom rows show some summary statistics of the replication experiment. Results in this table are a summary of results from \cite{Prelim}. This table shows that a 
  seemingly arbitrary choice of matching method can heavily impact the results.}
  \label{Tab:Prelim}
\end{table}


On average, two methods alone agree on rejecting vs$.$ not rejecting the same null hypothesis \textit{only 40.5\% of times} but achieve similar balance 70.5\% of times. These numbers drop dramatically if any group of three methods is considered, and \textit{are virtually zero if agreement among all six methods at once is considered.} As stated before, we believe that this happens because different matching methods choose among matches of similar quality arbitrarily and in different ways. According to these results, \textit{seemingly arbitrary choices of match assignment affect conclusions}. This is true in practical scenarios that involve real-world data, and that addressing it is an important step towards fully robust data analysis.



\section{Proposed Approach}\label{sec:approach}

Throughout this paper we adopt the potential outcomes framework \citep[see][]{Holland86,Rubin74}.
For each unit $i = 1, \dots, N$, we have potential outcomes $Y_i(t) \in \mathbb{R}$, where $t \in \{0, 1\}$. As is standard in causal inference settings, there are $N^t$ units that receive the treatment and $N^c$ units that receive the control condition, with $N^c + N^t = N$. We denote the condition administered to each unit with $T_i \in \{0, 1\}$. We never observe realizations of $Y_i(1)$ and $Y_i(0)$ for each unit at the same time, but only of $Y_i = Y_i(1)T_i + Y_i(0)(1-T_i)$. We also observe a $P$-dimensional vector of covariates $X_i$ taking value in some set $\mathcal{C}$ for every unit. We will denote the complete set of random variables representing the data with $\D = \{X_i, T_i, Y_i\}_{i=1}^N$, with each element being an independent draw from some probability distribution. Analogously, we will denote the set of observed values for the data with the lowercase equivalents of the notation above: $\{x_i, t_i, y_i\}_{i=1}^N$. We assume that potential outcomes are related to the covariates as follows: $Y_i(t) = \mu_t(x_i) + \nu_i$, with $\E[\nu_i|X_i=x_i] = 0$. To conduct our hypothesis test we will also notate our observations as: $(x_1^t,y_1^t),...,(x_{N^t}^t,y_{N^t}^t)$, and $(x_1^c,y_1^c),...,(x_{N^c}^c,y_{N^c}^c)$, where $x_i^t, y_i^t$ and $x_i^c, y_i^c$ represents observed values of $X$ and $Y$ for the $i^{th}$ treated and control units respectively. We make the classical assumptions of Conditional Ignorability of treatment assignment and Stable Unit Treatment Value (SUTVA):
\begin{assumption}{(Strong Conditional Ignorability)}\label{As:CondIgnorability}
For any unit in the sample $i = 1, \dots, N$ treatment allocation is independent of potential outcomes conditional on the observed covariates, that is: $T_i \indep (Y_i(1), Y_i(0)) |X_i$.
\end{assumption}
\begin{assumption}{(SUTVA)}\label{As:SUTVA}
A unit's potential outcomes depend only on that unit's treatment level, that is: for all units, $i$: $Y_i(t_1, \dots, t_N) = Y_i(t_i)$, where $Y_i(t_1, \dots, t_N)$ is the potential outcome for unit $i$ under all units' treatment assignments.
\end{assumption}
Under these two assumption, the Sample Average Treatment Effect (SATE) is our quantity of interest and is defined as: 
\begin{align}
\tau = \frac{1}{N}\sum_{i=1}^N\E_{Y_i|X_i=x_i}[Y_i(1) - Y_i(0)|X_i=x_i].
\end{align}
Under SUTVA and Conditional Ignorability, $\tau$ is identified -- it can be consistently estimated with the observed data -- and hypothesis tests based on it can be conducted in a variety of ways. In general, we will consider testing a null hypothesis of 0 average treatment effect in the observed sample, which is canonically defined as follows:
\begin{align}
\Hzero: \tau = 0 |X_1 = x_1, \dots, X_N = x_N.
\end{align}
Note that the defining feature of the SATE is that the contextual covariates, $\bX$ are only considered at the observed values, as implied by the definition of $\Hzero$. As it is done in much of the causal inference literature, we would like to take advantage of matching in order to test $\Hzero$. A matching operator determines which control is assigned to which treatment. In this paper we focus on one-to-one matching (though our results could be generalized to many-to-one or many-to-many matching), therefore we define the matching operator as follows.
\begin{definition}{(Matching Operator)} A matching operator $\mathbf{a}:\{1,...,N^t\}\rightarrow \{1,...,N^c,\emptyset\}$ obeys the following:
 if $i\neq k$ and $\mathbf{a}(i)\neq\emptyset$ then $\mathbf{a}(i)\neq \mathbf{a}(k)$. That is, no two treatment units $i$ and $k$ are assigned to the same control unit.
\end{definition}
We define the size of the matching, \textit{i.e.:} the number of matched pairs, to be $M = \sum_{i=1}^{N_t} \ind(\mathbf{a}(i) \neq \emptyset)$, with $\ind(E)$ representing the indicator function for the event $E$ throughout the paper. The set of all matching operators is $\A$: the set of all possible assignments $\{\mathbf{a}\}$. Throughout the rest of the paper we also use $\a$ to represent a $N^t \times N^c$ matrix of matches such that $a_{ij} = 1$ if $i$ and $j$ are matched together.

How should we make our matches? Choosing a matched set that leads to the most reliable estimates of $\tau$ is a canonical problem in observational causal inference, and indeed many good methods to find such an assignment have been proposed in the existing literature. One of the contributions of this paper is to not restrict consideration to a single good assignment, but, instead, to consider a set of potentially many equally good match assignments. We define such a set as $\Agood \subset\mathcal{A}$: the set of match assignments that satisfy some user-defined criterion of quality. Many criteria for defining a good match assignment already exist, e.g.: moment balance in the covariate distributions of treated and control samples after matching, low aggregate distance between units in covariate space, or similarity in propensity scores among matched units. Ultimately, $\Agood$ should be viewed as the set of match assignments that produce the most reliable estimates of $\tau$ with the given data. To represent the fact that one assignment is chosen among the many in $\Agood$, we will use the random variable $\bA$, having domain equal to $\Agood$. The probability distribution of $\bA$ is determined purely by human analysts' choice of matching algorithm.

Consider, for now, testing $\Hzero$ against a left-tailed alternative: $\mathbb{H}_1: \tau < 0$ with a generic test statistic $\Psia$, dependent on both the data, $\D$, and the chosen match assignment, $\a$, and denote the observed value of $\Psia$ with $\psia$. Virtually all hypothesis tests done on data matched by any of the existing methods require the following assumption, that we also maintain: 
\begin{assumption}{(Test statistic distribution)} \label{As:distribution}
Under Assumption \ref{As:CondIgnorability} and Assumption \ref{As:SUTVA}, we assume that for any assignment $\a \in \Agood$: $\Pr(\Psia \leq \psia|\Hzero, \bA=\a) \approx F(\psia)$, for some known cumulative density function $F$ that does not depend on $\a$. \end{assumption}
This assumption means that, if the matches are good enough, then the CDF of $\Psia$ under the null can be at least approximated well by some known function, $F$. This assumption is implicitly made in all studies that perform a hypothesis test on matched data. It is often justified by appealing to asymptotic arguments \citep{abadie2011}, or by making parametric assumptions about the data \citep{Rubin2007}. For example, if a researcher performs a z-test on matched data, they implicitly assume that $F$ is the CDF of the standard normal distribution. 


The key problem we wish to highlight in this paper can be stated as follows: existing matching procedures wishing to test $\Hzero$ under Assumption \ref{As:distribution} use $\Pr(\Psia \leq \psia|\Hzero, \bA=\a)$ to compute the p-value for $\Psia$ under $\Hzero$, however this p-value is both conditioned on $\Hzero$ \textit{and} the event $\bA=\a$, i.e., conditional on assignment $\a$ being the one chosen among the many in $\Agood$. The hypothesis being tested here is not $\Hzero$, but instead: $\Hzero|\bA=\a$:
$$\Hzero |\bA=\a: \frac{1}{M}\sum_{i=1}^N\E[Y(1) - Y(0)|X = x_i]\ind(\a(i) \neq \varnothing) = 0.$$
This is essentially a version of $\Hzero$ restricted to consider only those units that do receive a match under assignment $\a$, and, most importantly, \textit{there is no guarantee that we will fail to reject $\Hzero|\bA = \a$ if $\Hzero$ is true}. Thus, while testing $\Hzero|\bA=\a$ rather than $\Hzero$ seems not to be what the user ultimately desires, it is the calculation they typically perform. By overlooking the difference between these two quantities, they would fail to ask questions such as: how should we proceed if we find two assignments $\a_1 \in \Agood$ and $\a_2 \in \Agood$, that successfully reject $\Hzero | \bA = \a_1$ but fail to reject $\Hzero | \bA = \a_2$ at the same significance level? We now articulate our proposed solution to this problem. 

Since we have modeled which assignment is selected among those in $\Agood$ as a random variable, it would be natural, but ultimately impossible, to propose computing p-values in expectation over the entirety of $\Agood$ to handle the issue just introduced. This would be possible under Assumption \ref{As:distribution} by computing:
\begin{align*}
    \sum_{\a \in \Agood}\Pr(\Psia \leq \psia|\Hzero, \bA = \a)\Pr(\bA=\a|\Hzero) &=\sum_{\a \in \Agood}F(\psia)\Pr(\bA=\a|\Hzero)\\
    &= \E_{\bA|\Hzero}[F(\psi_{\D}({\bA}))].
\end{align*}
This quantity cannot be computed due to the presence of $\Pr(\bA=\a|\Hzero)$ in its definition: the probability of observing any match assignment among the ones in $\Agood$ cannot be modeled statistically. This is because this probability depends on factors such as which matching method is chosen by the analyst, which hyperparameter values the method is using, and how the method itself may determine how to choose one among many potentially equivalent matched sets. All these are variables that cannot be analytically studied in a general context, and will depend on the specific analyst, data, and method. 

Since directly modeling $\Pr(\bA=\a|\Hzero)$ is nonsensical, we propose to instead \textit{bound} this quantity using the assumption that all analysts choose reasonably good match assignments. This assumption leads to the following simple relationship:
\begin{align*}
\E_{\bA|\Hzero}[F(\psi_{\D}({\bA}))] &= \sum_{\a \in \Agood}F(\psia)\Pr(\bA=\a|\Hzero) \leq \max_{\a \in \Agood}F(\psia). 
\end{align*}
In words, we bound the expected p-value over all good assignments with the largest p-value found among all good assignments. Bounding approaches to handle situations without clear probability distributions are common in robust optimization and robust statistics. 

To further reduce the problem, we note that the CDF of the test statistic, $F(\psia)$, is monotonically increasing in $\psia$ (e.g., the larger the z-statistic, the larger the CDF of the normal), which implies that $\max_{\a \in \Agood}F(\psia) = F(\max_{\a \in \Agood} \psia)$. Letting $\psimax = \max_{\a \in \Agood} \psia$ be the largest observed test statistic obtainable with a good match assignment, we have our final proposed bound for the robust p-value:
\begin{align}
    \E_{\bA|\Hzero}[F(\psi_{\D}({\bA}))] \leq F(\psimax). \label{eq:pvalupper}
\end{align}
Clearly, if $\Hzero$ is rejected at a chosen level of significance by $F(\psimax)$, then it will also be rejected under $\E_{\bA|\Hzero}[F(\psi_{\D}({\bA}))]$. It is in this sense that our proposed tests are \textit{robust}, as they target minimization of the probability of Type-I error: incorrect rejection of $\Hzero$. That is, our tests are less likely to reject the null hypothesis when it is true.

While the bound in Eq. \eqref{eq:pvalupper} is a robust p-value against a left-sided alternative, the same set of arguments shown above can be applied to derive the following upper bound on the p-value for a right-sided alternative: $\Hone:\, \tau > 0$:
\begin{align}
    \E_{\bA|\Hzero}[1-F(\psi_{\D}({\bA}))] &\leq \max_{\a \in \Agood}(1-F(\psia)) = 1 - F(\min_{\a \in \Agood} \psia) = 1-F(\psimin). \label{eq:pvallower}
\end{align}
Finally, it follows from the bounds above that a robust p-value for testing $\Hzero$ against a two-sided alternative: $\Hone:\, \tau \neq 0$ can be found by applying the canonical definition of p-value to the bounds introduced above:
\begin{align}
    2\min\left\{\E_{\bA|\Hzero}\left[F(\psi_{\D}({\bA}))\right], \E_{\bA|\Hzero}\left[1-F(\psi_{\D}({\bA}))\right]\right\} \leq 2\min\{F(\psimax), 1-F(\psimin) \}. \label{eq:pvaltwo}
\end{align}
The problem we face is now one of finding, for a given dataset, the values of $\psimax$ and $\psimin$. Fortunately, this issue can be solved by adapting modern Mixed-Integer-Programming tools to finding the maximal and minimal matches that satisfy the set of constraints that define $\Agood$. In the rest of the paper, we propose MIP formulations and algorithms for computing these values for two of the most widely used test statistics.

\subsection{Motivation for Using a Robust Approach}
This robust approach is justified as follows: 

\paragraph{Why modeling $\Pr(\bA = \a|\Hzero)$ is not possible.}
First, we do not want to consider how likely a certain assignment $\a$ is to appear. This would involve modeling human behavior of the experimenter, or arbitrary and complex choices of different matching algorithms. We do not want to place a distribution over the choice of algorithms or matches that an experimenter would choose. As \citet{MorganWi2007} note, there is no clear guidance on the choice of matching procedure. We do not presuppose a distribution over these procedures. In reality most researchers tend to use popular and widely-cited matching software packages, and they make this choice independently of the specific problem being studied. 

\paragraph{Why the bounding approach is not too extreme.}
It is inaccurate to assume that our tests might lead to results that are ``too'' extreme. This is because there is no reason why maxima or minima over $\Agood$ should be considered outlying or extreme, even if most of the matches in $\Agood$  are different than those at the extrema of the set. Note also that considering a result output by one of the extrema as an ``outlier'' in the sense that it is believed that most other results with good matches should be concentrated elsewhere in the space of $\psia$ is circular logic: if a set of assignments is included in $\Agood$, then it must be thought to be good enough according to the analyst's criteria. Excluding this assignment because it produces values that are too extreme is tantamount to excluding observations from the dataset because they are not producing the desired result. 

\paragraph{It is unlikely that there is only one good assignment.} 
It would be tempting to say that matches chosen are often the unique maximizers of some metric of quality, and that, because of this, there never is uncertainty over choice of matches in practice. %
This is not true in practice, as researchers often start with a pre-defined level of quality that matches must meet to be considered acceptable, and only report results from the assignment that maximizes that quality criterion if the maximum achieved is above that pre-defined quality threshold. Clearly, all assignments above that threshold should be considered, and that is what we propose in this work. 

\paragraph{Asymptotic theory does not solve the problem.} Our robust method is also necessary because existing asymptotic arguments used to derive valid p-values and confidence intervals for matched data \citep[see, e.g.,][]{abadie2006large, abadie2011}. As argued previously, the defining feature of a ``good'' match assignment is that it leads to at least well approximated p-values for the desired test statistic. In the case of asymptotic approximation, it is assumed that $\Pr(\Psia \leq \psia | \Hzero, \bA = \a) \rightarrow F(\psia)$ as $N \rightarrow \infty$. This is assumed true for all assignments $\a \in \Agood$: while it might be the case that the p-values produced under all good assignments will converge to the same as the sample size grows, this is demonstrably not necessarily true in finite samples, hence the need for a robust procedure.

\section{Robust McNemar's Test}\label{SectionMcNemar}
We consider the problem of hypothesis testing for binary outcomes, that is $y(t) \in \{0,1\}$. After matching, treated and control units will be paired together, and since the outcome is binary there can exist four types of pairs: $A$ is the number of pairs such that  $y_t = 0,\, y_c=0$, $B$ is  the number of pairs such that $y_t = 1,\, y_c=0$, $C$ is the number of pairs such that $y_t = 0,\, y_c=1$, $D$ is  the number of pairs such that $y_t = 1,\, y_c=1$. More formally, let $B(\a) = \sumT\sumC a_{ij}y_i^t(1 - y_j^c)$ be the count of matched pairs in which the treated unit has outcome 1 and the control unit has outcome 0 under assignment $\a$, and let $C(\a) = \sumT\sumC a_{ij}y_j^c(1 - y_i^t)$ be the number of matched pairs in which the treated unit has outcome 0 and the control unit has outcome 1. 
We use the following test statistic for all null hypotheses:
\begin{equation} \label{eu_eqn}
          \chi = \frac{B(\a)-C(\a)-1}{\sqrt{B(\a)+C(\a)}} = \frac{\sumT\sumC a_{ij}(y_i^t - y_j^c) - 1}{\sqrt{\sumT\sumC a_{ij}(y_i^t + y_j^c - 2y_i^ty_j^c)}}.
\end{equation}
This is the classical formulation of McNemar's statistic with a continuity correction at the numerator \citep[see][]{tam}. Our proposed robust test statistic can be defined as the pair:
\begin{align*}
\chi^+ &= \max_{\a \in \Agood} \frac{B(\a) - C(\a) - 1}{\sqrt{B(\a) + C(\a)}},\quad \chi^- = \min_{\a \in \Agood}\frac{B(\a) - C(\a) - 1}{\sqrt{B(\a) + C(\a) }}.
\end{align*}
In what follows, we outline a strategy to compute $\chichi$ subject to general constraints that define $\Agood$. Theorem \ref{Thm:FastMCN} in Section \ref{Sec:FastMCN} shows that, in a special case where the constraints on $\Agood$ explicitly define strata for the data, the optimal statistic can be computed in linear time. In Section \ref{Sec:Theory}, we provide theoretical statements of the finite-sample behavior of the optimized statistics. 

\subsection{Optimizing McNemar's statistic with general constraints}
In this section we give an ILP formulation that optimizes $\chi$ with a pre-defined number of matches.

One can show that the number of pairs where the same outcome is realized for treatment and control (the number of $A + D$ pairs) is irrelevant, so we allow it to be chosen arbitrarily, with no constraints or variables defining it. The total number of pairs is thus also not relevant for this test. Therefore, we are permitted to choose only the total number of untied responses ($B$ and $C$ pairs), denoted $m$, which must always be greater than 0 for the statistic to exist. The problem can be solved either for a specific $m$ or globally for all $m > 0$ by looping over all possible values of $m$ until the problem becomes infeasible. In most practical scenarios, the largest $m$ for which a solution is feasible is chosen.

\noindent\textbf{Formulation 1: General ILP Formulation for McNemar's Test}
\begin{equation*}
          \text{Maximize/Minimize}_{\mathbf{a}}\quad \chi(\mathbf{a})=\left[\frac{B(\a)-C(\a)-1}{\sqrt{m}}\right] \textrm{ subject to:}
      \end{equation*}
\begin{align}
       \sumT\sumC a_{ij}y_i^t(1-y_j^c)&=B(\a)  \qquad&& \textrm{(Total number of first type of discordant pairs)} \label{Eq:F1:defB}\\
       \sumT\sumC a_{ij}y_j^c(1-y_i^t)&=C(\a)  &&\textrm{(Total number of second type of discordant pairs)} \label{Eq:F1:defC}\\
         B(\a)+C(\a)&=m &&\textrm{(Total number of discordant pairs)} \label{Eq:F1:response}\\
       \sumT a_{ij} &\leq  1      && \forall j  \quad \textrm{(Match each control unit at most once)} \label{Eq:F1:onlyt1}\\
       \sumC a_{ij} &\leq  1     &&    \forall i  \quad \textrm{(Match each treatment unit at most once)}  \label{Eq:F1:onlyc1}\\
       a_{ij}&\in \{0,1\} &&    \forall i,j \quad \textrm{(Defines binary variable $a_{ij}$)}\\
        \lefteqn{\textrm{(Additional user-defined match quality constraints.)}}
      \end{align}

Equations  (\ref{Eq:F1:defB}) and (\ref{Eq:F1:defC})  are used to define variables $B$ and $C$. To control the total number of untied responses, we incorporate Equation  (\ref{Eq:F1:response}). Equations  (\ref{Eq:F1:onlyt1}) and (\ref{Eq:F1:onlyc1})  confirm that only one treated/control unit will be assigned in a single pair. As it is common practice in matching, it is possible to test a null hypothesis for the SATT by adding the constraint $ \sumT\sum_C{a}_{ij}=N^t $, only in the case where $N^t \leq N^c$, so that all treatment units are matched.

\subsection{Fast Optimization of McNemar's Test Under Special Constraints}\label{Sec:FastMCN}
\begin{figure}
    \centering
    \includegraphics[width=.8\textwidth]{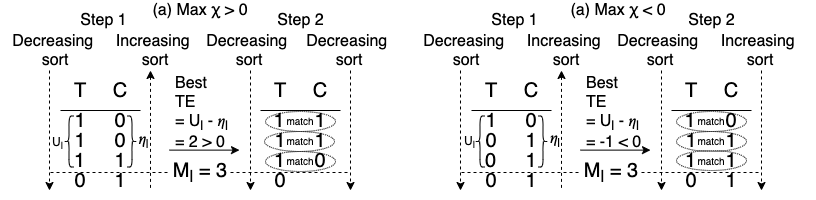}
    \caption{Example of linear-time optimization of McNemar's test in one stratum, in which exactly $M_l$ matches are made. Numbers `0' and `1' in the figure are unit outcome values. In each stratum $l$, $U_l$ denotes the sum of treated outcomes matched, $\eta_l$ is the sum of matched control outcomes, and $M_l$ is the number of matches made.}
    \label{Fig:McNExample}
\end{figure}

The general optimization problem stated in Formulation 1 can also be solved quickly when constraints have the following property: there exists a stratification of the data such that an assignment is a feasible solution if and only if it matches units exclusively within the same stratum. This is stated formally below:
\begin{definition}{(Exclusively Binning Constraints)}\label{Def:Binning}
The constraints on $\Agood$ are exclusively binning if there exists a partition $\mathcal{S}$ of $\{1, \dots, N\}$ such that for all $\a \in \A$ we have $\a \in \Agood$ if and only if: $\forall\; i \in \{ 1\dots, N\}$, $i \in S$, and $\a(i) \in S$ for some subset $S \in \mathcal{S}$. 
\end{definition}
This type of grouping is commonly referred to as blocking~\citep{Imai2008}, and there already exist matching methods that openly adopt it as a strategy to construct good quality matches~\citep{Iacus2012}. Blocking can also occur naturally in data; there need only exist natural subcommunities. In practice, several types of constraints on the quality of matches, particularly balance constraints, as defined in Equation \eqref{Eq:ATEBalance}, can be implemented by making coarsening choices on the covariates~\citep{Iacus2011}, and as such, these coarsening choices are exclusively binning constraints.
Under exclusively binning constraints, solving the optimization problem to find $\chichi$ becomes much simpler, as shown in the following theorem:
\begin{theorem} \label{Thm:FastMCN}
Let the constraints on $\Agood$ be Exclusively Binning Constraints. Then either the max or the min optimization problem in Formulation 1 can be solved in linear time, in the number of units $N$.
\end{theorem}
Proof of the theorem can be found in the appendix, where we explicitly state these linear-time algorithms, together with a proof of their correctness. These algorithms are not difficult, though stating them and providing the proof requires several pages. The algorithms work by first determining the sign of the optimal test statistic, and then by matching units in each stratum separately to optimize the test statistic once the sign is known. Maximizing (minimizing) the test statistic in each stratum is then accomplished by matching as many treated (control) units as possible with outcome 1 to control (treated) units with outcome 0. This procedure is intuitively linear in $N$, the total number of units.  Figure \ref{Fig:McNExample} gives a simple example of how our linear-time algorithms optimize McNemar's test in each stratum. 

\section{Robust Z-Test}\label{SectionZtest}
In this section we consider the canonical $z$-test for estimating whether the difference in mean of the treatment and control populations is sufficiently greater than 0, when outcomes are real-valued, that is: $y(t) \in \mathbb{R}$. Again we will compute extreme values of $z$ that could occur from the set of feasible match assignments.

Again, after computing the extreme values of the z-statistics, p-values for them can be obtained with the canonical Normal asymptotic approximation. At that point, we can determine whether the hypothesis test result is robust to the choice of match assignment.

In what follows, $M$ is the total number of pairs, and $\hat{\sigma}$ is the sample standard deviation of the differences,  $y^t_i-y_{\a(i)}^c$.
The $z$-score is:
\begin{align}
z_{\mathbf{y}}(\a)&=\frac{\bar{d}_{\a}\sqrt{M}}{\hat{\sigma}_\a}, \textrm{ where: }\label{Eq:Zscore}
\bar{d}_{\a} = \frac{1}{M}\sumT\sumC a_{ij}(y_i^t - y_{j}^c),\textrm{ and }\hat{\sigma}_{\a} = \sqrt{\displaystyle\frac{1}{M}\sum_{ i=1 }^{N^t}\sum_{j=1}^{N^c}a_{ij}(y_i^t - y_j^c)^2 - \bar{d}_{\a}^2}. 
\end{align}
Our robust statistic in this case is defined as the pair:
\begin{align}\label{eq:zrobust}
z^+ := \max_{\a \in \Agood} \frac{\bar{d}_{\a}\sqrt{M}}{\hat{\sigma}_\a}, z^- := \min_{\a \in \Agood} \frac{\bar{d}_{\a}\sqrt{M}}{\hat{\sigma}_\a}.
\end{align}
Below, we provide general ILP formulations for computing the robust statistic under (balance and other types of) constraints on $\Agood$ by devising a linearized formulation of the $z$-statistic optimization problem that allows it to be solved with any ILP solver. ILP formulations that are slightly different from each other (that we will discuss) can handle testing of $\Hzero$ for the ATE and ATT.

\subsection{Computing $\zz$ Under General Constraints}\label{SubsectionForm}
The $z$ statistic is clearly not linear in the decision variables (the match assignments). If one were to optimize it directly, a solution could be approximated using a MINLP (mixed-integer nonlinear programming solver) but guarantees on the optimality of the solution might take an incredibly long time. In what follows, we show how this problem can be simplified to be solved by an algorithm that solves several linear integer programming problems instead. This algorithm benefits from the computational speed of ILP solvers, compared to MINLP solvers, and has a guarantee on the optimality of the solution.

To create the ILP formulation, we note that the objective is increasing in the average of the differences (this term appears both in the numerator and denominator), and it is decreasing in the sum of the squared differences (this term is the first term of $\hat{\sigma}$). We then replace the nonlinear objective in \eqref{eq:zrobust}, expanded in \eqref{Eq:Zscore}, as follows:
\begin{equation} \label{ztestobjlinear}
          \text{Maximize/Minimize}_{\mathbf{a}}\quad \displaystyle\sumT\sumC(y_i^t-y_j^c){a}_{ij},
      \end{equation}
which is now linear in the decision variables. The quantity in \eqref{ztestobjlinear} is the estimated treatment effect. Simultaneously, we will limit the sum of squared differences term in the denominator by $b_l$, which is now a parameter rather than a decision variable. Thus, we will optimize treatment effect subject to a bound on the variance. We accomplish this by introducing a new constraint:
\begin{equation} \label{ztestlinearconstraint2}
      \sumT\sumC(y_i^t-y_j^c)^2{a}_{ij} \leq b_l.
      \end{equation}
Putting this together, the new formulation is an ILP. We simply need to solve it for many values of $b_l$. \\

\noindent\textbf{Formulation 3: ILP formulation for $z$-test}
\begin{equation}
          \text{Maximize/Minimize}_{\mathbf{a}}\quad \sumT\sumC(y_i^t-y_j^c)\emph{a}_{ij} \quad  \textrm{(Treatment effect),\;\;\;\;\; subject to}\label{Eq:LinearObj}
      \end{equation}
\begin{eqnarray}
      \sumT\sumC(y_i^t-y_j^c)^2{a}_{ij} &\leq &b_l \quad \quad \textrm{(Upper bound on sample variance)}  \label{Eq:VarBound}\\
       \sumT\sumC a_{ij}&=&M  \quad \quad \quad \quad \textrm{(Choose $M$ pairs)} \label{Eq:MMatches}\\
       \sumT a_{ij} &\leq & 1      \quad  \quad  \quad     \forall j  \quad \textrm{(Match each control unit at most once)} \\
       \sumC a_{ij} & \leq & 1      \quad  \quad  \quad     \forall i  \quad \textrm{(Match each treatment unit at most once)}  \\
       a_{ij}& \in& \{0,1\}  \quad  \quad  \quad     \forall i,j \quad \textrm{(Defines binary variable $a_{ij}$)}   \\
        \lefteqn{\textrm{(Additional user-defined covariate balance constraints.)}}
      \end{eqnarray}
This formulation optimizes treatment effect, subject to the variance of the treatment effect being small. This formulation can be used by itself to find the range of reasonable treatment effects, given a fixed bound $b_l$ on the variance. The problem of testing $\Hzero$ for the ATT and under full matching can be formulated by setting $M=N^t$ (where $N^t$ is the number of treatment points) in Formulation 3. 

\noindent \textbf{Modification 1:} Formulation 3 can also be modified to solve the problem of choosing both treatment and control populations simultaneously. This is also the setting of \citet{Rosenbaum2012}. In that case, the mean is taken over the region of overlap between the control and treatment populations, removing extreme regions. This setting can be handled by looping over increasing values of $M$ until the program becomes infeasible. We would choose the solutions corresponding to the largest values of $M$. 

\noindent \textbf{Modification 2:} For testing a sharp null hypothesis of 0 treatment effect, a simplified formulation is possible, because the sample variance is fixed and known in this case. We would use a special case of Formulation 3, where the variance constraint in \eqref{Eq:VarBound} is replaced with an equality constraint, requiring the variance of the solution to be equal to the known sample variance. (The formula for the variance computed under the sharp null is available in textbooks, e.g., \citealt{imbens2015}.)

\noindent \textbf{Algorithm for Optimizing z-score:} Let us get back to optimizing the z-score. Our algorithm will solve this formulation for many different values of $b_l$ to find the optimal z-scores and p-values.
Let us denote the solution of the maximization problem as $\a_l$. Here, $\a_l$ is an optimal match assignment for a specific value of $b_l$. The indices of the match assignment $\a_l$ are, as usual, $ij$, which are pairs of treatment and control units. Using $\a_l$, we will then be able to bound the value of $z$. Shortly we will use Theorem \ref{Thm:Bound} to prove bounds on the z-score as follows:
\begin{equation}\label{ub}
\max_l \frac{\bar{d}_{\a_l}\sqrt{M}}{\sqrt{\frac{1}{M}b_{l} - (\bar{d}_{\a_l})^2 }} \leq \max_\mathbf{a} z(\a) \leq \max_l\frac{\bar{d}_{\a_l}\sqrt{M}}{\sqrt{\frac{1}{M}b_{l-1} - (\bar{d}_{\a_l})^2 }}.
\end{equation}
That is, when $b_{l-1}$ is close to $b_{l}$, we have little uncertainty about $\max_\mathbf{a} z(\a)$.

Using this bound, we can now formulate an algorithm (Algorithm \ref{Alg:MaxZILP} in Appendix \ref{App:Sec:MaxZ}) to choose progressively finer meshes for $b_l$ to maintain the guarantee on the quality of the solution for  $\max_\mathbf{a} z(\a)$, repeatedly solving the ILP Formulation. An analogous algorithm (with some signs flipped) will allow us to compute  $\min_\mathbf{a} z(\a)$.
Algorithm \ref{Alg:MaxZILP} in Appendix \ref{App:Sec:MaxZ} works as follows:
\begin{itemize}
\item It first solves the ILP Formulation by relaxing (removing) the first constraint (upper bound on sample standard deviation). 
\item It then uses the resulting matches to compute the first upper bound on the standard deviation, $b^{(0)}_L$ (Line 2 in Algorithm \ref{Alg:MaxZILP}). 
\item The algorithm then creates a coarse mesh $b_1^{(iter)},\dots,b_L^{(iter)}$ where $b_1^{(iter)}<b_l^{(iter)}<b_L^{(iter)}$ (Line 3) new refined mesh will be created at each iteration, and we denote iterations by $iter$. We want the interval $[b_1^{(iter)},b_L^{(iter)}]$ to be wide enough to contain the true value of $f_2(\mathbf{a}^*):={\displaystyle\sumT\sumC(y_i^t-y_j^c)^2 a^*_{ij}},$ where $\mathbf{a}^*\in\argmax z(\mathbf{a})$, which we do not know and are trying to obtain. Determining which procedure to use to create this mesh is left up to the user, the $b_l^{(iter)}$ could be chosen evenly spaced, though they do not need to be. Note that the choice of $(b_1^{(iter)}, \dots, b_L^{(iter)})$ at each iteration does not affect the optimality of the solution, only the speed at which it is obtained. 
\item The algorithm then determines the sign of the maximal $z$-statistic, if negative (Line 17), it will try to maximize the denominator (variance) of the $z$ statistic, and if positive it will try to minimize it (Line 5). 
\item The algorithm then computes the solution to the ILP Formulation as well as upper and lower bounds for the solution using (\ref{ub}) (Lines 7-8 and 21-22).
\item We then do some re-indexing. We take the union of all grid points $\bigcup_{l,(iter)} b_l^{(iter)}$ created over all iterations, order them and create a ordered single vector $\mathbf{b}=[b_1,...,b_{L}]$. This is a single 1--D grid. On each point of this grid, we have match assignment $\a_l$ that maximizes $z$, subject to a constraint $b_l$ on the variance. 




\item For each $l$, the algorithm determines whether the interval $[b_{l-1},b_{l}]$ can be excluded because it provably does not contain a $f_2(\mathbf{a})$ value corresponding to the maximum value of $z(\mathbf{a})$ (Lines 9-12 or 22-25). In particular, we know from the bounds in (\ref{ub}) and from Theorem \ref{Thm:Bound} that if the upper bound on the objective for a particular $b_{l^{'}}$ is lower than all lower bounds for the optimal solution $\mathbf{a}^*$ then $l^{'}$ cannot equal $l^*$ and the interval $[b_{l^{'}-1},b_{l^{'}}]$ can be excluded from further exploration. Specifically, we check for each $l$ whether
$$\frac{\bar{d}_{\a_l}\sqrt{M}}{\sqrt{\frac{1}{M}b_{l-1} - (\bar{d}_{\a_l})^2 }}<\max_\ell \frac{\bar{d}_{\a_\ell}\sqrt{M}}{\sqrt{\frac{1}{M}b_{\ell} - (\bar{d}_{\a_\ell})^2 }}.$$
If this holds for some $l$, it means $l$ cannot equal $l^*$ and the interval $[b_{l-1},b_{l}]$ can be excluded from further exploration. 

\item The intervals that remain included after this procedure are then refined again at each iteration, thus creating finer and finer meshes, and the process repeats on these finer meshes until most intervals are excluded and the desired tolerance ($\epsilon$) is achieved.  
\item The output is a binary matrix of match assignments, $\mathbf{a}$, and the optimal value of $z(\mathbf{a})$. 
\end{itemize}

Correctness of the algorithm follows directly from optimality of the solutions at the bounds mentioned earlier and through use of the following theorem:

\begin{theorem}{(Optimal Solution of ILP with Upper Bound on Variance)}\label{Thm:Bound}
Let the functions $f_1$ and $f_2$ be real-valued functions of $x\in X$, and let $F(f_1(x), f_2(x))$ be monotonically increasing in $f_1$ and monotonically decreasing in $f_2$. Consider the optimization problem of finding $x^* \in \argmax_x F(f_1(x), f_2(x)),$ and assume we are given $[b_1, b_2,...,b_l,..,b_L]$ that span a wide enough range so that $x^*$ obeys: $b_{l^*-1}\leq f_2(x^*) \leq b_{l^*}\textrm{ for some } l^*\in \{1,...,L\}$.\\
\textbf{(1)}For $x_l\in \argmax\limits_{x:f_2(x)\leq b_l} F(f_1(x),b_l) = \argmax\limits_{x:f_2(x)\leq b_l} f_1(x),$ where the equality follows because $F$ monotonically increases in $f_1$, we have:
\begin{eqnarray*}
\max_l F(f_1(x_l),b_l)\leq \max_x F(f_1(x),f_2(x))\leq F(f_1(x_{l^*}),b_{l^*-1})\leq \max_l F(f_1(x_{l}),b_{l-1}).
\end{eqnarray*}
\textbf{(2)}For $x_l\in \argmin\limits_{x:f_2(x)\leq b_l} F(f_1(x),b_l) = \argmin\limits_{x:f_2(x)\leq b_l} f_1(x),$ we have:
\begin{eqnarray*}
\min_l F(f_1(x_l),b_l)\geq \min_x F(f_1(x),f_2(x))\geq F(f_1(x_{l^*}),b_{l^*-1})\geq \min_l F(f_1(x_{l}),b_{l-1}).
\end{eqnarray*}
\end{theorem}
\proof{Proof} See Appendix \ref{Section:BoundProof}. \endproof
This theorem bounds the optimal value of $F$ along the whole regime of $x$ in terms of the values computed at the $L$ grid points.
Note that the objective function of Formulations 1 and 2 is exactly of the form of Theorem \ref{Thm:Bound}, where
$$f_1(\mathbf{a})=\displaystyle\frac{1}{M}\sumT\sumC(y_i^t-y_j^c)a_{ij}, \;\;f_2(\mathbf{a})=\displaystyle\sumT\sumC(y_i^t-y_j^c)^2{a}_{ij}, \textrm{ and}$$
$$F(f_1(\mathbf{a}), f_2(\mathbf{a}))=\displaystyle\frac{f_1(\mathbf{a})\sqrt{M}}{\sqrt{\frac{1}{M}f_2(\mathbf{a})-(f_1(\mathbf{a}))^2}}.$$ Note that monotonicity of $F$ in $f_2$ is ensured in the algorithm by the conditions on the sign of the treatment effect at lines 5 and 17 of the algorithm. The extra constraints on $\mathbf{a}$ in the ILP Formulation are compatible with Theorem \ref{Thm:Bound}. Thus, the bounds in (\ref{ub}) are direct results of Theorem \ref{Thm:Bound} applied to the $z$ statistic as an objective function. Finally, minimization of $z$ can be achieved with Algorithm \ref{Alg:MaxZILP} by flipping the treatment indicator, i.e., calling control units as treated and treated units as control in the input to the algorithm, then running the procedure as is.

\section{Approximate and Exact Distributions of Robust Test Statistics}\label{Sec:Theory}

As introduced in Section \ref{sec:approach}, our method aims to find an upper bound on the p-value of a desired test statistic over the set of good matches. While not directly used, a quantity relevant to our robust tests is the distribution of the test statistics under the maximal and minimal match assignments themselves, that is the distribution of $\Psimax = \max_{\a \in \Agood} \Psia$, and $\Psimin = \min_{\a \in \Agood}\Psia$ which are random variables denoting a generic test statistic ($\Psi$) under the match assignment that maximizes and minimizes it respectively. While we do not employ these distributions to derive p-values for our tests, we still choose to study them here to better understand the statistical behavior of our robust tests. 

Specifically, we will study the distribution of the robust version of McNemar's test taking into account the randomness due to the choice of match assignment, in order to provide analysts with exact distributions for the robust statistics. 
Without Exclusively Binning Constraints as defined in Section \ref{SectionMcNemar}, the joint distributions can be extremely complex, but with these constraints, the calculation becomes much clearer. As a reminder, these are constraints that uniquely define a division of the units into strata, such that matches between two units in the same stratum always satisfy the quality constraints that define $\Agood$. To maintain exposition, we refrain from stating our results for this section formally (as the statements of these distributions are analytical but cumbersome), and instead give informal statements of our theorems. We refer to the appendix for rigorous versions of all our results and their proofs. This computation can be used to derive exact p-values for our robust McNemar's test.
Specifically, in the following sections, \textit{we show that analytical formulas exist for both null distributions we seek under Exclusively Binning Constraints, and that these formulas can be used to generate a lookup table for the distributions of $\chichi$ in polynomial time.}

\subsection{Exact Randomization Distribution of $\chichi$}
Let us compute the finite sample joint distribution of $\chichi$ under Fisher's sharp null hypothesis of no treatment effect, which is defined as, for all $i$: $$\Hsharp:= Y_i(1) = Y_i(0). $$ We compute this distribution for the case in which constraints are exclusively binning and $\mathcal{S}$  can be constructed prior to matching. In this case, potential outcomes for all of the units in our sample are fixed and nonrandom: the only randomness in our calculations will stem from the treatment assignment distribution.

Our approach to constructing this randomization distribution without conditioning on the match assignment is as follows: we assume that any units that could be matched together under a good assignment have the same probability of receiving the treatment, and that, after treatment is assigned, matches are made to compute $\chichi$ with the ILP in Formulation 1. This ILP is modified to include Exclusively Binning Constraints in addition to those in the formulation, and to make as many matches as possible within each of the strata defined by the constraints. In this setting, we assume that treatment assignment is random within the strata defined by the Exclusively Binning Constraint. This implies that Assumption~\ref{As:CondIgnorability} becomes:
\begin{assumption}{(Stratified Conditional Ignorability)}\label{As:StratIgnorability}
All units $i$ in stratum $l$ receive the treatment with fixed probability $e_l$. That is: for all $l = 1, \dots, L$: $\Pr(T_i=1|X = x_i, i \in S_l) = \Pr(T_i=1|i \in S_l) = e_l$, with $0 < e_l <1$.
\end{assumption}
Let us introduce some notation to represent outcome counts within strata. 
We use $U_l$ to denote the number of treated units in stratum $l$ with outcome equal to 1, $V_l$ to denote the number of treated units with outcome 0, $\eta_l$, the number of control units with outcome 1, and $\nu_l$ the number of control units with outcome 0. Note that, by $\Hsharp$, we have the number of units with outcome 1 and 0 fixed in each stratum, thus, for stratum $l$, we use $N_l^1$ and $N_l^0$ to denote the number of units with outcome 1 and 0 in stratum $l$ respectively. Under Assumptions \ref{As:SUTVA}-\ref{As:StratIgnorability}, these count variables follow the following generating process:
\begin{alignat}{1}
U_l|\Hsharp \overset{iid}{\sim} Bin(e_l, N_l^1),\quad &\eta_l = N_l^1 - U_l,\label{Eq:DGPSharp1a}\\
V_l|\Hsharp \overset{iid}{\sim} Bin(e_l, N_l^0),\quad &\nu_l = N_l^0 - V_l,\\
N_l^t = U_l + V_l,\quad & N_l^c = \nu_l + \eta_l. \label{Eq:DGPSharp1b}
\end{alignat}
One can view this data generation process as a results of our assumptions, including those of $\Hsharp$.
The distribution of $\chichi$ under $\Hsharp$ has the following properties:
\begin{theorem}{(Randomization Distribution of $\chichi$)}\label{Thm:Sharpa} Under Assumptions \ref{As:CondIgnorability}, \ref{As:SUTVA}, \ref{As:StratIgnorability}, and conditionally on $\Hsharp$:
\begin{itemize}
    \item For a given stratum, $l$, the joint distribution of $B_l$ and $C_l$ (the counts of optimal matched pairs) can be expressed as a function of the distribution of the count variables $U_l$, $V_l$, $\nu_l$, $N_l^t$, and $N_l^c$ defined in \eqref{Eq:DGPSharp1a}-\eqref{Eq:DGPSharp1b}.
    \item Given the signs of $\chichi$, for any two strata, $l$ and $l'$,  $B_l$ and $C_l$ are independent of  $B_{l'}$ and $C_{l'}$. 
    \item The joint distribution of $\chichi$ is the convolution of the joint distributions over $B_l, C_l$ with each other.
\end{itemize}

\end{theorem}
A precise formulation for the distribution of test statistics in the theorem above is given in Theorem \ref{Thm:Sharp} in Appendix \ref{App:Sec:RandDist}. 

\subsection{Exact Conditional Distribution of $\chichi$}
In this section, we seek a formulation for $\Pr(\chi^-=s, \chi^+ = r|X)$, the distribution of robust McNemar's statistics under exclusively binning constraints and a fixed number of treated units, $N_l^t$ in each stratum, but without conditioning on a specific match assignment. Potential outcomes are treated as random quantities in this case, and  treated and control outcomes are assumed to be equal only on average. This distribution permits testing of hypotheses other than $\Hsharp$, and is more general than the randomization distribution derived in the previous subsection.
Now denote with $N_l^t$ the number of treated units in the stratum $l$, and with $N_l^c$ the number of control units in that same stratum. In addition, let $p_l^t = \Pr(Y_i=1|i \in S_l, T_i=1)$, and $p_l^c = \Pr(Y_i=1|i \in S_l, T_i=0)$. For each stratum, $l=1, \dots, L$, the data are generated as follows:
\begin{alignat}{1}
U_l|N_l^t \overset{iid}{\sim} Bin(p_l^t, N_l^t)\label{Eq:ConditionalDGP1a}, &\quad V_l = N_l^t - U_l \\
\eta_l|N_l^c \overset{iid}{\sim} Bin(p_l^c, N_l^c), &\quad \nu_l = N_l^c - \eta_l.\label{Eq:ConditionalDGP2a}
\end{alignat}
As we did before, we can use the distributions of these count variables to arrive at a formulation for the joint distribution of $\chichi$ under the DGP just outlined.
\begin{theorem}{(Conditional Distribution of $\chichi$)}\label{Thm:Nulla}
Under Assumptions \ref{As:CondIgnorability}, \ref{As:SUTVA}, \ref{As:StratIgnorability}, and, for all strata, $l$, conditionally on $N_l^t, N_l^c$:
\begin{itemize}
    \item For a given stratum, $l$, the joint distribution of $B_l$ and $C_l$ (the counts of optimal matched pairs) can be expressed as a function of the distribution of the count variables $U_l$, $V_l$, $\nu_l$, $N_l^t$, and $N_l^c$ defined in \eqref{Eq:ConditionalDGP1a}--\eqref{Eq:ConditionalDGP2a}.
    \item Given the signs of $\chichi$, for any two strata, $l$ and $l'$,  $B_l$ and $C_l$ are independent of  $B_{l'}$ and $C_{l'}$. 
    \item The joint distribution of $\chichi$ is the convolution of the joint distributions over $B_l, C_l$ with each other.
\end{itemize}
\end{theorem}
A precise formulation for the distribution in the theorem above is in Theorem \ref{Thm:Null} in Appendix \ref{App:Sec:Null}. Note that this distribution can be used to test the null hypothesis of no effect in each stratum by requiring $p_l^t = p_l^c$ for all $l$ in the general formulation we provide in the theorem. 

\subsection{Fast Computation of Lookup Tables for $\chichi$}
Another important advantage of the theorems given in the last two sections is that they permit us to \textit{generate lookup tables for the finite-sample null distributions of interest in polynomial time}. Let $M = \sum_{l=1}^L M_l$ be the total number of matches made. A na\"ive procedure for computing a lookup table for the distribution in Theorems \ref{Thm:Sharpa} and \ref{Thm:Nulla} for a given dataset based on brute force enumeration  has computational complexity $O(2^{N^M + (N^M)^{2}})$, because naively computing the domain of these distributions is a version of the subset-sum problem. 

Fortunately, the complexity of computing the distribution of $\chichi$ under both $\Hzero$ and $\Hsharp$ can be significantly reduced because by Thms. \ref{Thm:Nulla} and \ref{Thm:Sharpa}, these distributions are convolutions of simpler distributions. We take advantage of this fact and of existing fast convolution algorithms to establish the following result:
\begin{theorem}\label{Thm:DFT}
There exists an algorithm that creates a probability table for the null distribution of $\chichi$ given in Theorems \ref{Thm:Sharpa} and \ref{Thm:Nulla} in $O(4M N^{4M}\log N)$.
\end{theorem}
We give proof of this theorem and outline the algorithm in question in the appendix. Note that the theorem directly implies that the worst-case time of computing the exact distributions of $\chichi$ is polynomial in $N$. Figure \ref{Fig:Distributions} in Appendix \ref{App:Sec:SimDist} shows marginal distributions of $\chichi$ computed with the algorithm in Theorem \ref{Thm:DFT}, and shows that there is a location and scale difference between the distribution of $\chichi$  when the null is true and when it is not. This demonstrates our tests' capacity to detect full-sample treatment effects.


\section{Simulations}\label{Sec:Simulations}
We present a set of experiments on simulated datasets that demonstrated the usefulness and performance of our method in a series of settings. We compare the performance of both our robust tests against several popular matching methods and find that our proposed tests tend to display better performance (lower Type-I error) in several cases, without sacrificing statistical power.

In all our simulations, we pre-specify the number of units $N$, number of treated units $N^t$, and number of covariates $P$. We will vary $N$ throughout, however we will keep $N^t$ at 20\% of the whole sample and $P=20$ unless otherwise specified (simulations with varying $P$ and constant $N$ are available in the appendix). We then generate data as follows, for $i=1,\dots,N$:
\begin{align*}
    x_{i1}, \dots, x_{iP} &\sim \text{Uniform}(-1, 1),\quad \epsilon_i \sim \text{Normal}(0, 1).
\end{align*}
Propensity scores are then simulated according to: 
\begin{align*}
    e_i &= \frac{1}{1 + \exp(- x_i\beta_0)},
\end{align*}
where $\beta_0$ is a $P$-dimensional vector of coefficients, which are pre-specified before data generation. Treatment $t_i$ is then set to 1 for exactly $N_t$ units, such that each unit has $t_i=1$ with probability $e_i$. We then introduce two different data generating processes for our outcome variable:
\begin{align*}
\text{Linear: } & y_i^* = t_i \tau + x_i\beta + \epsilon_i\\
\text{Complex: } & y_i^* = t_i \tau + x_i\beta + x_i^2 \beta_2 + \sin(x_i)\beta_3 + \ind[x_{i1} > 0] \beta_4 + \epsilon_i, 
\end{align*}
To simulate binary data for McNemar's test we then set $y_i = \ind[y_i^* > 0]$, and to simulate continuous data for the Z-test we set $y_i = y_i^*$. The coefficient vectors $\beta_0, \dots, \beta_4$, as well as $\tau$, are calibrated at each simulation so that the population ATT has a Cohen's d-statistic (defined as: $\frac{\E[Y(1) - Y(0)|T=1]}{Var[Y(1) - Y(0)|T=1]}$) of a pre-specified value. In most settings, we will simulate data for $d = 0, 0.2, 0.5, 0.8, 1.2, 2$, which are commonly understood to correspond to null, tiny, weak, moderate, strong, and very strong treatment effects. For each setting, we simulate 1000 datasets and compute the proportion in which each method rejects the null hypothesis of 0 treatment effect at the 5\% level as our estimate of the rejection rate of each method. 

We compare against several commonly employed matching methods, as well as benchmark approaches such as a completely na\"ive test without any matching, and an idealized hypothesis test conducted on both potential outcomes, which is never possible in practice. We summarize the methods included in our simulation studies in Table \ref{tab:simmethods}.

\begin{table}[!htbp]
    \centering
    \resizebox{\textwidth}{!}{
    \begin{tabular}{r|l|l}
        \hline\hline
        Name & Description & Citation\\
        \hline
        robust & Robust test     & This paper    \\
        \hline
        true & True potential outcomes & Benchmark \\
        naive & No matching at all & Benchmark \\
        \hline
        pscore & Propensity score matching  & \citep{Rosenbaum1984}\\
        l2 & Matching on the $L_2$ distance of the covariates & \citep[e.g.,][]{abadie2006large}\\
        optimal & Optimal matching via Mixed Integer Programming & \citep{Zub2012}\\
        \hline\hline
    \end{tabular}}
    \caption{Methods used in the simulation studies.}
    \label{tab:simmethods}
\end{table}

To ensure the closest possible comparison, we constrain all methods to match all treated units to one control unit without replacement, that is $M=N^t$. Given that one of the main points of our paper is that there is not clear guidance on how to select hyperparameters for any given matching method, we try to use default or author-recommended values whenever possible for all the other methods tested. When this is not possible, we set hyperparameters to mimic those of the robust test as much as possible. 

In all simulations, the set of good matches considered by our tests will be defined by all those matches that satisfy a mean balance constraint, as well as a caliper on the propensity score distance of the units matched. These constraints are implemented in our MIPs as follows:
\begin{itemize}
     \item \textit{Mean Balance}: $|\frac{1}{M}\sum_{i=1}^{N^t}\sum_{j=1}^{N^c}a_{ij}(x^t_{ip} - x^c_{jp})| \leq \epsilon \times (\sigma(x_p^t)/2 + \sigma(x_p^c)/2)$, where $\sigma_p(x_{p}^t)$, $\sigma_p(x_{p}^c)$ are the standard deviation of covariate $x_p$ in the treated and control set respectively.  ($P$ constraints total).
    \item \textit{Caliper}: $a_{ij}|e^t_i - e^c_j| \leq \epsilon$ for $i = 1, \dots N^t$, $j = 1, \dots, N^c$, where $e^t_i$ is the propensity score for treated unit $i$, and $e^c_j$ the analogue for control unit $j$ ($N^t \times N^c$ constraints total).
\end{itemize}
The value of $\epsilon$ is chosen to be the smallest feasible one on a grid between 0.01 and 1. 

\subsection{Results}

We first present results from comparison of our robust McNemar's test against the other methods. For all methods, p-values were computed after matching using the conventional formula for exact p-values for McNemar's test \citep[see][]{tam}. 

\begin{figure}[!htbp]
    \centering
    \includegraphics[width=\textwidth]{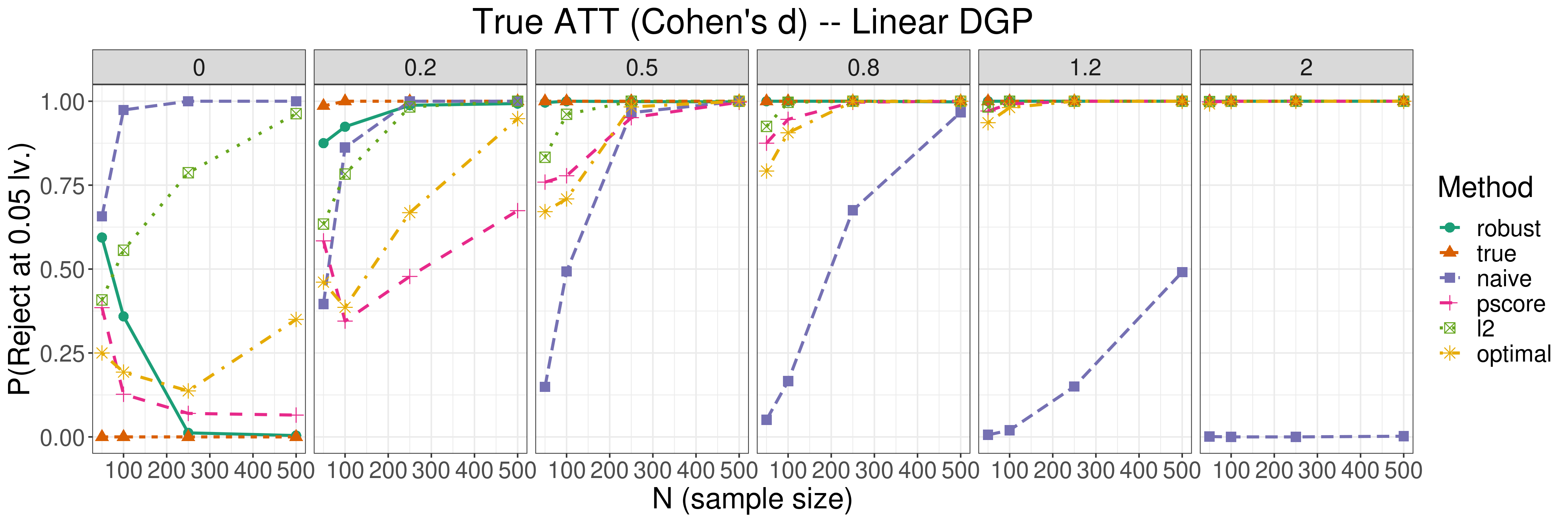}
    \includegraphics[width=\textwidth]{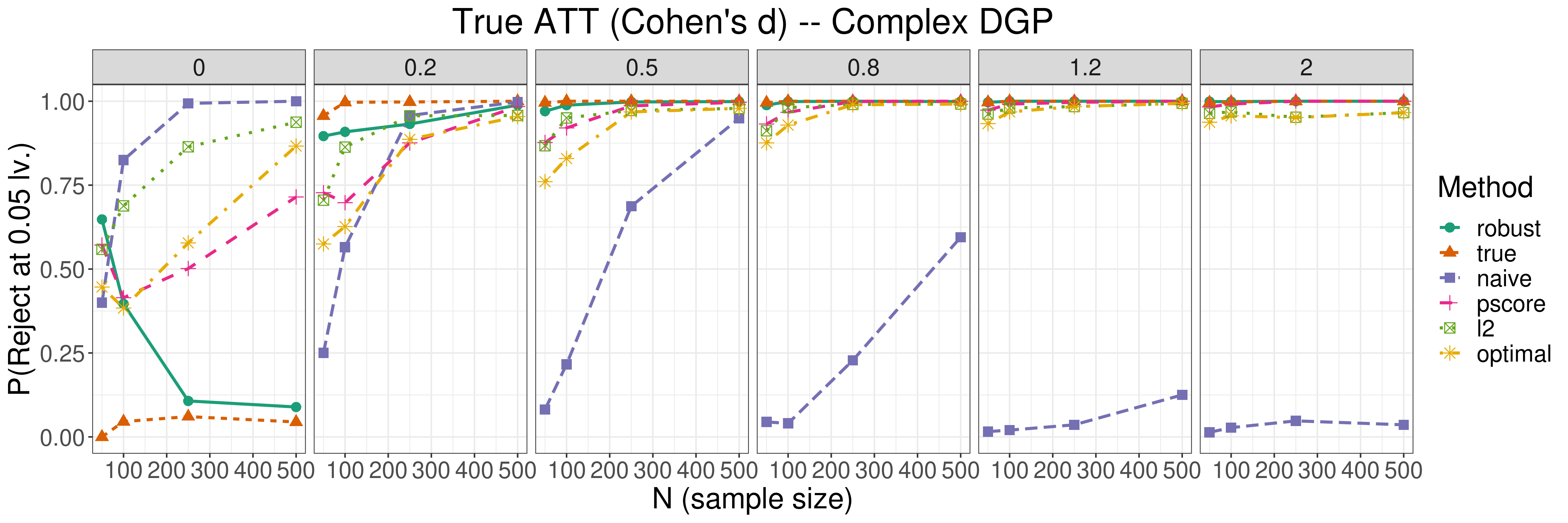}
    \caption{Comparison of performance of robust McNemar's test against several other matching methods. First row: linear outcome DGP, second row: complex outcome DGP. Ideally, a test would not reject the null when the ATT is 0 (leftmost panel), and the probability of rejection would increase as the ATT becomes stronger (four rightmost panels). The ideal method is labelled ``true'' and represented by the orange dashed line in the figure. }
    \label{fig:McNComp}
\end{figure}

Results are reported in Figure \ref{fig:McNComp}. We see that the robust test performs well as sample size grows: it displays low probability of rejection when the treatment effect is 0, and probability of rejection grows as the strength of the treatment also grows. When there is no ATT in the population, the robust test does display an error rate, but this is only when the sample size is low ($\leq$100), and the error rate decreases as the sample size grows, eventually  converging with the one for the idealized McNemar's test performed with the true potential outcomes. This is not true for methods we compare against, which seem to display an \textit{increase} in error rate (incorrect rejection) as the sample size increases. When an ATT is present in the population, the robust test displays statistical power comparable to or better than that of other matching methods, indicating that our bounding approach is not too extreme, given that matches are chosen well. 
\begin{figure}[!htbp]
    \centering
    \includegraphics[width=\textwidth]{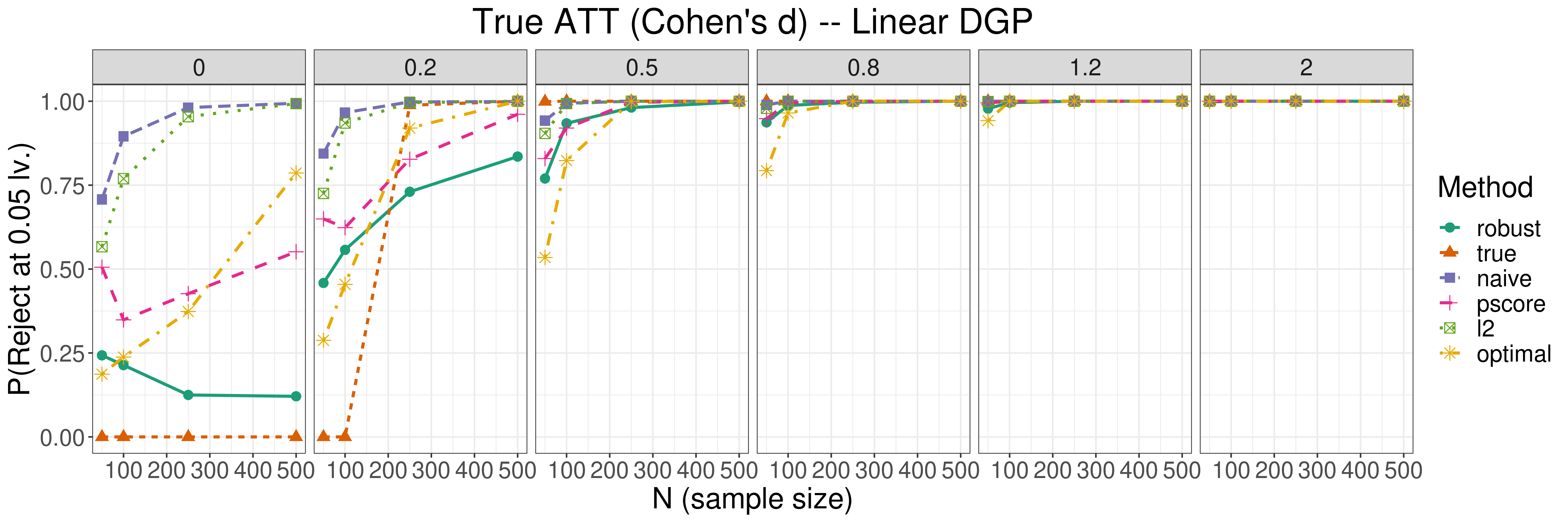}
    \includegraphics[width=\textwidth]{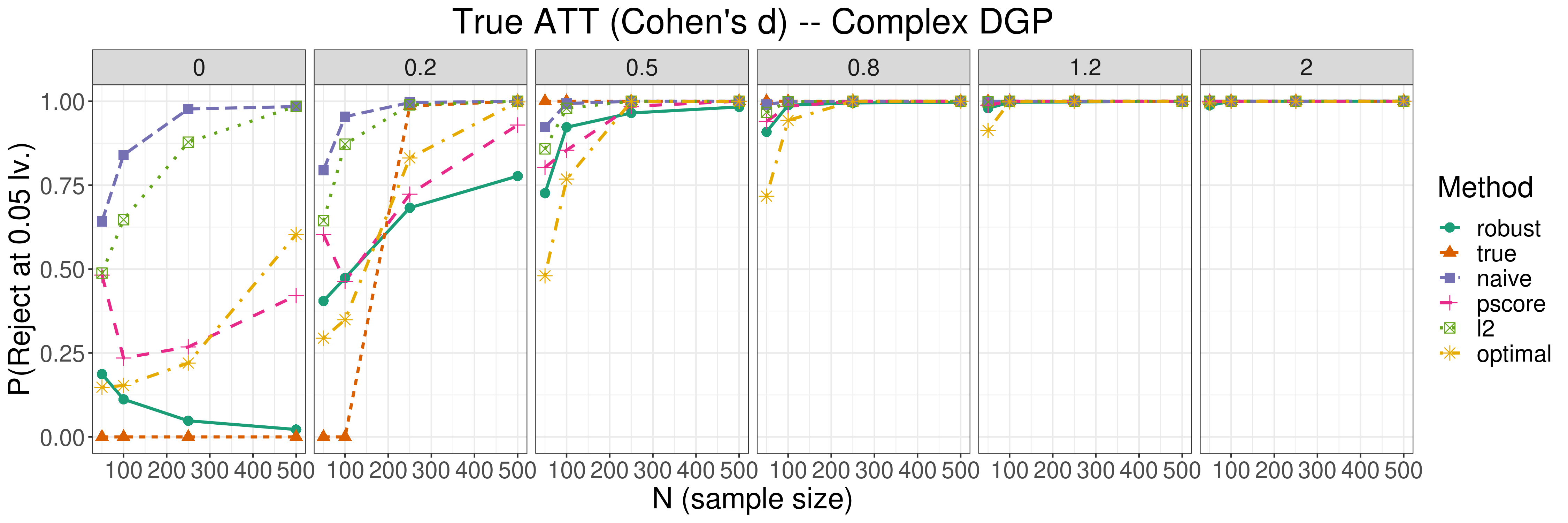}
    \caption{Comparison of performance of robust Z test against several other matching methods. First row: linear outcome DGP, second row: complex outcome DGP. The ideal method is labelled ``true'' and represented by the dark orange dashed line in the figure. }
    \label{fig:ZComp}
\end{figure}
Figure \ref{fig:ZComp} reports results for the robust Z-test in the simulated data settings introduced previously. This version of our robust test performs even more strongly than the robust McNemar's test, compared to other prominent matching methods. In this case, the robust z-test displays a good rejection rate even at low sample values ($N=50$) when the true ATT is 0, and is well powered as the ATT increases in strength, as well as when sample size grows. 

\section {Case Studies}\label{SectionCaseStudies}
In this section we apply our proposed methods to two real world datasets. We show that our test statistics can produce robust results in both datasets, which suggests that our methods can have wide practical applicability. Some of our results show that null hypotheses can be rejected robustly, while in some other cases they show that there is not enough evidence to reject the null hypothesis once the additional uncertainty from the choice of matching procedure is quantified via our robust tests.

\subsection{Case Study 1: The Effect of Smoking on Osteoporosis in Women}
In this case study we used GLOW (\emph {Global Longitudinal study of Osteoporosis in Women}) data used in the study of \cite{glow}. Each data point represents a patient and the outcome is whether or not the person developed a bone fracture. The treatment is smoking. We match on several pre-treatment covariates: age, weight, height, and BMI. As there are several more control than treated units, we test $\Hzero$ for the ATT with the general formulation of McNemar's test introduced before, by matching almost all of the treated units.

\begin{figure}[htbp]
\begin{center}
\includegraphics[width=0.4\linewidth]{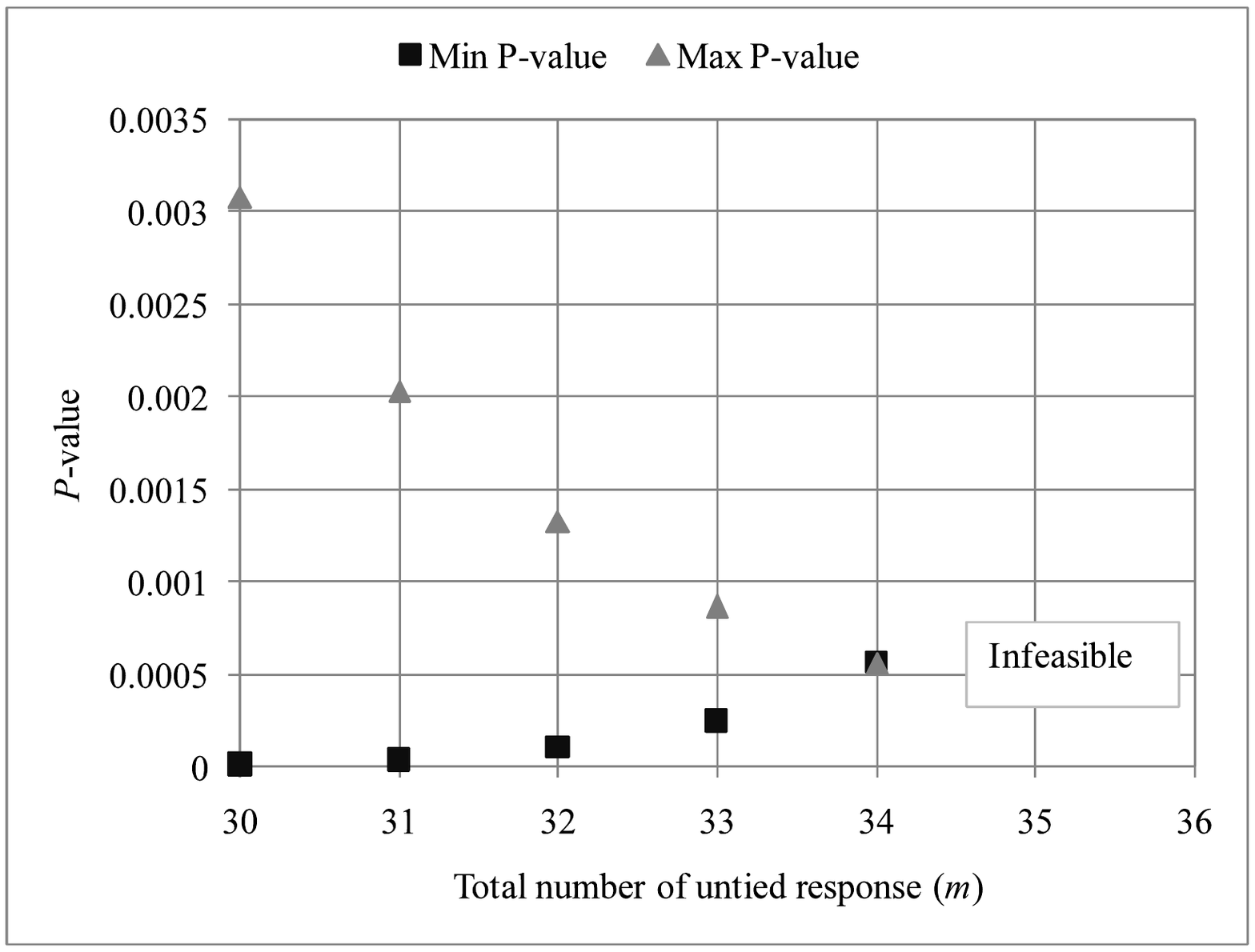}
\caption{(Case Study 1) Variation of McNemar's test p-values for different $m$. As long as there are enough matched groups, then for \textit{any} reasonable choice of match assignment, the null hypothesis is rejected.}
\label{Fig:GLOWMaP}
\end{center}
\end{figure}

Figure \ref{Fig:GLOWMaP} shows results for testing $\Hzero$ with the general program in Formulation 1, without binning constraints. We include a balance constraint in the formulation, by requiring that matched units $i,j$ respect $dist_{ij} \leq 0$, where $dist_{ij}$ is 0 if the sum of the differences of all the covariates is 6 or less and 1 otherwise. We choose the value 6 because this is the smallest caliper on the absolute distance between units that still permits all treated units to be matched. This figure lends evidence to the fact that $\Hzero$ can be rejected robustly when the match assignment induces better balance in the data. We can conclude this because \textit{both min and max p-values are below 0.01 at each value of $M$, signifying that results would be statistically significant and positive using any good quality match assignment}. 

This case study shows the ability of our tests to detect effects when they are robust and present, even with a small number of observations.

\subsection{Case Study 2: The Effect of Mist on Bike Sharing Usage}
In our second case study, we used 2 years (2011-2012) of bike sharing data from the Capital Bike Sharing (CBS) system \citep[see][]{cbs} from Washington DC. We study the effect of misty weather on the number of bikes rented. Control covariates include Season, Year, Workday, Temperature, Humidity and Wind Speed. Additional information on this case study is available in Appendix \ref{App:Sec:CS2}.

Figures \ref{Fig:BsMax} and \ref{Fig:BsMin} show the upper and lower bounds for the maximum objective function value for different $b_l$ with \emph{n}=30 for the maximization problem in Figure \ref{Fig:BsMax} and minimization problem in Figure \ref{Fig:BsMin}. These figures illustrate the meshes at different scales within the algorithm. We computed p-values for $Z^+$ and $Z^-$ under several counts of matched pairs: \textit{M}=30, 50, 70, 90, 110. For $M = 30$ through $M=90$, the p-value for $Z^+$ was 0 and the p-value for $Z^-$ was 1, while both p-values become 1 when the number of matches is 110. This is shown in Figure \ref{Figure7}. The problem becomes infeasible for a larger number of matches. This illustrates that there is a lot of uncertainty associated with the choice of match assignment -- that is, \textit{a reasonable experimenter choosing 90 matched pairs can find a p-value of $\sim$~0 and declare a statistically significant difference while another experimenter can find a p-value of $\sim$~1 and declare the opposite}. In this case it is truly unclear whether or not mist has an effect on the total number of rental bikes. Figure \ref{Figure7} shows robust p-values for different amounts of matched units. 

\section{Conclusion}

Believing hypothesis test results conducted from matched pairs studies on observational data can be perilous. These studies typically ignore the uncertainty associated with the choice of matching method, and in particular, how the experimenter or an algorithm chooses the matched groups: we have given both simulated and real-data evidence of this problem. We want to know that for \textit{any} reasonable choice of experimenter who chooses the assignments, the result of the test would be the same.
In this work, we have addressed the issue above by introducing robust test statistics that consider extrema over all possible good matches. This is justified because p-values obtained for the extrema must, by definition, include p-values obtained under any other possible good match. We have provided practical implementations of this principle for both discrete and continuous data, as well as theoretical and empirical analyses of the performance of our methods. 


\begin{figure}[!bp]
\centering
\begin{subfigure}{.43\textwidth}
  \centering
  \includegraphics[width=\linewidth]{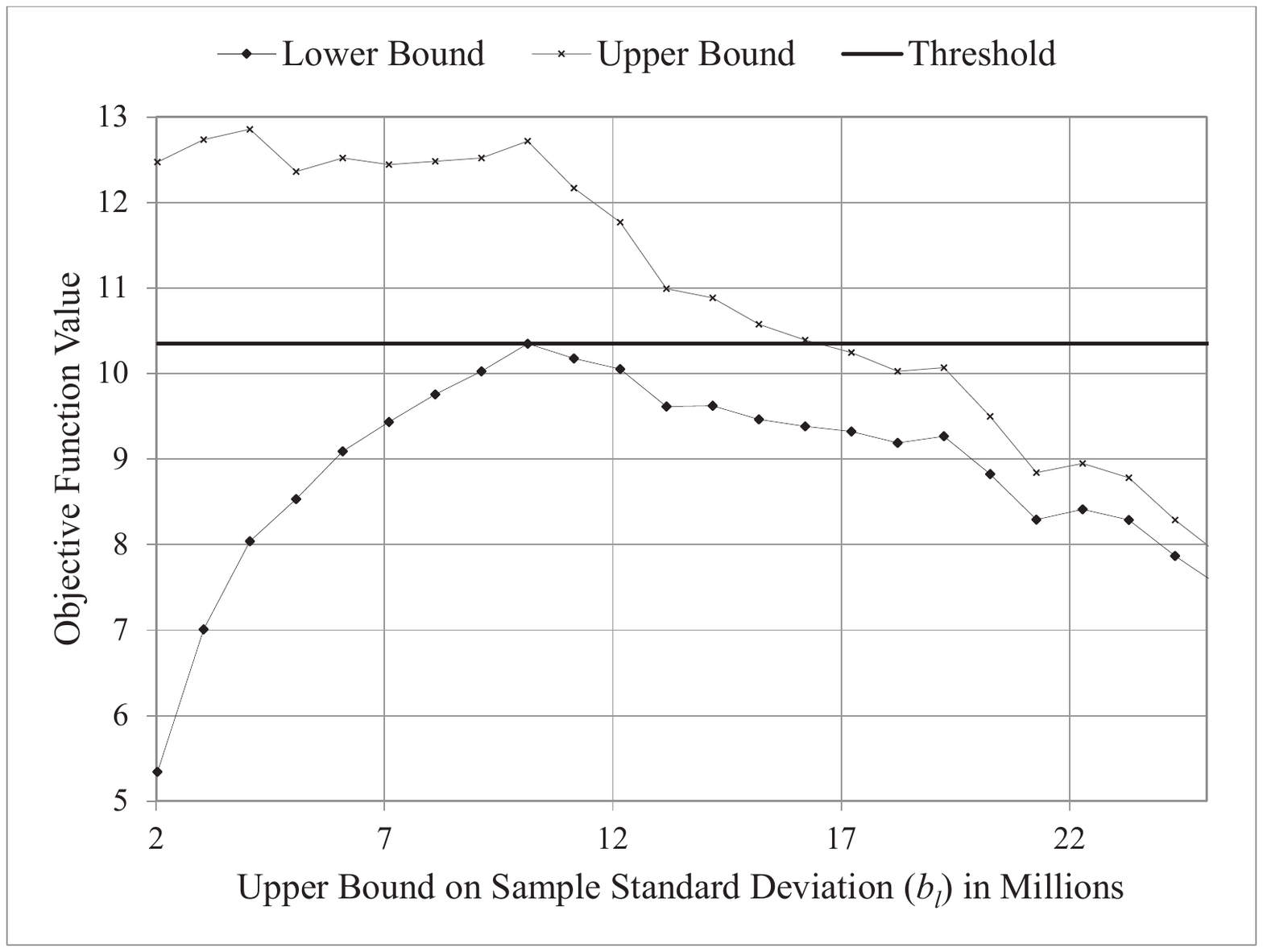}
  \caption{Initial Mesh}
  \label{Fig:BsMax:P1}
\end{subfigure}%
\begin{subfigure}{.43\textwidth}
  \centering
  \includegraphics[width=\linewidth]{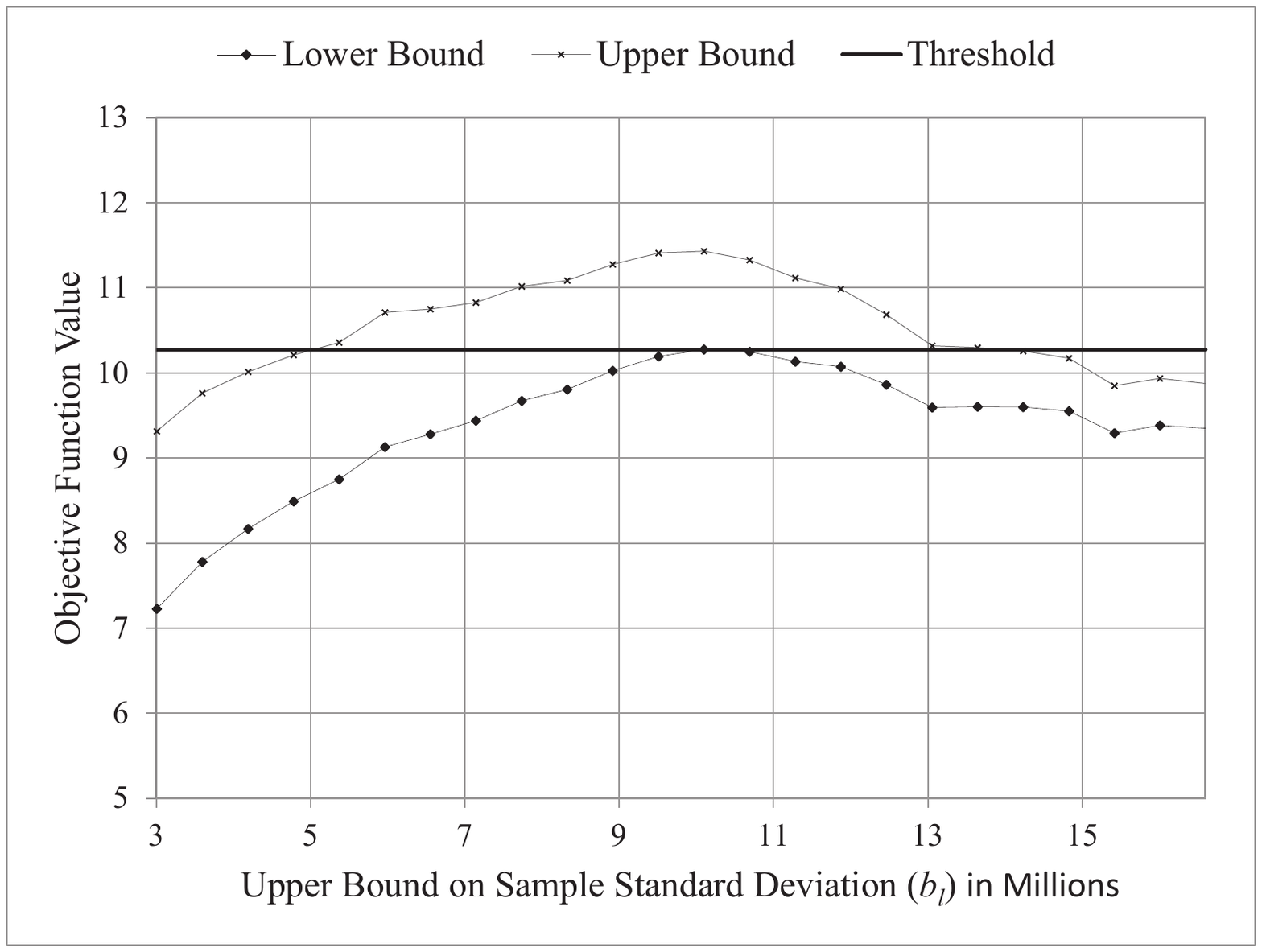}
  \caption{Refined Mesh}
  \label{Fig:BsMax:P2}
\end{subfigure}
\begin{subfigure}{.43\textwidth}
  \centering
  \includegraphics[width=\linewidth]{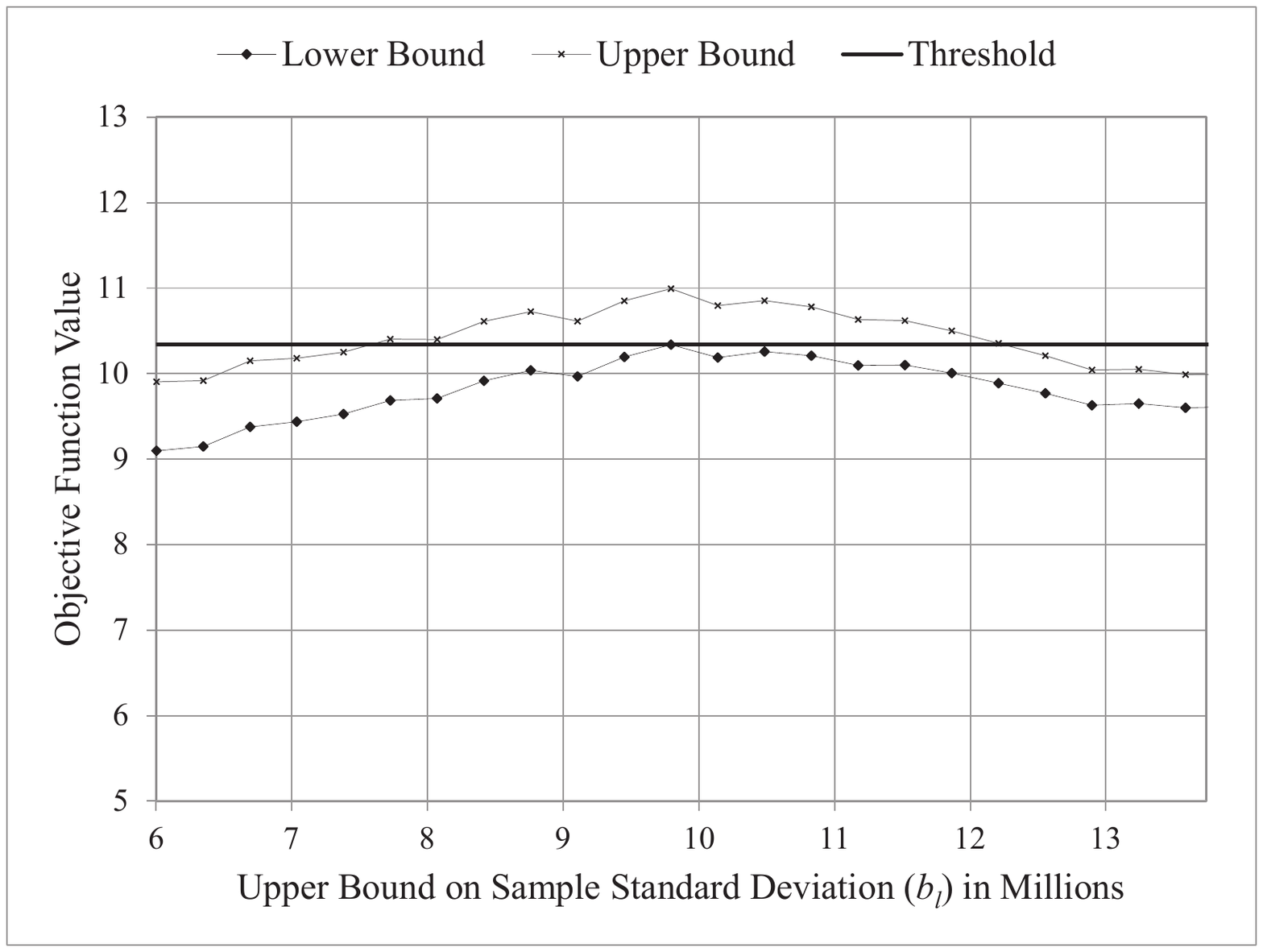}
  \caption{Finest Mesh}
  \label{Fig:BsMax:P3}
\end{subfigure}
\caption{Upper and lower bounds for maximum z-test objective function value over a range of $b_l$ (Bike Sharing data, N=45), illustrating the optimum search range at various steps in Algorithm \ref{Alg:MaxZILP}. The final value with 3 iterations is shown in Panel \ref{Fig:BsMax:P3}. The final value for the maximization problem is between 10.34 and 10.99.}
\label{Fig:BsMax}
\end{figure}

\begin{figure}[!bp]
\centering
\begin{subfigure}{.43\textwidth}
  \centering
  \includegraphics[width=\linewidth]{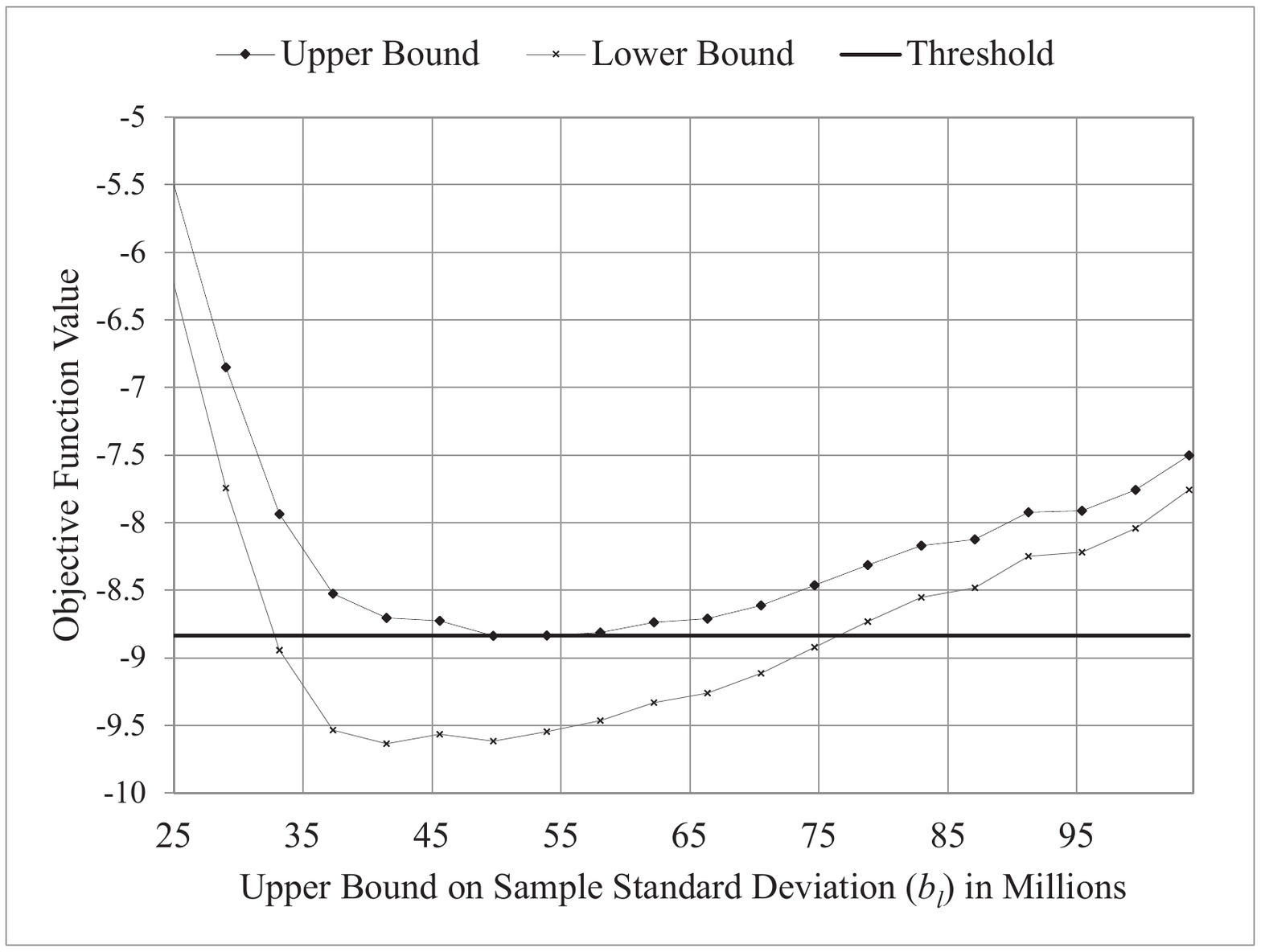}
  \caption{Initial Mesh}
  \label{Fig:BsMin:P1}
\end{subfigure}%
\begin{subfigure}{.43\textwidth}
  \centering
  \includegraphics[width=\linewidth]{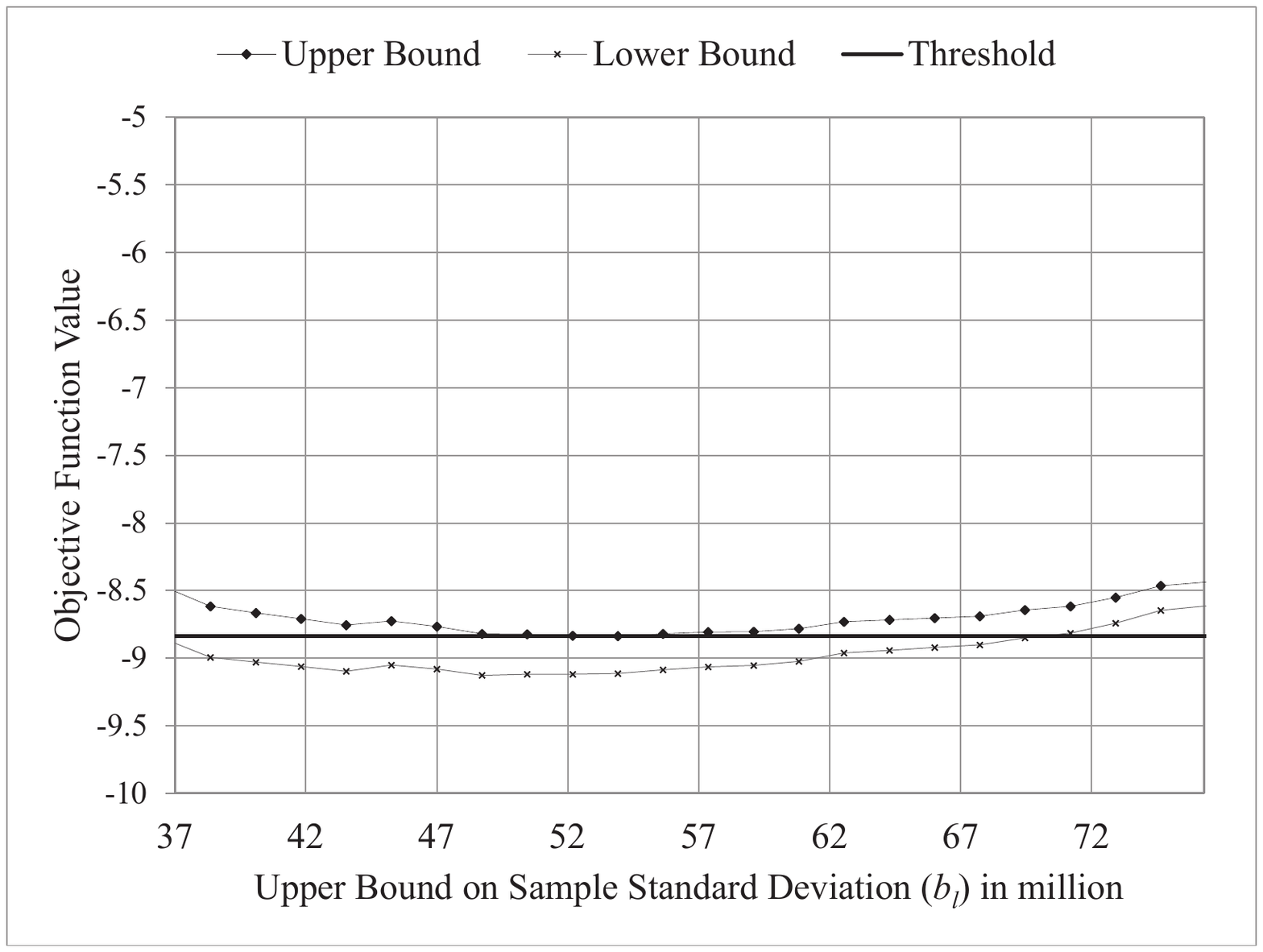}
  \caption{Refined Mesh}
  \label{Fig:BsMin:P2}
\end{subfigure}
\begin{subfigure}{.43\textwidth}
  \centering
  \includegraphics[width=\linewidth]{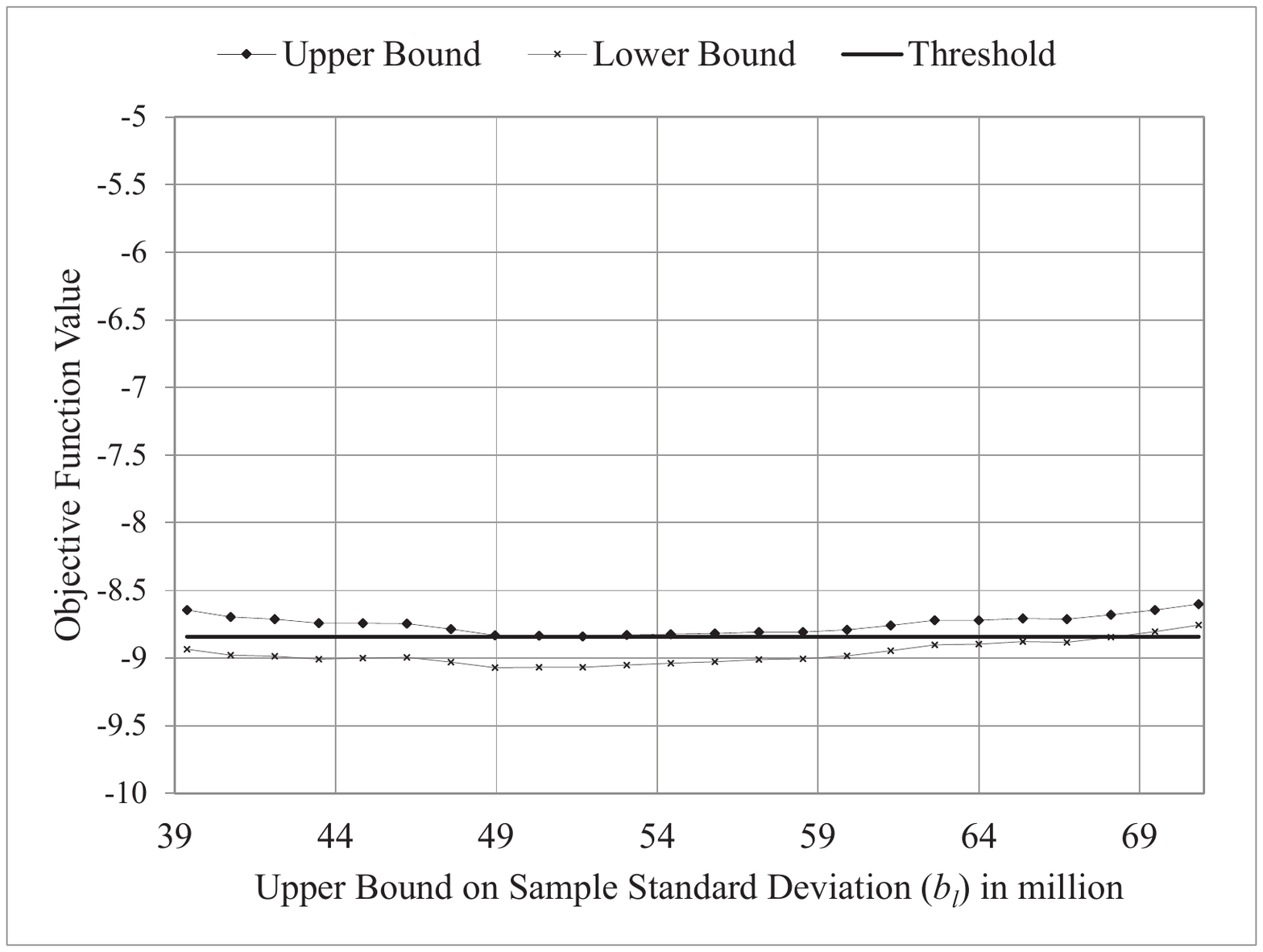}
  \caption{Finest Mesh}
  \label{Fig:BsMin:P3}
\end{subfigure}
\caption{Upper and lower bounds for minimum z-test objective function value over a range of $b_l$ (Bike Sharing data, N=90), illustrating the optimum search range at various steps in Algorithm \ref{Alg:MaxZILP}. The final value with 3 iterations is found in Panel \ref{Fig:BsMax:P3}. The final value for the minimization problem is between  -8.84 and -9.06.}
\label{Fig:BsMin}
\end{figure}

\begin{figure}[htbp]
\begin{center}
\includegraphics[height=2in]{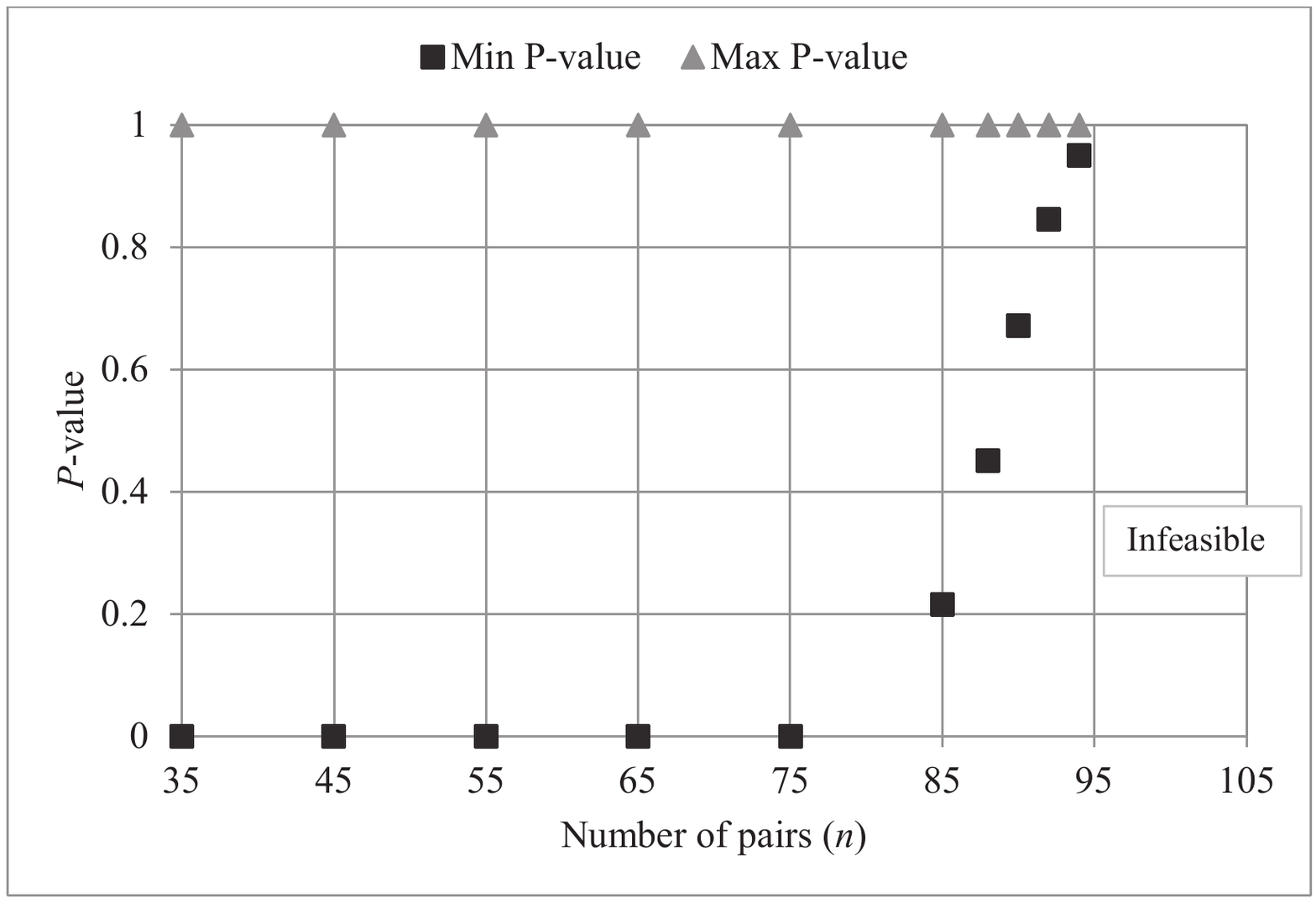}
\caption{Variation of $z$-test optimum \emph{P}-values for different \emph{N}. (Case Study 2) \label{Figure7}}
\end{center}
\end{figure}

%

\textbf{Acknowledgement} The authors express their gratitude to the Natural Sciences and Engineering Research Council of Canada (NSERC) for partial financial support of this research.

%
%
%

\bibliographystyle{plainnat} 
\bibliography{reference} 

\pagebreak

\part*{Appendices}
\renewcommand{\thesection}{Appendix \Alph{section}: }
\setcounter{section}{0}

\section{Constraining the Quality of the Matches}\label{App:Sec:Constraints}
Determining the quality requirements that matches should satisfy to be usable is ultimately up to the experimenter, however there are several general types of constraints that matches in the uncertainty set $\Agood$ should most often obey. Let $dist(x, x')$ be a metric on the space of $x$, some of these constraints are:
\begin{itemize}
\item (Calipers) When $\a(i)\neq \emptyset$ then $\textrm{dist}(x_i^t,x_{\a(i)}^c)\leq\epsilon$.
\item (Covariate balance, mean of chosen treatment units similar to mean of control group) $\forall \textrm{ covariates } p$ we have:
\begin{eqnarray}
\biggl|\frac{1}{M}\sum_{\{i:\a(i)\neq\emptyset\}} x_{ip}^{t} -  \frac{1}{M}\sum_{\{j:\a(j)\neq\emptyset\}}x^{c}_{jp}\biggr|\leq\epsilon_p.\label{Eq:ATEBalance}
\end{eqnarray}
\item (Maximizing the fitness of the matches) In general, one can optimize any measure of user-defined fitness for the assignment, and then constrain $\Agood$ to include all other feasible assignments at or near that fitness level, by including the following constraints:
\[
\textrm{Fitness}(\a,\{x_i^t\}_{i=1}^{N^t},\{x_i^c\}_{i=1}^{N^c})\geq \textrm{Maxfit}-\epsilon,
\]
where Maxfit is precomputed as: $\textrm{Maxfit}=\max_{\a\in\A}\textrm{Fitness}(\a,\{x_i^t\}_{i=1}^{N^t},\{x_i^c\}_{i=1}^{N^c}). $
If one desires the range of results for all maximally fit pairs and no other pairs, $\epsilon$ can be set to 0. \end{itemize}
We note that, to constitute a ILP formulation, the chosen type of fitness function must be expressed as some linear function of the matching indicators. While it is technically true that not all fitness functions can be encoded in this way, many popular constraints for matching can be. For example, \cite{Zub2012} gives linear constraint formulations for a number of popular match quality constraints. However, despite the fact that many match assignments may exist that satisfy these constraints, these methodologies produce only one, and it is not chosen uniformly at random from any known distribution.

In what follows, we provide special cases of the Robust Procedure for two specific hypothesis tests, McNemar's test for binary outcomes and the $z$-test for real-valued outcomes. We outline strategies for computing these statistics as well as their distributions under the hypotheses of interest. We begin with McNemar's test.

\section{Proof of Theorem \ref{Thm:Bound}}\label{Section:BoundProof}
\proof{Proof} We prove the result for statement (1), as the proof of statement (2) is exactly symmetrical. Because the $b_l$'s are defined on a pre-specified grid, we know the maximum value of $F$ may not occur at one of the grid points. Since by definition of $x_l$ we know that $f_2(x_l)\leq b_l$, and since $F$ is decreasing in its second argument, we have $F(f_1(x_l),b_l)\leq F(f_1(x_l),f_2(x_l))$ for each $l$, and taking a max over all $l$:
\begin{equation*}
\max_{l\in 1...L} F(f_1(x_l),b_l)
\leq \max_{l\in 1...L} F(f_1(x_l),f_2(x_l))\leq \max_x F(f_1(x),f_2(x)).
\end{equation*}
This is the left inequality of the bound. The rest of the proof deals with the right inequalities.
First, it is true that:
\begin{equation}
\displaystyle f_1(x^*)=\max_{x:f_2(x)=f_2(x^*)}f_1(x).\label{fstareqn}
\end{equation}
If this were not true then either $\displaystyle f_1(x^*)<\max_{x:f_2(x)=f_2(x^*)}f_1(x)$ or $\displaystyle f_1(x^*)>\max_{x:f_2(x)=f_2(x^*)}f_1(x)$.
If the first inequality were true then
\begin{equation*}
F(f_1(x^*),f_2(x^*))< F\left(\max_{x:f_2(x)=f_2(x^*)} f_1(x), f_2(x^*)\right),
\end{equation*}
which contradicts the definition of $x^*$. The second option also cannot be true as we know there exists a solution $x^*$ so that the maximum is attained with $f_1$ and $f_2$ values $f_1(x^*)$ and $f_2(x^*)$. So we can say that (\ref{fstareqn}) holds.

From (\ref{fstareqn}) and using $l^*$ defined in the statement of the theorem, we can derive:
\begin{eqnarray}\label{xstareqn}
f_1(x^*)=\max_{x:f_2(x)=f_2(x^*)}f_1(x) \leq \max_{x:f_2(x)\leq f_2(x^*)} f_1(x) \leq \max_{x:f_2(x)\leq b_{l^*}} f_1(x)=f_1(x_{l^*}) ,
\end{eqnarray}
where we used that the set $\{x:f_2(x)=f_2(x^*)\}$ is smaller than the set $\{x:f_2(x)\leq f_2(x^*)\}$ which is smaller than $\{x:f_2(x)\leq b_{l^*}\}$, since $f_2(x^*)\leq b_{l^*}$ by definition of $l^*$. Thus, $f_1(x^*)\leq f_1(x_{l^*})$.
Now,
\begin{eqnarray*}
F(f_1(x^*),f_2(x^*)) &\leq & F(f_1(x^*),b_{{l^*}-1})\\
&\leq & F(f_1(x_{l^*}),b_{l^*-1}) \\
&\leq & \max_l F(f_1(x_{l}),b_{l-1}).
\end{eqnarray*}
Here the first inequality above follows from the definition of $l^*$, $b_{l^*-1}\leq f_2(x^*)$, and the fact that $F$ decreases in the second argument. The second inequality comes from (\ref{xstareqn}) and the fact that $F$ is increasing in its first argument. The third inequality follows from taking a maximum over all $l$ rather than using $l^*$.
The proof is complete. \endproof

\section{Algorithms that maximize and minimize $\chi$ under Exclusively Binning Constraints}\label{App:Sec:MaxMcN}

In the following sections, we offer exact formulas and computational algorithms for the distribution of the test statistic values obtained with the results in this section, and when matches are considered as part of the statistic.

If the constraints on $\Agood$ are exclusively binning, then $\chi$ can be optimized quickly and easily in both directions if the partition $\mathcal{S}$ is constructed first. Before showing this, we introduce an IP formulation for McNemar's test with exclusively binning constraints.

As a reminder, we are in the following situation: we have $N^t$ treated units and $N^c$ control units measured on P features, $X$ that take value in a finite set, $\mathcal{X}$. All these units also have an outcome $Y \in \{0,1\}$ and treatment $T \in \{0, 1\}$. Since the constraints on the optimization problem are exclusively binning, then based on the values of $X$ we group the units into $L$ strata such that each stratum $S_1, \dots, S_L$ contains $N_l^t$ and $N_l^c$ units. In each stratum there will be $U_l$ treated units with outcome $Y_i=1$, $\eta_l$ control units with outcome $Y_j=1$, $V_l = N_l^t-U_l$ treatment units with outcome 0, and $\nu_l = N_l^c - \eta_l$ control units with outcome 1.  We then would like to create pairs of units within each stratum, such that each pair contains exactly one treated unit ($T_i=1$) and one control unit ($T_j=0$). Once we have created the pairs we compute:
\begin{itemize}
\item $A_l$: the number of matched pairs in stratum $l$ such that the treated unit has $Y=1$ and the control unit has $Y=0$. We refer to these pairs as $A$-pairs.
\item $B_l$: the number of matched pairs in stratum $l$ such that the treated unit has outcome 1 and the control unit 0. We refer to these pairs as $B$-pairs.
\item $C_l$, the number of matched pairs in stratum $l$ such that the treated unit has outcome $0$ and the control unit 1. We refer to these pairs as $C$-pairs.
\item$D_l$, the number of matched pairs in stratum $l$ such that both the treated and control units have outcome $0$, referred to as $D$-pairs.
\end{itemize}
The following tables summarize the data and the statistics we are interested in:\\
\begin{figure}[h]
\centering
\begin{tabular}{c|cccc}
&  \multicolumn{3}{c}{$Y$}\\
\hline
 \multirow{4}{*}{$T$} & & 1 & 0 \\
&  1 & $U_l$ & $V_l$\\
&  0 & $\eta_l$ & $\nu_l$\\
\hline
\end{tabular} $\qquad$
\begin{tabular}{c|cccc}
&  \multicolumn{3}{c}{$Y_c$}\\
\hline
 \multirow{4}{*}{$Y_t$} & & 1 & 0 \\
&  1 & $A_l$ & $B_l$\\
&  0 & $C_l$ & $D_l$\\
\hline
\end{tabular}
\end{figure}
We then take the sums of $B_l$ and $C_l$ across all strata to obtain $B = \sum_{l=1}^L B_l$ and $C = \sum_{l=1}^LC_l$. Finally, we use these two quantities to compute:
$$\chi = \frac{TE - 1}{\sqrt{SD + 1}} = \frac{B - C - 1}{\sqrt{B + C + 1}}.$$
We would like to make the matches within each stratum such that $\chi$ is either maximized or minimized, assuming that we must match as many units as possible. Throughout the rest of this document we use a $+$ superscript to denote values of $A$, $B$, $C$, $D$ and $\chi$ output by maximizing $\chi$ and a $-$ superscript to denote the corresponding values obtained by minimizing it. We limit our analysis to the case in which the maximum number of feasible matches needs to be achieved, that is, exactly $M = \sum_{l=1}^L\min(N_l^t, N_l^c)$ must to be made:\\

\noindent\textbf{Formulation 2: IP formulation for McNemar's test with exclusively binning constraints}
\begin{equation*}
          \text{Maximize/Minimize}_{\mathbf{a}}\quad \chi(\mathbf{a})=\left[\frac{B(\a)-C(\a)-1}{\sqrt{B(\a)+C(\a)+1}}\right]
      \end{equation*}
\noindent subject to:
\begin{align}
       \sumT\sumC a_{ij}y_i^t(1-y_j^c)&=B(\a)  &&\textrm{(Total number of first type of discordant pairs)} \label{defB}\\
       \sumT\sumC a_{ij}y_j^c(1-y_i^t)&=C(\a)  && \textrm{(Total number of second type of discordant pairs)} \label{defC}\\
       \sumT a_{ij} &\leq 1     &&     \forall j  \quad \textrm{(Match each control unit at most once)} \label{onlyt1}\\
       \sumC a_{ij} & \leq 1     &&   \forall i  \quad \textrm{(Match each treatment unit at most once)}  \label{onlyc1}\\
       \sum_{i \in S_l}\sum_{j \in S_l}a_{ij} &= \min(N_l^t, N_l^c)\qquad&&\forall l \quad \textrm{(Make as many matches as possible)}\label{(maxmatches)}\\
       a_{ij}& \in \{0,1\}  &&   \forall i,j \quad \textrm{(Defines binary variable $a_{ij}$)}\\
        \lefteqn{\textrm{(Additional user-defined \textbf{exclusively binning} constraints.)}}
      \end{align}
Note that analysts can remove this part of the formulation and introduce a fixed number of matches as constraint in a way similar to Constraint \eqref{Eq:F1:response} if desired.

This problem can be solved in linear time and without one of the canonical IP solution methods by using the fact that the strata defined by the exclusively binning constraints can each be optimized separately, once the direction of the resulting statistic is known. In stratum $S_l$ there will be $U_l$ treated units with outcome $Y_i=1$, $\eta_l$ control units with outcome $Y_j=1$, $V_l = N_l^t-U_l$ treatment units with outcome 0, and $\nu_l = N_l^c - \eta_l$ control units with outcome 1. This is summarized in Figure~\ref{Tab:Summary}.
\begin{figure}
\centering
\begin{tabular}{c|cccc}
&  \multicolumn{3}{c}{$Y$}\\
\hline
 \multirow{4}{*}{$T$} & & 1 & 0 \\
&  1 & $U_l$ & $V_l$\\
&  0 & $\eta_l$ & $\nu_l$\\
\hline
\end{tabular} $\qquad$
\begin{tabular}{c|cccc}
&  \multicolumn{3}{c}{$Y_c$}\\
\hline
 \multirow{4}{*}{$Y_t$} & & 1 & 0 \\
&  1 & $A_l$ & $B_l$\\
&  0 & $C_l$ & $D_l$\\
\hline
\end{tabular}
\caption{Summary of pair counts in stratum $l$.}
\label{Tab:Summary}
\end{figure}
To ensure that as many units as possible are matched, within each stratum we make exactly $M_l = \min(N_l^t, N_l^c)$ matches. We would like to make the matches within each stratum such that $\chi$ is either maximized or minimized. Algorithm \ref{Alg:MaxMcN} maximizes $\chi$,
\begin{algorithm}[!htbp]
\small
\caption{Maximize $\chi$ with Exclusively Binning Constraints}
\label{Alg:MaxMcN}
  \SetAlgoLined
  \KwData{Positive integer vectors $(U_1, \dots, U_L)$, $(V_1, \dots, V_L)$, $(\eta_1, \dots, \eta_L)$, $(\nu_1, \dots, \nu_L)$}
  \KwResult{Maximal $\chi$ statistic value over all possible matches in $\Agood$.}
  \For{$l=1, \dots,L$}{
    $M_l = \min(N_l^t, N_l^c)$\\
    $U^+_l := U_l - \max(U_l - N_l^c, 0)$\\
    $V^+_l := M_l - U_l^+$\\
    $\eta_l^+ :=\max(\eta_l - \max( N_l^c - N_l^t, 0), 0)$\\
    $\nu_l^+ := M_l - \eta_l^+$
  }
  $TE = \sum_{l=1}^L U_l^+ - \eta_l^+$\\
  \If{$TE \geq 0$}{
    \For{$l=1, \dots,L$}{
        $A_l^+ := \min(U_l^+,  \eta_l^+)$\\
        $D_l^+ := \min(\nu_l^+, V_l^+)$\\
        $B_l^+ := \min(U_l^+- A, \nu_l^+ -  D)$\\
        $C_l^+ := \min(\eta_l^+ - A, V_l^+ - D)$\\
    }
  }
  \Else{
    \For{$l=1, \dots,L$}{
        $B_l^+ := \min(U_l^+, \nu_l^+)$\\
        $C_l^+ := \min(\eta_l^+, V_l^+)$\\
        $A_l^+ := \min(U_l^+ - B,  \eta_l^+ - C)$\\
        $D_l^+ := \min(\nu_l^+ - B, V_l^+ - C)$\\
    }
  }
  \Return $\frac{\sum_{l=1}^L B_l^+ - C_l^+ - 1}{\sqrt{\sum_{l=1}^L B_l^+ + C_l^+ + 1}}$
\end{algorithm}

\begin{algorithm}[!htbp]
\small
  \SetAlgoLined
  \KwData{Positive integers: $U, V, \eta, \nu$}
  \KwResult{Positive integers: $A, B, C, D$}
  Make $B := \min(U, \nu)$ $B$-pairs.\\
  Make $C := \min(\eta, V)$ $C$-pairs.\\
  Make $A := \min(U - B,  \eta - C)$ $A$-pairs.\\
  Make $D := \min(\nu - B, V - C)$ $D$-pairs.\\
  Return $A, B, C, D$
  \caption{ComputeMaximizedSD}
  \label{App:Alg:MaxSD}
\end{algorithm}

\begin{algorithm}[!htbp]
\small
  \SetAlgoLined
  \KwData{Positive integers: $U, V, \eta, \nu$}
  \KwResult{Positive integers: $A, B, C, D$}
  Make $A := \min(U,  \eta)$ $A$-pairs.\\
  Make $D := \min(\nu, V)$ $D$-pairs.\\
  Make $B := \min(U- A, \nu -  D)$ $B$-pairs.\\
  Make $C := \min(\eta - A, V - D)$  $C$-pairs.\\
  Return $A, B, C, D$.
  \caption{ComputeMinimizedSD}
  \label{App:Alg:MinSD}
\end{algorithm}

\begin{algorithm}[!htbp]
\small
  \SetAlgoLined
  \KwData{Positive integer vectors $(U_1, \dots, U_L)$, $(V_1, \dots, V_L)$, $(\eta_1, \dots, \eta_L)$, $(\nu_1, \dots, \nu_L)$}
  \KwResult{Maximal $\chi$ statistic value}
  \For{$l=1, \dots,L$}{
    $M_l := \min(N_l^t, N_l^c)$\\
    $U^+_l := U_l - \max(U_l - N_l^c, 0)$\\
    $V^+_l := M_l - U_l^+$\\
    $\eta_l^+ :=\max(\eta_l - \max( N_l^c - N_l^t, 0), 0)$\\
    $\nu_l^+ := M_l - \eta_l^+$
  }
  $TE^+ = \sum_{l=1}^L U_l^+ - \eta_l^+$\\
  \If{$TE^+ \geq 1$}{
    \For{$l=1, \dots,L$}{
      $(A_l^+, B_l^+, C_l^+, D_l^+) := ComputeMinimizedSD(U_l^+, V_l^+, \eta_l^+, \nu_l^+)$
    }
  }
  \Else{
    \For{$l=1, \dots,L$}{
      $(A_l^+, B_l^+, C_l^+, D_l^+) := ComputeMaximizedSD(U_l^+, V_l^+, \eta_l^+, \nu_l^+)$
    }
  }
  \Return $\chi^+ = \frac{\sum_{l=1}^L B_l^+ - C_l^+ - 1}{\sqrt{\sum_{l=1}^L B_l^+ + C_l^+ + 1}}$
  \caption{Maximize $\chi$ with exclusively binning constraints.}
  \label{App:Alg:MaxMcN}
\end{algorithm}
\pagebreak
\begin{algorithm}[!htbp]
\small
  \SetAlgoLined
  \KwData{Positive integer vectors $(U_1, \dots, U_L)$, $(V_1, \dots, V_L)$, $(\eta_1, \dots, \eta_L)$, $(\nu_1, \dots, \nu_L)$}
  \KwResult{Maximal $\chi$ statistic value}
  \For{$l=1, \dots,L$}{
    $M_l := \min(N_l^t, N_l^c)$\\
    $U^-_l := \max(U_l - \max(N_l^t - N_l^c, 0), 0)$\\
    $V^-_l := M_l - U_l^-$\\
    $\eta_l^- := \eta_l - \max(\eta_l - N_l^t, 0)$\\
    $\nu_l^- := M_l - V_l^-$
  }
  $TE^- = \sum_{l=1}^L U_l^- - \eta_l^-$\\
  \If{$TE^- \geq 1$}{
    \For{$l=1, \dots,L$}{
      $(A_l^-, B_l^-, C_l^-, D_l^-) := ComputeMaximizedSD(U_l^-, V_l^-, \eta_l^-, \nu_l^-)$
    }
  }
  \Else{
    \For{$l=1, \dots,L$}{
      $(A_l^-, B_l^-, C_l^-, D_l^-) := ComputeMinimizedSD(U_l^-, V_l^-, \eta_l^-, \nu_l^-)$
    }
  }
  \Return $\chi^- = \frac{\sum_{l=1}^L B_l^- - C_l^- - 1}{\sqrt{\sum_{l=1}^L B_l^- + C_l^- + 1}}$
  \caption{Minimize $\chi$ with exclusively binning constraints.}
  \label{App:Alg:MinMcN}
\end{algorithm}

\subsection{Correctness of the Optimization Algorithms}\label{App:MaxMcNProof}
As it is clear from their definitions, these algorithms do not require any of the conventional MIP solving techniques and as such are much faster: the running time of Algorithm \ref{Alg:MaxMcN} is clearly linear in $N$, the number of units. This constitutes a substantial speed up over solving the problem with a regular MIP solver. The following theorem states the correctness of the algorithm for solving Formulation 2:
\begin{theorem}{(Correctness of Algorithm~\ref{Alg:MaxMcN})}\label{Thm:MaxMcN}
Algorithm~\ref{Alg:MaxMcN} globally solves Formulation 2 for the Maximal value of $\chi$.
\end{theorem}

What follows is a proof of the algorithms' correctness. The proof is structured into 4 different claims and a theorem equivalent to Theorem~\ref{Thm:MaxMcN} following directly from these claims. Before introducing these claims it is useful to summarize the various ways in which units can be matched and unmatched to lead to different pair types: the cells in  Table~\ref{TablePair} show what pairs can be created by unmaking two other pairs and matching their units across. For example, if we unmake an $A$-pair and a $D$-pair, we are left with a treated and a control unit with $Y=1$ and a treated and a control unit with $Y=0$: if we match them across we obtain a $B$ and a $C$-pair. In the rest of this document, we refer to this operation of unmaking two pairs and matching across their units as exchanging pair one with pair two.
\begin{table}[h]
\centering
\begin{tabular}{l|llll}
& A & B & C & D\\
\hline
A&A,A&A,B&A,C&B,C\\
B&A,B&B,B&A,D&B,D\\
C&A,C&A,D&C,C&C,D\\
D&B,C&B,D&C,D&D,D\\
\hline
\end{tabular}
\caption{What pairs can be created by exchanging matches. Cells are the resulting pairs when a pair in the left margin is exchanged with a pair in the top row.}
\label{TablePair}
\end{table}
Note also that, if the number of treated units to be matched equals the number of control units, and all units are matched in some way, the only way we can change those matches is by performing one or more of the operations detailed in the table. This is, of course, only possible if the required pairs are present among the existing matches for example, we cannot unmake a $A$-pair if there are none made already. For claims 1-3, assume that any two units can be matched together. Since we don't explicitly consider the different strata in these claims, we omit the $l$ subscript from the notation.

\begin{claim}\label{CDiscard}
Suppose we want to maximize $\chi$ subject to the constraint that we must make as many matches as possible, that is we must make $M = \min(N^t, N^c)$ matches, and suppose that $N^c > N^t$ so that $M = N^t$. Then it is always optimal to leave unmatched $\min(N^c - N^t, \eta)$ control units with outcome 1 and $\max(N^c - N^t - \eta, 0)$ control units with outcome 0. Suppose instead that $N^t \geq N^c$, then it is always optimal to leave unmatched $\min(N^t - N^c, V_l)$ treated units with outcome 0 and $\max(U_l - N_l^c, 0)$ treated units with outcome 1.
\end{claim}

\proof{Proof.} We will show that, if $N^c >N^t$ and we must make $N^t$ matches, then it is optimal to first leave unmatched as many control units with outcome 1 as possible, that is $\min(N^c - N^t, \eta)$, and, if after these units have been excluded, there still are more control than treated units, to leave unmatched the remaining control units with outcome 0. Suppose initially that $N^c = N^t + 1$, fix $\chi  = \frac{B - C -1}{\sqrt{B + C}}$ and assume that there are exactly $N^t-1$ already matched treatment units and at least two leftover control units. Since we must make exactly $N^t$ matches we can only match the leftover treatment unit with one of the two controls.\\
Assume that the two control units are $u_1$ with outcome $Y_{u_1} =1$ and $u_0$ with $Y_{u_0} = 0$. There are two possible scenarios: first, the currently unmatched treatment unit has outcome 1: if we match it with $u_0$ we get a $B$-pair with corresponding value of $\chi$:
$$\chi_{u_0} = \frac{B - C}{\sqrt{B + C + 2}},$$
if we match the treatment unit with $u_1$  we get a $A$-pair, and:
$$\chi_{u_1} = \frac{B - C - 1}{\sqrt{B + C + 1}}.$$
With some algebra we can see that, $\chi_{u_0} \geq \chi_{u_1}$, which implies that we always gain more from matching a treated unit with outcome 1 to a control unit with outcome 0. In this case, leaving $u_1$ unmatched is the optimal choice. 

Now suppose that the leftover treatment unit has outcome 0. Then, if we match it with $u_0$ we form a $D$-pair and get:
$$\chi_{u_0} = \frac{B - C - 1}{\sqrt{B + C +1}},$$
no change from the initial $\chi$. If we match this treatment unit with $u_1$ we have formed a $C$-pair instead get:
$$\chi_{u_1} = \frac{B - C - 2}{\sqrt{B + C + 2}},$$
again some algebra reveals that $\chi_{u_0} \geq \chi_{u_1}$: the value of $\chi$ is maximized by choosing to match the treatment unit with $u_0$ in this case as well. This shows that, when there is a choice of multiple control units to match with a treatment unit the control with outcome 1 should always be left out if we wish to maximize $\chi$.

We now show that, if $N^t = N^c + 1$ then it is always optimal to leave unmatched a treated unit with outcome 0 instead of one with outcome 1 if $\chi$ is to be maximized. Let there be one unmatched control unit and two candidate treatment units for it to be matched with, $u_1$ such that $Y_{u_1}^t = 1$ and $u_0$ such that $Y_{u_0} = 0$. Let the value of $\chi$ without those units be $\chi = \frac{B + C  -1}{\sqrt{B + C + 1}}$. If the unmatched control has outcome 1 and we match it to $u_0$ we end up with a $C$-pair and an updated value of $\chi$ equal to:
$$\chi_{u_0} = \frac{B - C - 2}{\sqrt{B + C + 2}}.$$
If we instead match the control unit with outcome 1 to $u_0$ we produce an $A$-pair, and the following value of $\chi$:
$$\chi_{u_1} = \chi.$$
Then, with some algebra we can see that $\chi_{u_0} \leq \chi_{u_1}$, implying that, in order to maximize $\chi$, the optimal strategy is to leave $u_0$ unmatched. If the control unit to be matched has outcome 0 and we match it with $u_0$ we get a $D$-pair and
$\chi_{u_0} = \chi.$
If we instead match the control unit with outcome 0 to $u_1$ we get a $B$-pair and:
$$\chi_{u_1} = \frac{B - C}{\sqrt{B + C + 2}}.$$
Again, with some algebra we see that $\chi_{u_1} \geq \chi_{u_0}$, implying that leaving $u_0$ unmatched is optimal in this case as well. This shows that leaving treated units with outcome 0 unmatched if $N^t = N^c + 1$ is the optimal strategy to maximize $\chi$.

For the case in which $N^t = N^c + k$ we can proceed by induction on the number of matched control units: assume inductively that the optimal choice at $k - 1$ is to leave unmatched a treated unit with outcome 0. At match $k$ we will have a value $\chi_k$ and one unmatched control unit and two candidate treatments. By the above, the value of $\chi_k$ that we would get by leaving the treated unit with outcome 0 unmatched is always larger than the one we would get by leaving the unit with outcome 1 unmatched. Because of this it is optimal to leave the treated unit with outcome 0 unmatched also for the $k^{th}$ match. This proves that, regardless of the difference between the number of treated and control units, it is optimal to leave as many treated units with outcome 0 as possible unmatched. If $N^t \geq N^c$, the largest possible number of treated units with outcome 0 that can be left unmatched is clearly $\min(N^t - N^c, V)$, either we exhaust the difference between $N^t$ and $N^c$ by not matching treatment units with outcome 0, or we exhaust all $V$ treated units with outcome 0 and still have leftover treated units that don't have a match in the control group. In this second case, we will have to make up this difference by leaving unmatched treated units with outcome 1 in excess: the precise amount of which is $N^t - N^c - V = N^t - N^c - N^t + U = U - N^c$. A symmetrical argument shows that, if $N^c = N^t + k$ it is always optimal to leave as many control units with outcome 1 as possible unmatched, before leaving units with outcome 0 unmatched, and that the largest possible amount of control units with outcome 1 that can be left unmatched in this case is $\min(N^c - N^t, \eta)$, and the amount of control units with outcome 0 is $N^c - N^t - \eta$, in case all $\eta$ control units with outcome 1 are left unmatched without being able to exhaust the difference between unmatched units. 

Finally, note that the proof for the case in which we wish to minimize $\chi$ is exactly symmetrical to this one.\endproof

\begin{claim}\label{CTE}
If $N^t = N^c = M$, and exactly $M$ matches must be made, then $B^+ - C^+ = B^- - C^-  = U - \eta$ independently of how units are matched.
\end{claim}
\proof{Proof.} Let $W$ be the matching in which all units are paired in a way such that $B=\min(U, \nu)$ and $C=\min(\eta, V)$. Since $M = U + V = \eta + \nu$, then $U \geq \nu \iff \eta \geq V$, so it must be that either: $B = U, C=\eta$ or $B = V, C=\nu$. By definition of $V$ and $\nu$ we have that in both cases: $B - C = U - \eta$. Now we will prove that this equality must hold for any other match in which all units are matched. Let $B$ be the number of $B$-pairs created with that matching and $C$ the number of $C$-pairs. Consider any other match $W' \neq W$ also satisfying the fact that all $M$ treatment and control units are matched and let $B'$ and $C'$ be the counts of $B$ and $C$-pairs generated by $W'$. Since exactly the same number of units are matched in $W$ and $W'$ it must be that there exists some sequence of exchange operations that, if applied to $W$ generates $W'$. We now proceed by induction on $k$, the number of operations applied to $W$ to get to $W'$, starting with $k=1$. Note first, by Table~\ref{TablePair} that the only operations that can alter the number of $B$ and $C$-pairs are: exchanging an $A$-pair with a $D$-pair, obtaining a $B$-pair and a $C$-pair, and exchanging a $B$-pair with a $C$-pair, obtaining an $A$-pair and a $D$-pair. In the first case, we unmake an $A$-pair and a $D$-pair in $W$ and use those units to make a $B$ and a $C$-pair in $W'$ so the respective counts are now $B' = B+1$ and $C'=C+1$, which implies that $B'-C' = B-C$. In the other case we have $B' = B-1$, $C' = C-1$ and $B'-C' = B-C$. Finally, suppose inductively that $B^{(k)} - C^{(k)} = B-C$  after the $k$th exchange operation. Again, by Table~\ref{TablePair} the only operations that can alter the counts of $B$ and $C$ are exactly the two discussed in the base case ; and by the same reasoning, we conclude that $B^{(k+1)} - C^{(k+1)} = B - C$. \endproof

\begin{claim}\label{CSD}
If $N^t = N^c = M$ and exactly $M$ matches must be made,  then Algorithms~\ref{App:Alg:MaxSD} and~\ref{App:Alg:MinSD} respectively make the matches that maximize and minimize $B + C$.
\end{claim}
\proof{Proof.} Consider Algorithm~\ref{App:Alg:MinSD} first. We will only show the statement for this algorithm as the proof of the correctness of the other algorithm is exactly symmetrical to this. Let $W^*$ be the match output by this algorithm and let $A^*, B^*, C^*, D^*$ be the respective pair counts under $W^*$. By lines 1-4 of Algorithm~\ref{App:Alg:MinSD} we know that:
\begin{align*}
A^* &= \min(U, \eta)\\
D^* &= \min(V, \nu)\\
B^* &= \min(U - A^*, \nu - D^*)\\
C^* &= \min(\eta - A^*, V - D^*)
\end{align*}
By definition of $U, V, \eta, \nu$, $A^*$ and $D^*$ are the largest possible number of $A$ and $D$-pairs that can be made with the given units. It can be seen from the definitions above that one of $C^*$ or $B^*$ will always be 0: in case $A^* = U$ then $U - A^* = 0$ and $B^* = \min(U - A^*, \nu - D^*) = 0$. In case $A^* = \eta$ we have $\eta - A^* = 0$ and $C^* = \min(\eta - A^*, V-D^*) = 0$ by consequence. Since exactly $M$ units must be matched, and $N^t = N^c = M$ by assumption, then the only operations allowed to change the matches from $W^*$ are those in table~\ref{TablePair}, as there are no leftover units unmatched. Note that the only operation in the table that would allow for a decrease in $B$ and $C$ is exchanging$B$ with $C$, but this operation can never be performed on $W^*$ as either it contains no $C$-pairs or no $B$-pairs. Then it must be that $B^*$ and $C^*$ are the smallest number of $B$ and $C$-pairs that can be made with the existing units and thus that they minimize $B+C$. \endproof

\subsubsection{Proof of Theorem \ref{Thm:MaxMcN}}
\proof{Proof.} Consider first the problem of maximizing $\chi$. The observations are divided into $l$ strata such that $N_l^t$ treated units and $N_l^c$ control units are in each stratum: we can only match units that are in the same stratum. If we denote the count of $B$-pairs and $C$-pairs in stratum $l$ with $B_l$ and $C_l$ then the objective function for the problem is $$\chi = \frac{TE}{\sqrt{SD}} = \frac{\sum_{l=1}^L(B_l - C_l) - 1}{\sqrt{\sum_{l=1}^L B_l + C_l + 1}} .$$ As $TE$ is a separable function of $B_l$ and $C_l$, optimizing the former equals maximizing each of the latter individually, the same is true for $SD$. Note first, that, by the form of the objective function, we must make as many matches as possible. This number of matches is exactly $M = \sum_{i=1}^L M_l = \sum_{i=1}^L \min(N_l^t, N_l^c)$ because, by constraints \eqref{Eq:F1:onlyt1} and \eqref{Eq:F1:onlyc1}, each treatment and control unit can only be matched once. By Constraint \eqref{(maxmatches)} we know that exactly $M_l = \min(N_l^t, N_l^c)$ matched pairs must be constructed in each stratum, therefore, units in excess of $M_l$ must be discarded in each stratum. By Claim~\ref{CDiscard} if there are more control units to be matched than treated units in one stratum it is always optimal to leave unmatched the amounts of units described in Claim~\ref{CDiscard} in each stratum, independently of how matches are made in other strata by separability of $TE$. The algorithm does this explicitly when defining $U_l^+, V_l^+, \eta_l^+, \nu_l^+$ at lines 2-5. These are updated counts of units to be matched, such that, for all $l$,  $U_l^+ + V_l^+ = \eta_l^+ + \nu_l^+ =M_l$ by definition of these quantities, therefore, the count of treated and control units to be matched is equal in all strata and equal to $M_l$. Note that definitions of the counts given in lines 2-5 of Algorithm \ref{Alg:MaxMcN} are precisely the initial number of each outcome-treatment pair minus the optimal amount of units to be left unmatched given in Claim \ref{CDiscard}. By Claim~\ref{CTE} we know that if the number of treated units to be matched is equal to the number of control units to be matched, as it is the case after line 5 of the algorithm, then $TE = \sum_{l=1}^L U_l^+ - \eta_l^+$ independently of how matches are made. Because of this, $TE$ can be considered fixed at this point, and maximizing $\chi$ equates with maximizing $SD$ if $TE <0$ and minimizing it if $TE \geq 0$: this is checked explicitly by the algorithm at line 8. Lastly, if the algorithm calls $ComputeMinimizedSD$ if $TE \geq 0$ and $ComputeMinimizedSD$ if $TE < 0$: these two procedures are shown to correctly maximize and minimize $B_l + C_l$ in each stratum by Claim~\ref{CSD}. Note that $SD = \sum_{l=1}^L B_l + C_l$ is also separable in the strata, and therefore it can be optimized globally by separately optimizing $B_l + C_l$ in each stratum. Since $\sqrt{SD}$ is monotonic in $B_l + C_l$, we know that this quantity is also maximized or minimized in this way. This shows that Algorithm~\ref{App:Alg:MaxMcN} globally maximizes $\chi$ over the set of allowed matches. Finally, note that constraints \eqref{Eq:F1:defB}, and \eqref{Eq:F1:defC} are all not violated by definition of these quantities, and constraints \eqref{Eq:F1:onlyt1} and \eqref{Eq:F1:onlyc1} are not violated because no two units are matched more or than once. Finally, the additional exclusively binning constraints are obeyed by definition, as matches are made exclusively within each stratum. The proof of the correctness of Algorithm~\ref{App:Alg:MinMcN} is exactly symmetrical to this one: this symmetry is made apparent by the fact that running Algorithm~\ref{App:Alg:MinMcN} on the data is equivalent to flipping the treatment indicator and then running Algorithm~\ref{App:Alg:MaxMcN} on the resulting data. \endproof

\section{Algorithm for Maximization of $Z$ statistic}\label{App:Sec:MaxZ}
Algorithm \ref{Alg:MaxZILP} is useful for maximization of $Z$. Minimization of $Z$ can be accomplished with a symmetric algorithm.

\begin{algorithm}[!htbp]
  \footnotesize
  \caption{Maximize $z$ with general constraints.}
  \label{Alg:MaxZILP}
    \SetAlgoLined
    \KwData{Set of real vectors $\{(y_{l_{1}}^t, \dots, y^t_{l_{N_l^t}})\}_{l=1}^L$
    and $\{(y_{l_{1}}^c, \dots, y^c_{l_{N_l^c}})\}_{l=1}^L$,\\
    Initial lower bound on the variance, $b_1$,
    Additional data parameters for optimization, such as covariates, $\mathbf{D}$.}
    \KwResult{$N^t \times N^c$ binary matrix of matches, $\a^*$.}
    Initialize: maximize Formulation 3 by removing the upper bound constraint in \eqref{Eq:VarBound}, denote solution by $\mathbf{a}^{(0)}$;\\
    Compute initial upper bound on variance: $b^{(0)}_L := \sumT\sumC (y_i^t - y_j^c)^2a_{ij}^{(0)}$;\\
    Create initial grid of length $L$: $\mathbf{b} = (b_{1}^{(0)}, \dots, b_{L}^{(0)})$;\\
    Start iteration counter, $iter = 1$\\
    
    \If{$\bar{d}_{\mathbf{a}^{(0)}} \geq 0$}{
    \While{$\max_l\frac{\bar{d}_{\mathbf{a}_l^{(iter)}}\sqrt{M}}{\sqrt{\frac{1}{M}b_{l-1}^{(iter)} - (\bar{d}_{\mathbf{a}_l^{(iter)}})^2 }} - \max_l \frac{\bar{d}_{\mathbf{a}_l^{(iter)}}\sqrt{M}}{\sqrt{\frac{1}{M}b_{l}^{(iter)} - (\bar{d}_{\mathbf{a}_l^{(iter)}})^2 }}\geq \epsilon$}{
        Increment iteration counter: $iter = iter + 1$\\
        \lFor{$l = 1, \dots, L$}{Maximize Formulation 3, using $b_l^{(iter-1)}$ as upper bound on variance. Denote solution with $\mathbf{a}_l^{(iter)}$}
        Compute lower bound on z-score $z_{LB} := \max_{l \in 1, \dots, L} \frac{\bar{d}_{\mathbf{a}_l^{(iter)}}\sqrt{M}}{\sqrt{\frac{1}{M}b_l^{(iter-1)} - (\bar{d}_{\mathbf{a}_l^{(iter)}})^2}}$\\
        \For{$l = 2, 4, 6, \dots, L$}{
          Apply Theorem \ref{Thm:Bound}:  \lIf{$\frac{\bar{d}_{\mathbf{a}_l^{(iter)}}\sqrt{M}}{\sqrt{\frac{1}{M}b_{l-1}^{(iter-1)} - (\bar{d}_{\mathbf{a}_l^{(iter)}})^2}} > z_{LB}$}{refine the grid between $b_{l-1}^{(iter-1)}$ and $b_l^{(iter-1)}$, obtaining a new set of grid points}
        }
        Concatenate all refined grid points into a new grid, denoted by $\mathbf{b}^{(iter)} = (b_1^{(iter)}, \dots, b_L^{(iter)})$, and sorted in increasing order.
    }
    \Return  $\mathbf{a}^* = \argmax_{l=1,\dots,L} \frac{\bar{d}_{\mathbf{a}_l^{(iter)}}\sqrt{M}}{\sqrt{\frac{1}{M}b_l^{(iter)} - (\bar{d}_{\mathbf{a}_l^{(iter)}})^2}}$    }
    
    \If{$\bar{d}_{\mathbf{a}^{(0)}} < 0$}{
        Flip the treatment indicator (1 becomes 0, 0 becomes 1) for convenience in handling only positive treatment effects.\\
        \While{$\min_l\frac{\bar{d}_{\mathbf{a}_l^{(iter)}}\sqrt{M}}{\sqrt{\frac{1}{M}b_{l}^{(iter)} - (\bar{d}_{\mathbf{a}_l^{(iter)}})^2 }} - \min_l \frac{\bar{d}_{\mathbf{a}_l^{(iter)}}\sqrt{M}}{\sqrt{\frac{1}{M}b_{l-1}^{(iter)} - (\bar{d}_{\mathbf{a}_l^{(iter)}})^2 }}\leq \epsilon$}{
            Increment iteration counter: $iter = iter + 1$\\
            \lFor{$l = 1, \dots, L$}{Minimize Formulation 3, using $b_l^{(iter-1)}$ as upper bound on variance. Denote solution with $\mathbf{a}_l^{(iter)}$}
            Compute upper bound on z-score $z_{UB} := \min_{l \in 1, \dots, L} \frac{\bar{d}_{\mathbf{a}_l^{(iter)}}\sqrt{M}}{\sqrt{\frac{1}{M}b_l^{(iter-1)} - (\bar{d}_{\mathbf{a}_l^{(iter)}})^2}}$\\
            \For{$l = 2, 4, 6, \dots, L$}{
              Apply Theorem \ref{Thm:Bound}:  \lIf{$\frac{\bar{d}_{\mathbf{a}_l^{(iter)}}\sqrt{M}}{\sqrt{\frac{1}{M}b_{l-1}^{(iter-1)} - (\bar{d}_{\mathbf{a}_l^{(iter)}})^2}} < z_{UB}$}{refine the grid between $b_{l-1}^{(iter-1)}$ and $b_l^{(iter-1)}$, obtaining a new set of grid points}
            }
            Concatenate all refined grid points into a new grid, denoted by $\mathbf{b}^{(iter)} = (b_1^{(iter)}, \dots, b_L^{(iter)})$, and sorted in increasing order.
        }
    \Return  $\mathbf{a}^* = \argmin_{l=1,\dots,L} \frac{\bar{d}_{\mathbf{a}_l^{(iter)}}\sqrt{M}}{\sqrt{\frac{1}{M}b_l^{(iter)} - (\bar{d}_{\mathbf{a}_l^{(iter)}})^2}}$.
    }
  \end{algorithm}

\section{Randomization Distribution of $\chichi$ Under Exclusively Binning Constraints}\label{App:Sec:RandDist}

In this section we use the algorithm above to derive a randomization distribution for $\chichi$ under the null hypothesis of no treatment effect. This allows us to test $\Hsharp$ under two types of uncertainty: in the data itself, and in the choices made by the analyst in a MaE model. As a reminder, under $\Hsharp$, potential outcomes are assumed to be fixed and invariant between treatment regimes. Under exclusively binning constraints, the units are divided into $S_1, \dots S_l$ strata, and matches are allowed only in the same stratum.

\begin{figure}[!htbp]
\centering
\begin{subfigure}{.5\textwidth}
  \centering
  \includegraphics[width=.6\linewidth]{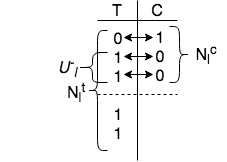}
  \caption{$\chi$ Minimization. }
  \label{fig:sub1}
\end{subfigure}%
\begin{subfigure}{.5\textwidth}
  \centering
  \includegraphics[width=.6\linewidth]{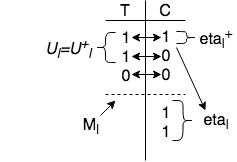}
  \caption{$\chi$ Maximization.}
  \label{fig:sub2}
\end{subfigure}
\caption{Matching procedures within one stratum. Matches are made above the dotted line and represented by the double arrow. Units below the line are discarded. }
\label{fig:expl}
\end{figure}

Given $L$ different levels in which the covariates are grouped, and that there are $N_l^t$ treatment units and $N_l^c$ control units in each stratum $l$. From the two assumptions before it follows that the data are generated as follows:
\begin{align}
U_l|\Hsharp &\overset{iid}{\sim} Bin(e_l, N_l^1), \label{Eq:DGPSharp1}\\
\eta_l &= N_l^1 - U_l,\\
V_l|\Hsharp &\overset{iid}{\sim} Bin(e_l, N_l^0),\\
\nu_l &= N_l^0 - V_l,\\
N_l^t &= U_l + V_l,\\
N_l^c &= \nu_l + \eta_l, \label{Eq:DGPSharp2}
\end{align}
As stated in the paper, all these quantities have interpretations in our matching framework. Specifically, within one stratum, I.E., at one level of $X$, called $x_l$: $U_l$ is the number of treated units with outcome 1,$V_l$ is the number of units with $T=1$ and $Y=0$, $\eta_l$ is the number of units with $T=0$ and $Y=1$ and $\nu_l$ is the number of units with $T=0$ and $Y=0$. The null hypothesis of no treatment effect is encoded in the fact that the distributions of $U_l$ and $\eta_l$ differ only in the number of trials and not in the probability of a success. Note now that these pair counts are random variables: the algorithms make pairs to purposefully obtain optimal values of $\chi$, which in turn depends on the random data. Recall that $B$ and represents the number of pairs such that the treated unit has outcome 1 and the control unit has outcome 0 and $C$ the total number of pairs in which the control unit has outcome 1 and the treated unit 0.\\
Recall that the test statistic is defined as follows:
\begin{align}
\chi^+ = \frac{TE^+ - 1}{\sqrt{SD^+ +1}} = \frac{B^+ - C^+ -1}{\sqrt{B^+ + C^+ + 1}}, \quad \chi^- = \frac{TE^- - 1}{\sqrt{SD^- + 1}} = \frac{B^- - C^- -1}{\sqrt{B^- + C^- + 1}},\label{Eq:Zdef}
\end{align}
where $B^+$ is the count of matched pairs produced by Algorithm~\ref{App:Alg:MaxMcN} such that the treated unit has outcome 1 and the control unit in the pair has outcome 0, $C^+$ is the count of pairs where the opposite is true, and $B^-$ and $C^-$ are the analogues produced by the minimization algorithm. For convenience, we also introduce "truncated" versions of the variables above, letting $G_l^- = \max(N_l^t - N_l^c, 0)$ and $G_l^+ = \max(N_l^c - N_l^t, 0)$:
\begin{align}
U_l^+ &= U_l - \max(U_l - N_l^c, 0)\label{Eq:Trunc1}\\
V_l^+ &= M_l - U_l^+\label{Eq:Trunc2}\\
U_l^- &= \max(U_l - G_l^-, 0), 0)\label{Eq:Trunc3}\\
V_l^- &= M_l - U_l^-\label{Eq:Trunc4}\\
\eta_l^+ &= \max(\eta_l - G_l^+, 0)\label{Eq:Trunc5}\\
\nu_l^+ &= M_l - \eta_l^+\label{Eq:Trunc6}\\
\eta_l^- &= \eta_l - \max(\eta_l - N_l^t, 0)\label{Eq:Trunc7}\\
\nu_l^- &= M_l - \eta_l^-.\label{Eq:Trunc8}
\end{align}
These definitions correspond to those introduced at lines 2-5 of Algorithms~\ref{App:Alg:MaxMcN} and~\ref{App:Alg:MinMcN}. Figure~\ref{fig:expl} summarizes the variables in our framework as well as how the maximization and minimization algorithms operate.

\subsection{Simplified Representations for McNemar's Statistic}
The following statement gives us simplified forms for the values of $TE$ and $SD$ output by the maximization and minimization algorithms respectively, we state it as a claim as it requires a simple yet nontrivial proof:
\begin{claim}\label{CDef} Let $TE^+$, $SD^+$ and $TE^-$ and $SD^-$ represent the values of the numerator and denominator of $\chi$ as defined in Eq~\eqref{Eq:Zdef}, and output by algorithms~\ref{App:Alg:MaxMcN} and ~\ref{App:Alg:MinMcN} respectively. Let $M_l = \min(N_l^t, N_l^c)$ denote the number of matches that are made in each stratum. Then they can be written as follows:
\begin{align}
TE^+ &= \sum_{l=1}^L TE_l^+ = \sum_{l=1}^L U_l^+ - \eta_l^+ \label{App:Eq:DefTEp}\\
TE^- &= \sum_{l=1}^L TE_l^- = \sum_{l=1}^L U_l^- - \eta_l^- \label{App:Eq:DefTEm}\\
SD^+ &= \begin{cases} S^+ = \sum_{l=1}^L|U_l^+ - \eta_l^+| \quad &\mbox{if } TE^+ \geq 1 \\
R^+ = \sum_{l=1}^LM_l -|U_l^+ + \eta_l^+ - M_l| \quad &\mbox{if } TE^+ < 1\end{cases} \label{App:Eq:DefSDp}\\
SD^- &= \begin{cases} S^- = \sum_{l=1}^L|U_l^- - \eta_l^-| \quad &\mbox{if } TE^- < 1 \\
R^- = \sum_{l=1}^LM_l -|U_l^- + \eta_l^- - M_l| \quad &\mbox{if } TE^- \geq 1\end{cases} \label{App:Eq:DefSDm}.
\end{align}
\end{claim}
\proof{Proof.} Note first that the definition of $TE^+$ and $TE^-$ follows directly from lines 3-5 and 8 of Algorithms~\ref{App:Alg:MaxMcN} and~\ref{App:Alg:MinMcN} respectively. As for the definition of $SD$ we will prove only the claim for $SD^+$ as the proof for $SD^-$ is exactly symmetrical. Proving the claim implies showing that, denoting with $B^+$ and $C^+$ the count of $B$ and $C$ pairs that maximize $\chi$:
$$B^+ + C^+ = SD^+.$$
Now define:
\begin{align}
SD_l^+ &=
\begin{cases}
S_l^+ = |U_l^+ - \eta_l^+|&\mbox{If } TE^+ \geq 1\\
R_l^+ = M_l - |U_l^+ + \eta_l^+ - M_l| &\mbox{If } TE^+ < 1\\
\end{cases}.\label{App:Eq:SDlp}
\end{align}
It is clear from this definition that $SD^+ = \sum_{l=1}^L SD_l^+$, and since $SD_l^+$ has the same definition in all strata, it suffices to prove that $B_l^+ + C_l^+$ can be written as in Eq. \eqref{App:Eq:SDlp} for one stratum to prove the equality in Eq. \eqref{App:Eq:DefSDp}. The rest of the proof is concerned with establishing this result.\\

Consider the computation for $SD^+$ occurring in Alg. \ref{App:Alg:MaxMcN}: the algorithm checks explicitly if $TE \geq 1$ at line 8 and, if true it calls Alg. \ref{App:Alg:MinSD} on inputs $(U_l^+, V_l^+, \eta_l^+, \nu_l^+)$ to generate the matched pair counts $A_l^+, B_l^+, C_l^+, D_l^+$, if false it generates the same quantities by calling Alg. \ref{App:Alg:MaxSD} on the same inputs.

Consider now the case in which $TE^+ \geq 1$, and Algorithm \ref{App:Alg:MinSD} is called with $(U_l^+, V_l^+, \eta_l^+, \nu_l^+)$  as inputs. We know that the algorithm returns the following counts for $B$ and $C$ pairs:
\begin{align}
B_l^{(\ref{App:Alg:MinSD})} &= \min(U_l^+ - A, V_l^+ - D)&&\mbox{(By line 3 of Alg. \ref{App:Alg:MinSD})}\nonumber\\
&= \min(U_l^+ - \min(U_l^+, \eta_l^+), \nu_l^+ - \min(\nu_l^+, V_l^+)),&&\mbox{(By lines 1 and 2 of Alg. \ref{App:Alg:MinSD})}\label{App:Eq:Blmin}
\intertext{and}
C_l^{(\ref{App:Alg:MinSD})} &= \min(\eta_l^+ - A, V_l^+ - D) &&\mbox{(By line 4 of Alg. \ref{App:Alg:MinSD})}\nonumber\\
&= \min(\eta_l^+ - \min(U_l^+, \eta_l^+), V_l^+ - \min(\nu_l^+, V_l^+)).&&\mbox{(By lines 1 and 2 of Alg. \ref{App:Alg:MinSD})}\label{App:Eq:Clmin}
\end{align}

\noindent\textbf{Case 1 for $TE^+ \geq 1$:} $U_l^+ \geq \eta_l^+ $.\\
Note first that, in this case:
\begin{align}
&U_l^+ \geq \eta_l^+\\
\implies &M_l - \eta_l^+ \geq M_l - U_l^+ &&\mbox{(Add $M_l$ to both sides)}\\
\implies &\nu_l^+ \geq V_l^+. && \mbox{(By definition of $\nu_l^+, V_l^+$)}\label{App:Eq:NulGeqVl}
\end{align}
Because of this we can see that:
\begin{alignat*}{3}
&B_l^{(\ref{App:Alg:MinSD})} + C_l^{(\ref{App:Alg:MinSD})} \\
&= \min(U_l^+ - \min(U_l^+, \eta_l^+), \nu_l^+- \min(\nu_l^+, V_l^+))&&\\
&+ \min(\eta_l^+ - \min(U_l^+, \eta_l^+), V_l^+ - \min(\nu_l^+, V_l^+))&&\mbox{(By \eqref{App:Eq:Blmin} and \eqref{App:Eq:Clmin})}\\
&= \min(U_l^+ - \eta_l^+, \nu_l^+ - V_l^+) + \min(\eta_l^+ - \eta_l^+, V_l^+ - V_l^+)\quad&&\mbox{(By assumption of this case and result in \eqref{App:Eq:NulGeqVl})}\\
&= \min(U_l^+ - \eta_l^+, M_l - U_l^+ - (M_l - \eta_l^+))&&\mbox{(By definition of $V_l^+$ and $\nu_l^+$)}\\
&= U_l^+ - \eta_l^+ &&\\
&= |U_l^+ - \eta_l^+| &&\mbox{(By assumption of this case)}\\
&= S_l^+.&&\mbox{(Definition of $S_l^+$)}
\end{alignat*}
\textbf{Case 2 for $TE^+ \geq 1$}: $U_l^+ < \eta_l^+$.\\
Note that in this case Eq. \eqref{App:Eq:NulGeqVl}, implies that: $\nu_l^+ < V_l^+$. Because of this we have:
\begin{alignat*}{3}
&B_l^{(\ref{App:Alg:MinSD})} + C_l^{(\ref{App:Alg:MinSD})} \\
&= \min(U_l^+ - \min(U_l^+, \eta_l^+), \nu_l^+ - \min(\nu_l^+, V_l^+))\\
&+ \min(\eta_l^+ - \min(U_l^+ - \eta_l^+), V_l^+ - \min(\nu_l^+, V_l^+))&&\mbox{(By \eqref{App:Eq:Blmin} and \eqref{App:Eq:Clmin})}\\
&= \min(U_l^+ - U_l^+, \nu_l^+ - \nu_l^+) + \min(\eta_l^+ - U_l^+, V_l^+ - \nu_l^+) \quad&&\mbox{(By assumption of this case and result in \eqref{App:Eq:NulGeqVl})}\\
&= \min(\eta_l^+ - U_l^+, M_l - U_l^+ - (M_l - \eta_l^+))&&\mbox{(By definition of $V_l^+$ and $\nu_l^+$)}\\
&= \eta_l^+ - U_l^+&&\\
&= |U_l^+ - \eta_l^+| &&\mbox{(By assumption of this case)}\\
&= S_l^+.&&\mbox{(Definition of $S_l^+$)}
\end{alignat*}
This shows $SD_l^+ = S_l^+$ in the case in which $TE \geq 1$, the first of Eq. \eqref{App:Eq:DefSDp}. 

The proof is similar for the case in which $TE < 1$. Now we must show that, in this case $SD_l^+ = R^+$. In this case Algorithm \ref{App:Alg:MaxSD} is called with $U_l^+, V_l^+, \eta_l^+, \nu_l^+$ as inputs at line 16 of Algorithm \ref{App:Alg:MaxMcN}. On those inputs, Algorithm \ref{App:Alg:MaxSD} will return the following counts of $B$ and $C$ pairs:
\begin{align}
B_l^{(\ref{App:Alg:MaxSD})} &= \min(U_l^+ , \nu_l^+)&&\mbox{(By line 1 of Alg. \ref{App:Alg:MaxSD})}\label{App:Eq:Blmax}
\intertext{and}
C_l^{(\ref{App:Alg:MaxSD})} &= \min(\eta_l^+, V_l^+). &&\mbox{(By line 1 of Alg. \ref{App:Alg:MaxSD})}\label{App:Eq:Clmax}
\end{align}
We proceed as before with two separate cases.\\
\noindent\textbf{Case 1 for $TE < 1$}: $U_l^+ \geq \nu_l^+$. \\
First note that:
\begin{align}
\nu_l^+ &= M_l - \eta_l^+ &&\mbox{(By definition of $\nu_l^+$)}\nonumber\\
&\leq U_l^+ &&\mbox{(By assumption of this case)}\nonumber\\
\implies \eta_l^+ &\geq M_l - U_l^+ &&\nonumber\\
\implies \eta_l^+ &\geq V_l^+.&&\mbox{(By definition of $V_l^+$ )}\label{App:Eq:Nul}
\end{align}
Second, the assumption of this case also implies:
\begin{align}
\nu_l^+ &= M_l - \eta_l^+ &&\mbox{(By definition of $\nu_l^+$)}\nonumber\\
&\leq U_l^+ &&\mbox{(By assumption of this case)}\nonumber\\
\implies & U_l^+ + \eta_l^+ \geq M_l\nonumber\\
\implies & U_l^+ + \eta_l^+ - M_l \geq 0\nonumber\\
\implies & U_l^+ + \eta_l^+ - M_l = |U_l^+ + \eta_l^+ - M_l|.&&\mbox{(Def. of absolute value)}\label{App:Eq:RpAbsGeq}
\end{align}
Putting these results together we obtain:
\begin{align}
B_l^{(\ref{App:Alg:MaxSD})} + C_l^{(\ref{App:Alg:MaxSD})}&= \min(U_l^+, \nu_l^+) + \min(\eta_l^+, V_l^+)&&\mbox{(By \eqref{App:Eq:Blmax} and \eqref{App:Eq:Clmax})} \\
&=  \nu_l^+ + V_l^+ &&\mbox{(By assumption of this case and \eqref{App:Eq:Nul})} \\
&= M_l - U_l^+ + M_l - \eta_l^+ &&\mbox{(By definition of $V_l^+$ and $\nu_l^+$)}\\
&= M_l - (U_l^+ + \eta_l^+ - M_l) \\
&= M_l - |U_l^+ + \eta_l^+ - M_l|&&\mbox{(By \eqref{App:Eq:RpAbsGeq})}\\
&= R_l^+.&&\mbox{(Definition of $R_l^+$)}
\end{align}
\textbf{Case 2 for $TE < 1$}: $U_l^+ < \nu_l^+$. \\
This assumption of this case together with \eqref{App:Eq:Nul} implies $\eta_l^+ < V_l^+$. Note also that:
\begin{align}
\nu_l^+ &= M_l - \eta_l^+ &&\mbox{(By By definition of $\nu_l^+$)}\nonumber\\
&> U_l^+ &&\mbox{(By assumption of this case)}\nonumber\\
\implies & U_l^+ + \eta_l^+ - M_l < 0\nonumber\\
\implies & M_l - U_l^+ - \eta_l^+ > 0\nonumber\\
\implies & M_l - U_l^+ - \eta_l^+ = |M_l - U_l^+ - \eta_l^+|. &&\mbox{(Def. of absolute value)}\label{App:Eq:RpAbsLeq}
\end{align}
With the two results above we obtain:
\begin{align}
B_l^{(\ref{App:Alg:MaxSD})} + C_l^{(\ref{App:Alg:MaxSD})}&= \min(U_l^+, \nu_l^+) + \min(\eta_l^+, V_l^+)&&\mbox{(By \eqref{App:Eq:Blmax} and \eqref{App:Eq:Clmax})} \\
&= U_l^+ + \eta_l^+ &&\mbox{(By assumption of this case and \eqref{App:Eq:Nul})}\\
&= M_l - M_l + U_l^+ + \eta_l^+&&\mbox{(Add and subtract $M_l$)}\\
&= M_l - (M_l - U_l^+ - \eta_l^+)\\
&= M_l - |M_l - U_l^+ - \eta_l^+|&&\mbox{(By \eqref{App:Eq:RpAbsLeq})}\\
&= M_l - |U_l^+ + \eta_l^+ - M_l|&&\mbox{(By def. of absolute value)}\\
&= R_l^+. &&\mbox{(Definition of $R_l^+$)}
\end{align}
This proves that $SD_l^+ = R_l^+$ in the case in which $TE^+ < 1$, the second case in Equation \eqref{App:Eq:DefSDp}. \endproof

\subsection{Lemma~\ref{Lem:TruncatedSharp}}\label{App:Sec:TruncatedSharp}
\begin{lemma}{(Randomization Distribution of Truncated Variables)}\label{Lem:TruncatedSharp} For $l = 1, \dots, L$ let $(N_l^1, N_l^0, e_l)$ be fixed and known and let $\mathbb{D} = (U_l, V_l, \eta_l, \nu_l, N_l^t, N_l^c)$ be drawn from the data generating process of equations \eqref{Eq:DGPSharp1} -- \eqref{Eq:DGPSharp2}. Let $a, b, c, d, m$ be elements of $\{0, \dots, \min(N_l^0, N_l^2)\}$ and let $m \in \{0, \dots, N_l\}$. The variables ($U_l^+$, $U_l^-$, $\eta_l^+$, $\eta_l^-$, $M_l$) have the following joint distribution:
\begin{align}
\Pr(U_l^+ = a, &U_l^- = b, \eta_l^+ = c, \eta_l^- = d, M_l = m|\Hsharp)\nonumber \\
= \sum_{j=0}^{N_l^1}\sum_{k=0}^{N_l^0}&\Pr(U_l = j|\Hsharp)\Pr(V_l = k|\Hsharp)\ind(U_l^+ = a|U_l = j, V_l = k)\ind(U_l^- = b|U_l = j, V_l = k)\nonumber\\
\times &\ind(\eta_l^+ = c|U_l = j, V_l = k)\ind(\eta_l^- = d|U_l = j, V_l = k)\ind(M_l = m|U_l = j, V_l = k),\label{Eq:TruncatedSharp}
\end{align}
where:
\begin{align}
\ind(U_l^+ = a|U_l = j, V_l = k) &= \begin{cases} 1 &\mbox{if } j = N_l - k - a,\; k \leq N_l - 2a\\
1 &\mbox{if } j = a,\; k \leq N_l - 2a\\
0 &\mbox{otherwise.}
\end{cases}\label{Eq:UlpSharp}\\
\ind(U_l^-=b|U_l=j, V_l=k) &= \begin{cases}
1 &\mbox{if }j = N_l - 2k - b,\; k \leq N_l - 2b\\
1 &\mbox{if }j = b,\; k \leq \frac{N_l - 2b}{2}\\
1 &\mbox{if }b=0,\;j \geq \frac{N_l}{2} - k,\; k \geq \frac{N_l - j}{2}\\
1 &\mbox{if }b = 0,\;j = 0,\;k\leq \frac{N_l}{2}\\
0 &\mbox{otherwise.}
\end{cases}\label{Eq:UlmSharp}\\
\ind(\eta_l^+=c|U_l=j, V_l=k) &= \begin{cases}
1 &\mbox{if }j = c + N_l^0 - 2k,\;k \geq \frac{N_l^0 - N_l^1 + 2c}{2}\;\\
1 &\mbox{if }j = N_l^1 - c,\; k \geq \frac{N_l^0 - N_l^1 + 2c}{2}\;\\
1 &\mbox{if }c = 0,\; j \leq  N_l^0 - 2k,\; k \geq \frac{N_l - 2j}{2}\\
1 &\mbox{if }c = 0,\; j = N_l^1,\; k \geq \frac{N_l^0 - N_l^1}{2}\\
0 &\mbox{otherwise.}
\end{cases}\label{Eq:ElpSharp}\\
\ind(\eta_l^-=d|U_l=j, V_l=k) &= \begin{cases}
1 &\mbox{if }j = d - k,\; k \geq 2d - N_l^1\\
1 &\mbox{if }j = N_l^1 - d, k \geq 2d - N_l^1\\
0 &\mbox{otherwise.}
\end{cases}\label{Eq:ElmSharp}\\
\ind(M_l = m|U_l=j, V_l = k) &= \begin{cases}
1 &\mbox{if }j = m - k,\; m \leq \frac{N_l}{2}\\
1 &\mbox{if }j = N_l - k - m,\; m \leq \frac{N_l}{2}\\
0 &\mbox{otherwise.}
\end{cases}\label{Eq:MSharp}
\end{align}
and $\Pr(U_l = j|\Hsharp) = Bin(j, N_l^1, e_l)$ and $\Pr(V_l = k|\Hsharp) = Bin(k, N_l^0, e_l)$.
\end{lemma}

Note that even though the domain of the distribution above cannot be expressed by a simpler formula, it is simple and computationally fast to enumerate it for finite $N_l$. 
\proof{Proof}
Throughout the proof we maintain Assumption \ref{As:StratIgnorability} and all of the probability statements to follow are conditional on it holding. The proof is simple and follows by inspecting the definitions of the truncated variables. Note first that the form for $\Pr(U_l^+ = a, U_l^- = b, \eta_l^+ = c, \eta_l^- = d, M_l = m|\Hsharp)$ given in Eq. \eqref{Eq:TruncatedSharp} follows from the law of total probability, independence of $U_l$ and $V_l$ and conditional independence of $U_l^+, V_l^+, \eta_l^+, \eta_l^-, M_l$ given $U_l$ and $V_l$. It remains to show that the forms for the indicator functions in Equations \eqref{Eq:UlpSharp}--\eqref{Eq:MSharp} are those in the theorem. This can be done by inspecting the definitions of the truncated variables and by expanding them out into conditions on $U_l$ and $V_l$. In what follows we refer to the following definitions of the quantities employed, listed here with reference to where they are introduced in the paper:
\begin{alignat*}{3}
U_l &\sim Bin(N_l^1, e_l)\\
\eta_l &= N_l^1 - U_l\\
V_l &\sim Bin(N_l^0, e_l)\\
\nu_l &= N_l^0 - V_l\\
N_l^t &= U_l + V_l\\
N_l^c &= \eta_l + \nu_l\\
U_l^+ &= U_l - \max(U_l - N_l^c, 0)\\
U_l^- &= \max(U_l - G_l^-, 0)\\
\eta_l^+ &= \max(\eta_l - G_l^+, 0)\\
\eta_l^- &= \eta_l - \max(\eta_l - N_l^t, 0)\\
M_l &= \min(N_l^t, N_l^c)\\
G_l^+ &= \max(N_l^c - N_l^t, 0)\\
G_l^- &= \max(N_l^t - N_l^c, 0).
\end{alignat*}
We now derive conditions on $U_l$ and $V_l$ that lead to the realization of the event $U_l^+ = a$, we do so by expanding the definition of $U_l^+$ and considering all the cases it entails separately. Starting with the definition of $U_l^+$ we have:
\begin{alignat}{3}
U_l^+ &= U_l - \max(U_l - N_l^c, 0) &&\mbox{(By definition of $U_l$)}\nonumber\\
&= U_l - \max(U_l - \eta_l - \nu_l, 0)&&\mbox{(By definition of $N_l^c$)}\nonumber\\
&=U_l - \max(U_l - N_l^1 + U_l - N_l^0 + V_l, 0)\qquad &&\mbox{(By definition of $\eta_l, \nu_l$)} \nonumber\\
&= U_l - \max(2U_l + V_l - N_l, 0).&&\mbox{(Because $N_l^1 + N_l^0 = N_l$)}\label{App:Eq:UlExpanded}
\end{alignat}
There are now two cases for the event $U_l^+ = a$, depending on how the max function in the definition of $U_l^+$ is resolved: \\

\noindent\textbf{Case 1 for $U_l^+ = a$}:
\begin{align}
\max(2U_l + V_l - N_l, 0) &= 2U_l + V_l - N_l. \label{App:Eq:Ulp1}
\end{align}
Because of this our event of interest can be written as:
\begin{alignat}{3}
a &= U_l^+ = U_l - 2U_l - V_l + N_l \qquad&&\mbox{(By \eqref{App:Eq:UlExpanded})}\nonumber\\
\implies U_l &= N_l - V_l - a\label{App:Eq:Ulp11},
\end{alignat}
which is the first condition in the first case of Equation \eqref{Eq:UlpSharp}. Note that the condition of this case given in Eq. \eqref{App:Eq:Ulp1} implies that $2U_l + V_l - N_l \geq 0$, and we can use the result in \eqref{App:Eq:Ulp11} to expand this condition as follows:
\begin{alignat*}{3}
0 &\leq 2U_l + V_l - N_l \qquad&&\mbox{(By \eqref{App:Eq:Ulp1})}\\
&= 2(N_l - V_l - a) + V_l - N_l\qquad&&\mbox{(By \eqref{App:Eq:Ulp11})}\\
&= N_l - V_l - 2a \\
\implies V_l &\leq N_l - 2a,
\end{alignat*}
which is the second condition in the first case of Equation \eqref{Eq:UlpSharp}.\\

\noindent\textbf{Case 2 for $U_l^+ = a$}:
\begin{align}
\max(2U_l + V_l - N_l, 0) = 0.\label{App:Eq:Ulp2}
\end{align}
Using this case and the result in \eqref{App:Eq:UlExpanded}, we can write the event of interest as
\begin{align}
a = U_l^+ = U_l\label{App:Eq:Ulp21},
\end{align}
the first condition in the second case of Equation \eqref{Eq:UlpSharp}. Second, the fact that the $\max$ function equals 0 in Case 2, implies that $2U_l + V_l - N_l \leq 0$ is another condition for this case. Using the result in \eqref{App:Eq:Ulp21} this can be simplified as:
\begin{align*}
0 &\geq 2U_l + V_l - N_l &&\mbox{(By \eqref{App:Eq:Ulp2})}\\
&= 2a + V_l - N_l\qquad&&\mbox{(By \eqref{App:Eq:Ulp21})}\\
\implies V_l &\leq N_l - 2a,
\end{align*} which is the second condition in the second case of Equation \eqref{Eq:UlpSharp}.\\

We proceed in the same manner for the event $U_l^- = b$. In Equation \eqref{Eq:UlmSharp} we start by expanding the definition of $U_l^-$:
\begin{alignat}{3}
U_l^- &= \max(U_l - \max(N_l^t - N_l^c, 0), 0)&&\mbox{(By definition of $U_l^-$)} \nonumber\\
&= \max(U_l - \max(U_l + V_l - \eta_l - \nu_l, 0),0)&&\mbox{(By definition of $N_l^t, N_l^c$)}\nonumber\\
&= \max(U_l - \max(U_l + V_l - N_l^1 + U_l - N_l^0 + V_l,0),0)\qquad&&\mbox{(By definition of $\eta_l, \nu_l$)}\nonumber\\
&= \max(U_l - \max(2(U_l + V_l) - N_l, 0), 0). &&\mbox{(Because $N_l = N_l^1 + N_l^0$)}\label{App:Eq:UlmExp}
\end{alignat}
Because of the two $\max$ functions we have four different possibilities for the value of $U_l^-$, all dependent on how the $\max$ functions resolve. As we did before, we can derive conditions on $U_l$ and $V_l$ by studying these four cases separately. We start each of the four cases by listing the ways the inner max and the outer max resolve.\\

\noindent\textbf{Case 1 for $U_l^- = b$}:
\begin{align}
\max(2(U_l + V_l) - N_l, 0) &= 2(U_l + V_l) - N_l \label{App:Eq:Ulm11}\\
\max(U_l - \max(2(U_l + V_l) - N_l, 0), 0) &= U_l - \max(2(U_l + V_l)-N_l, 0). \label{App:Eq:Ulm12}
\end{align}
First, we can use both conditions to simplify the event $U_l^- =b$ as follows:
\begin{alignat}{3}
b &= U_l^- = U_l - 2(U_l + V_l) + N_l\qquad&&\mbox{(By \eqref{App:Eq:UlmExp}, \eqref{App:Eq:Ulm11}, \eqref{App:Eq:Ulm12})} \label{App:Eq:Ulm1Case}\\
\implies U_l &= N_l - 2V_l - b\label{App:Eq:Ulm1cond1},
\end{alignat}
which is the first condition in the first case of Equation \eqref{Eq:UlmSharp}. Note now that the condition in \eqref{App:Eq:Ulm12} implies that $U_l - \max(2(U_l + V_l) - N_l, 0) \geq 0$, combining this with the above we obtain:
\begin{alignat*}{3}
0 &\leq U_l - \max(2(U_l + V_l) - N_l, 0)&& \\
&= b&&\mbox{(By \eqref{App:Eq:Ulm1Case})}\\
&\geq 0,&&\mbox{(By definition)}
\end{alignat*}
so this condition is always satisfied because of how we restrict the domain of $b$.
The condition in \eqref{App:Eq:Ulm11} implies that $2(U_l + V_l) - N_l \geq 0$, again we use the result in \eqref{App:Eq:Ulm1cond1} to expand this as follows:
\begin{alignat*}{3}
0 &\leq 2(U_l + V_l) - N_l \\
&= 2(N_l - 2V_l - b + V_l) - N_l \qquad&&\mbox{(By \eqref{App:Eq:Ulm1cond1})}\\
&= N_l - V_l - 2b \\
\implies V_l &< N_l - 2b,
\end{alignat*} the second condition in the first case of Eq. \eqref{Eq:UlmSharp}. \\

\noindent\textbf{Case 2 for $U_l^- = b$}:
\begin{align}
\max(2(U_l + V_l) - N_l, 0) &= 0\label{App:Eq:Ulm21}\\
\max(U_l - \max(2(U_l + V_l) - N_l, 0), 0) &= U_l - \max(2(U_l + V_l)-N_l, 0). \label{App:Eq:Ulm22}
\end{align}
The first condition in the second case of Eq. \eqref{Eq:UlmSharp} follows from using these conditions with the event $U_l^- = b$:
\begin{alignat}{3}
b &= U_l^- = \max(U_l - \max(2(U_l + V_l) - N_l, 0), 0)\qquad&&\mbox{(By \eqref{App:Eq:UlmExp})}\nonumber\\
&= U_l - \max(2(U_l + V_l) - N_l, 0) &&\mbox{(By \eqref{App:Eq:Ulm22})}\nonumber\\
&= U_l.&&\mbox{(By \eqref{App:Eq:Ulm21})}\label{App:Eq:UlmCase2}
\end{alignat}
The condition in \eqref{App:Eq:Ulm22} implies that $U_l - \max(2(U_l + V_l) - N_l) \geq 0$, using the above we see that this condition is always satisfied in this case:
\begin{alignat*}{3}
0 &\leq U_l - \max(2(U_l + V_l) - N_l, 0)&& \\
&= U_l&&\mbox{(By \eqref{App:Eq:Ulm21})}\\
&= b &&\mbox{(By \eqref{App:Eq:UlmCase2})}\\
&\geq 0. &&\mbox{(By definition)}
\end{alignat*}
Because of this, this condition is omitted from the formulation in Eq. \eqref{Eq:UlmSharp}. Finally, condition \eqref{App:Eq:Ulm21} implies:
\begin{alignat*}{3}
0 &\geq 2(U_l + V_l) - N_l \\
&= 2(b + V_l) - N_l \qquad &&\mbox{(By \eqref{App:Eq:UlmCase2})}\\
\implies V_l &\leq \frac{N_l - 2b}{2},&&
\end{alignat*} the second condition of the case. \\

\noindent\textbf{Case 3 for $U_l^- = b$}:
\begin{align}
\max(2(U_l + V_l) - N_l, 0) &= 2(U_l + V_l) - N_l\label{App:Eq:Ulm31}\\
\max(U_l - \max(2(U_l + V_l) - N_l, 0), 0) &= 0.\label{App:Eq:Ulm32}
\end{align}
First, we have the event $U_l^- = b$ taking form:
\begin{align*}
b &= U_l^- = \max(U_l - \max(2(U_l + V_l) - N_l, 0), 0)\qquad&&\mbox{(By \eqref{App:Eq:UlmExp})}\nonumber\\
&= 0, &&\mbox{(By \eqref{App:Eq:Ulm32})}
\end{align*}
which leads us to the first condition in case 3 of Equation \eqref{Eq:UlmSharp}. For the second condition, start with \eqref{App:Eq:Ulm31}, then we have:
\begin{alignat}{3}
2(U_l + V_l) - N_l &\geq 0\qquad&&\mbox{(By \eqref{App:Eq:Ulm31})}\nonumber \\
\implies  U_l \geq \frac{N_l}{2} - V_l,\label{App:Eq:Ulm33}
\end{alignat}
which is the form of the second condition of Case 4 in \eqref{Eq:UlmSharp}. Finally, from Condition \eqref{App:Eq:Ulm32} we have:
\begin{alignat}{3}
0 &\geq U_l - \max(2(U_l + V_l) - N_l, 0)&&\mbox{(By \eqref{App:Eq:Ulm32})}\nonumber\\
&= U_l - 2(U_l + V_l) + N_l&&\mbox{(By \eqref{App:Eq:Ulm31})}\nonumber\\
\implies V_l &\geq \frac{N_l - U_l}{2}. \label{App:Eq:Ulm34}
\end{alignat}
Clearly, \eqref{App:Eq:Ulm34} is the last condition in the fourth case of Equation \eqref{Eq:UlmSharp}.\\

\textbf{Case 4 for $U_l^- = b$}:\\
\begin{align}
\max(2(Ul + V_l) - N_l, 0) = 0 \label{App:Eq:Ulm41}\\
\max(U_l - \max(2(U_l + V_l) - N_l, 0), 0) &= 0.\label{App:Eq:Ulm42}
\end{align}
First, the event $U_l^- = b$ takes the following form in this case:
\begin{alignat*}{3}
b &= U_l^- = \max(U_l - \max(2(U_l + V_l) - N_l, 0), 0)\qquad&&\mbox{(By definition of $U_l^-$)}\\
&= 0,&&\mbox{(By \eqref{App:Eq:Ulm42})}
\end{alignat*}
which is the first condition in the fourth case of Eq. \eqref{Eq:UlmSharp}. Second, we have:
\begin{alignat}{3}
0 &\geq U_l - \max(2(U_l + V_l) - N_l, 0)\qquad &&\mbox{(By \eqref{App:Eq:Ulm42})}\nonumber\\
&= U_l&&\mbox{(By \eqref{App:Eq:Ulm41})}\nonumber\\
\implies U_l &\leq 0,
\end{alignat}
but since $U_l$ is a binomial random variable its value can never be less than 0. Because of this the second condition in the case becomes:
\begin{align}
U_l = 0. \label{App:Eq:Ulm43}
\end{align}
Finally we have:
\begin{alignat}{3}
0 &\geq 2(U_l + V_l) - N_l \qquad&&\mbox{(By \eqref{App:Eq:Ulm41})}\nonumber\\
&= 2V_l - N_l  &&\mbox{(By \eqref{App:Eq:Ulm43})}\nonumber\\
\implies V_l &\leq \frac{N_l}{2},
\end{alignat}
the third condition in the fourth case of \eqref{Eq:UlmSharp}. \\

As we did before, we can now expand the definition of $\eta_l^+$ to derive conditions on $U_l$ and $V_l$ that lead to the realization of the event $\eta_l^+ = c$. Starting with the definition of $\eta_l^+$ we have:
\begin{alignat}{3}
\eta_l^+ &= \max(\eta_l - G_l^+, 0)&&\mbox{(By definition of $\eta_l^+$)} \nonumber\\
&= \max(\eta_l - \max(N_l^c - N_l^t, 0), 0)&&\mbox{(By definition of $G_l^+$)} \\
&= \max(\eta_l - \max(\eta_l + \nu_l - U_l - V_l, 0), 0)&&\mbox{(By definition of $N_l^t, N_l^c$)} \nonumber\\
&= \max(N_l^1 - U_l - \max(N_l^1 - U_l + N_l^0 - V_l - U_l - V_l, 0), 0)\qquad&&\mbox{(By definition of $\eta_l, \nu_l$)} \nonumber\\
&= \max(N_l^1 - U_l - \max(N_l - 2(U_l + V_l), 0), 0).\label{App:Eq:ElpExp}
\end{alignat}
Since there are two nested $\max$ functions in the definition of $\eta_l^+$, there will be 4 cases that correspond to how the max functions are resolved; each one of these cases is going to represent a different value of $\eta_l^+$. Below we study each case separately and show that they lead to the four cases in the indicator function of Eq. \eqref{Eq:ElpSharp}. \\

\noindent\textbf{Case 1 for $\eta_l^+ = c$}:
\begin{align}
\max(N_l^1 - U_l - \max(N_l - 2(U_l + V_l), 0), 0) &=  N_l^1 - U_l - \max(N_l - 2(U_l + V_l), 0)\label{App:Eq:Elp11}\\
\max(N_l - 2(U_l + V_l), 0) &= N_l - 2(U_l + V_l).\label{App:Eq:Elp12}
\end{align}
First, use both condition to derive a form for the event $\eta_l^+ = c$:
\begin{alignat}{3}
c &= \eta_l^+ = \max(N_l^1 - U_l - \max(N_l - 2(U_l + V_l), 0), 0)\qquad&&\mbox{(By \eqref{App:Eq:ElpExp})}\nonumber\\
&= N_l^1 - U_l - N_l + 2(U_l + V_l)&&\mbox{(By \eqref{App:Eq:Elp11} and \eqref{App:Eq:Elp12})}\nonumber\\
&= U_l + 2V_l - N_l^0 \nonumber\\
\implies U_l &= c + N_l^0 - 2V_l,\label{App:Eq:Elp13}
\end{alignat}
which is the first condition in the first case of Equation \eqref{Eq:ElpSharp}. Second, we use the result above to rewrite the first condition:
\begin{alignat*}{3}
0 &\leq N_l - 2(U_l + V_l) &&\mbox{(By \eqref{App:Eq:Elp12})}\\
&= N_l - 2(c + N_l^0 - 2V_l + V_l) \qquad&&\mbox{(By \eqref{App:Eq:Elp13})}\\
&= N_l^1 - N_l^0 - 2c + 2V_l\\
\implies V_l &\geq \frac{N_l^0 - N_l^1 + 2c}{2},
\end{alignat*}
which is the second condition in the first case in Equation \eqref{Eq:ElpSharp}.\\

\noindent\textbf{Case 2 for $\eta_l^+ = c$}:
\begin{align}
 \max(N_l^1 - U_l - \max(N_l - 2(U_l + V_l), 0), 0) &=  N_l^1 - U_l - \max(N_l - 2(U_l + V_l), 0)\label{App:Eq:Elp21}\\
 \max(N_l - 2(U_l + V_l), 0) &= 0. \label{App:Eq:Elp22}
\end{align}
First, use the second condition to find a form for the event $\eta_l^+ = c$
\begin{alignat}{3}
c &= \eta_l^+ = \max(N_l^1 - U_l - \max(N_l - 2(U_l + V_l), 0), 0)\qquad&&\mbox{(By \eqref{App:Eq:ElpExp})}\nonumber\\
&= N_l^1 - U_l&&\mbox{(By \eqref{App:Eq:Elp21} and \eqref{App:Eq:Elp22})}\nonumber\\
\implies U_l &= N_l^1 - c.\label{App:Eq:Elp23}
\end{alignat}
Now use the second condition together with the above to derive:
\begin{alignat*}{3}
0 &\geq N_l - 2(U_l + V_l)&&\mbox{(By \eqref{App:Eq:Elp22})}\\
&= N_l - 2(N_l^1 - c + V_l)\qquad&&\mbox{(By \eqref{App:Eq:Elp23})} \\
&= N_l^0 - N_l^1 + 2c - 2V_l \\
\implies V_l &\geq \frac{N_l^0 - N_l^1 + 2c}{2}.
\end{alignat*}
This, and Equation \eqref{App:Eq:Elp23} are the conditions in the second case of Equation \eqref{Eq:ElpSharp}.\\

\noindent\textbf{Case 3 for $\eta_l^+ = c$}:
\begin{align}
\max(N_l^1 - U_l - \max(N_l - 2(U_l + V_l), 0), 0) &= 0\label{App:Eq:Elp31}\\
\max(N_l - 2(U_l + V_l), 0) &= N_l - 2(U_l + V_l).\label{App:Eq:Elp32}
\end{align}

The first condition of Case 3 in Eq. \eqref{Eq:ElpSharp} is given by:
\begin{alignat*}{3}
c &= \eta_l^+ = \max(N_l^1 - U_l - \max(N_l - 2(U_l + V_l), 0), 0)\qquad&&\mbox{(By \eqref{App:Eq:ElpExp})}\nonumber\\
&= 0. &&\mbox{(By \eqref{App:Eq:Elp31})}
\end{alignat*}
Now use \eqref{App:Eq:Elp32} to rewrite the \eqref{App:Eq:Elp31}:
\begin{alignat}{3}
0 &\geq N_l^1 - U_l - \max(N_l - 2(U_l + V_l), 0)\qquad&&\mbox{(By \eqref{App:Eq:Elp31})}\nonumber\\
&= N_l^1 - U_l - N_l + 2(U_l + V_l)&&\mbox{(By \eqref{App:Eq:Elp32})}\nonumber\\
&= U_l - N_l^0 + 2V_l\nonumber\\
\implies U_l &\leq N_l^0 - 2V_l,\label{App:Eq:Elp33}
\end{alignat}
this is the second condition of the third case in Eq. \eqref{Eq:ElpSharp}. The final condition in the case is given by:
\begin{align*}
0 &\leq N_l - 2(U_l + V_l)\qquad&&\mbox{(By \eqref{App:Eq:Elp32})} \\
\implies V_l &\geq \frac{N_l - 2U_l}{2}.
\end{align*}

\noindent \textbf{Case 4 for $\eta_l^+ = c$}:
\begin{align}
\max(N_l^1 - U_l - \max(N_l - 2(U_l + V_l), 0), 0) &= 0\label{App:Eq:Elp41}\\
\max(N_l - 2(U_l + V_l), 0) &= 0.\label{App:Eq:Elp42}
\end{align}

The first condition in the fourth case of Eq. \eqref{Eq:ElpSharp} is obtained as follows:
\begin{alignat}{3}
c &= \eta_l^+ = \max(N_l^1 - U_l - \max(N_l - 2(U_l + V_l), 0), 0)\qquad&&\mbox{(By \eqref{App:Eq:ElpExp})}\nonumber\\
&= 0. &&\mbox{(By \eqref{App:Eq:Elp41})}\nonumber
\end{alignat}
The second condition in the same case can be obtained by starting from \eqref{App:Eq:Elp41}:
\begin{align}
0 &\geq N_l^1 - U_l - \max(N_l - 2(U_l + V_l), 0)\qquad &&\mbox{(By \eqref{App:Eq:Elp41})}\nonumber\\
&= N_l^1 - U_l&&\mbox{(By \eqref{App:Eq:Elp42})}\nonumber\\
\implies U_l &\geq N_l^1\nonumber.
\end{align}
Since $U_l$ is a binomial random variable with number of trials $N_l^1$, it can never be greater then $N_l^1$, so the condition above becomes:
\begin{align}
U_l = N_l^1,\label{App:Eq:Elp43}
\end{align}
which is the second condition in the fourth case of Eq. \eqref{Eq:ElpSharp}.
Now rearrange the terms in the second condition and use the result above to obtain the final condition in the case:
\begin{align*}
0 &\geq N_l - 2(U_l + V_l) \\
&= N_l - 2N_l^1 + 2V_l\qquad&&\mbox{(By \eqref{App:Eq:Elp43})}\\
&= N_l^0 - N_l^1  + 2V_l&&\mbox{(Because $N_l^1 + N_l^0 = N_l$)}\\
\implies V_l &\geq \frac{N_l^0 - N_l^1}{2}.
\end{align*}
This concludes the derivation of Eq. \eqref{Eq:ElpSharp}.\\

As before, we derive the conditions on $U_l$ and $V_l$ that lead to the realization of the event $\eta_l^- = d$ by expanding its definition:
\begin{align}
\eta_l^- &= \eta_l - \max(\eta_l - N_l^t, 0)&&\mbox{(By definition of $\eta_l^-$)}\nonumber \\
&= N_l^1 - U_l - \max(N_l^1 - U_l - U_l - V_l, 0)\qquad&&\mbox{(By definition of $\eta_l$ and $N_l^t$)}\nonumber\\
&= N_l^1 - U_l - \max(N_l^1 - 2U_l - V_l, 0).\label{App:Eq:ElmExp}
\end{align}
Using this expansion, we see that there are two possible definitions for the event $\eta_l^- = d$ in terms of $U_l$ and $V_l$, both depending on how the $\max$ function is evaluated. \\

\noindent\textbf{Case 1 for $\eta_l^- = d$}:
\begin{align}
\max(N_l^1 - 2U_l - V_l, 0) = N_l^1 - 2U_l - V_l.\label{App:Eq:Elm11}
\end{align}
First, we use the above to rewrite the event $\eta_l^- = d$ for this case:
\begin{align}
d &= \eta_l^- = N_l^1 - U_l - \max(N_l^1 - 2U_l - V_l, 0)\qquad &&\mbox{(By \eqref{App:Eq:ElmExp})}\nonumber\\
&=N_l^1 - U_l - N_l^1 + 2U_l + V_l &&\mbox{(By \eqref{App:Eq:Elm11})}\nonumber\\
&= U_l + V_l \nonumber\\
\implies U_l &= d - V_l.\label{App:Eq:Elm12}
\end{align}
 Second, we use the result just derived to rewrite the condition for the max in \eqref{App:Eq:Elm11} in this case:
\begin{align*}
0 &\leq N_l^1 - 2U_l - V_l\qquad&&\mbox{(By \eqref{App:Eq:Elm11})} \\
&= N_l^1 - 2(d - V_l) - V_l &&\mbox{(By \eqref{App:Eq:Elm12})} \\
\implies V_l &\geq 2d - N_l^1.
\end{align*}
This and \eqref{App:Eq:Elm12} are, respectively, the first and second conditions on the first case in Equation \eqref{Eq:ElmSharp}.\\

\noindent\textbf{Case 2 for $\eta_l^- = d$}:
\begin{align}
\max(N_l^1 - 2U_l - V_l, 0) &= 0.\label{App:Eq:Elm21}
\end{align}
First, we use the above to rewrite the event $d = \eta_l^-$ for this case:
\begin{align}
d &= \eta_l^- = N_l^1 - U_l - \max(N_l^1 - 2U_l - V_l, 0)\qquad &&\mbox{(By \eqref{App:Eq:ElmExp})}\nonumber\\
&= N_l^1 - U_l&&\mbox{(By \eqref{App:Eq:Elm21})}\nonumber\\
\implies U_l &= N_l^1 - d. \label{App:Eq:Elm22}
\end{align}
Second, we rewrite the condition on the max with the above result:
\begin{align*}
0 &\geq N_l^1 - 2U_l - V_l \qquad &&\mbox{(By \eqref{App:Eq:Elm21})}\\
&= N_l^1 - 2(N_l^1 - d)  - V_l &&\mbox{(By \eqref{App:Eq:Elm22})}\\
\implies V_l &\geq 2d - N_l^1.
\end{align*}
These are the condition in the second case of Equation \eqref{Eq:ElmSharp}.\\

Finally we have the event $M_l = m$: conditions on $U_l$ and $V_l$ that lead to its realization can again be established by expanding its definition, this will lead to the form of $M_l$ in Equation \eqref{Eq:MSharp}. Starting with the definition of $M_l$:
\begin{align}
M_l &= \min(N_l^t, N_l^c)&&\mbox{(By definition of $M_l$)}\nonumber\\
&= \min(U_l + V_l, \eta_l + \nu_l)&&\mbox{(By definition of $N_l^t, N_l^c$)}\nonumber\\
&= \min(U_l + V_l, N_l^1 - U_l + N_l^0 - V_l)\qquad&&\mbox{(By definition of $\eta_l, \nu_l$)}\nonumber\\
&= \min(U_l + V_l, N_l - (U_l + V_l)).&&\mbox{(Beacause $N_l^1 + N_l^0 = N_l$)}\label{App:Eq:MlExp}
\end{align}
Because of this we see that the definition of $M_l$ is composed of two cases that depend on how the $\min$ function is resolved. \\

\noindent\textbf{Case 1 for $M_l = m$}:
\begin{align}
\min(U_l + V_l, N_l - (U_l + V_l)) = U_l + V_l.\label{App:Eq:Ml11}
\end{align}
First, use the expanded definition of $M_l$ with the case above to rewrite the event $M_l = m$:
\begin{align}
m &= M_l = \min(U_l + V_l, N_l - (U_l + V_l))\qquad&&\mbox{(By \eqref{App:Eq:MlExp})}\nonumber\\
&= U_l + V_l&&\mbox{(By \eqref{App:Eq:Ml11})}\label{App:Eq:Ml12}\\
\implies U_l &= m - V_l.\nonumber
\end{align}
Second, we use the result just introduced to rewrite the condition on the min for this case:
\begin{align*}
U_l + V_l &\leq N_l - (U_l + V_l)\qquad&&\mbox{(By \eqref{App:Eq:Ml11})}\\
\implies 0 &\geq U_l + V_l - N_l + (U_l + V_l) \\
&= 2U_l + 2V_l  - N_l\\
&= 2m - N_l&&\mbox{(By \eqref{App:Eq:Ml12})}\\
\implies m &\leq N_l/2.
\end{align*}
leading us to both conditions in the first case of Equation \eqref{Eq:MSharp}.\\

\noindent\textbf{Case 2 for $M_l = m$}:
\begin{align}
\min(U_l + V_l, N_l - (U_l + V_l)) &= N_l - (U_l + V_l).\label{App:Eq:Ml21}
\end{align}
Again, we use the case above to rewrite the definition of the event $M_l = m$:
\begin{align}
m &= M_l = \min(U_l + V_l, N_l - (U_l + V_l))\qquad&&\mbox{(By \eqref{App:Eq:MlExp})}\nonumber\\
&= N_l - (U_l + V_l) &&\mbox{(By \eqref{App:Eq:Ml21})}\label{App:Eq:Ml22}\\
\implies U_l &= N_l - V_l - m. \label{App:Eq:Ml23}
\end{align}
Second, we use the above to rewrite the condition for this case:
\begin{align*}
U_l + V_l &\geq N_l - (U_l + V_l) &&\mbox{(By \eqref{App:Eq:Ml21})}\\
\implies N_l - V_l - m + V_l &\geq m &&\mbox{(By \eqref{App:Eq:Ml22} and \eqref{App:Eq:Ml23})}\\
\implies m &\leq N_l/2.
\end{align*}
These are the two conditions in the definition of $M_l = m$ in equation $\eqref{Eq:MSharp}$. \\

Finally, the forms in equation \eqref{Eq:TruncatedSharp} are simply obtained by conditioning on the event $U_l = j, V_l = k$ for any of the definitions above. This concludes the proof of the lemma.
 \endproof

\subsection{Theorem~\ref{Thm:Sharp}}\label{App:Sec:SharpNull}
The joint null distribution of $\chichi$ is given in the following theorem:
\begin{theorem}{(Randomization Distribution of $\chichi$)}\label{Thm:Sharp}. For $l = 1, \dots, L$ let $(N_l^1, N_l^0, e_l)$ be fixed and known and let data, $\mathbb{D} = (U_l, V_l, \eta_l, \nu_l, N_l^t, N_l^c)$, be drawn from the data generating process of Equations \eqref{Eq:DGPSharp1} -- \eqref{Eq:DGPSharp2}. Let $\chi^+$ be the maximum of Formulation 2 on $\mathbb{D}$, and let $\chi^-$ be the minimum of the problem on the same variables. Let $N^m = \sum_{l=1}^L \min(N_l^1, N_l^0)$ and define: $\mathcal{X}_{N^m} = \left\{\frac{b - c - 1}{\sqrt{b+c+1}}: b, c \in \{0, \dots, N^m\}\right\}$, Additionally define:
\begin{align}
\mathcal{A}(y) &= \left\{\mathbf{a}=(a_1, \dots, a_l): \sum_{l=1}^L a_l = y,\, a_l \in\{0, \dots, y\}\right\},\label{Eq:DefAset}\\
\mathcal{B}(y,s) &= \left\{\mathbf{b}=(b_1, \dots, b_l): \sum_{l=1}^L b_l = \left(\frac{y-1}{s}\right)^2 - 1,\, b_l \in\left\{0, \dots,\left(\frac{y-1}{s}\right)^2 -1\right\}\right\},\\
\mathcal{C}(x) &= \left\{\mathbf{c}=(c_1, \dots, c_l): \sum_{l=1}^L c_l= x,\, c_l \in \{0,\dots,x\}\right\},\\
\mathcal{D}(x, r) &= \left\{\mathbf{d}=(d_1, \dots, d_l): \sum_{l=1}^L d_l = \left(\frac{x-1}{r}\right)^2-1,\, d_l \in \left\{0, \dots, \left(\frac{x-1}{r}\right)^2 -1\right\}\right\}.
\end{align}
Let $\mathcal{H}(x, y, r, s) = \mathcal{A}(y) \times \mathcal{B}(y,s) \times \mathcal{C}(x) \times \mathcal{D}(x, r)$ be the Cartesian product of the sets above, such that each element of $\mathcal{H}(x, y, r, s)$ is a 4-tuple of vectors: $(\mathbf{a}, \mathbf{b}, \mathbf{c}, \mathbf{d})$. Let $N^1 = \sum_{l=1}^L N_l^1$;  the pmf of $\chichi$, for two values $(s,r) \in \mathcal{X}_{N^m} \times  \mathcal{X}_{N^m}$ is:
\begin{align}
&\Pr(\chi^- = s, \chi^+ = r|\Hsharp) =\nonumber\\
&\sum_{x=-{N^m}}^{N^m}\sum_{y=-N^m}^{N^m}\sum_{(\mathbf{a}, \mathbf{b}, \mathbf{c}, \mathbf{d}) \in \mathcal{H}(x,y,r,s)}\prod_{l=1}^L\begin{cases}
h_1(a_l,b_l,c_l,d_l) \quad & \mbox{if } x < 1, y < 1 \\
h_2(a_l,b_l,c_l,d_l) \quad & \mbox{if } x \geq 1, y < 1 \\
h_3(a_l,b_l,c_l,d_l) \quad & \mbox{if } x \geq 1, y \geq 1 \\
0 \quad & \mbox{otherwise,}\end{cases}\label{Eq:SharpRange}
\end{align}
where:
\begin{align}
h_1(a_l, b_l, c_l, d_l) &= \ind(|a_l| = b_l)\sum_{m=0}^{N_l/2}\sum_{j=0}^{m}\biggl\{\nonumber\\
&\Pr\left(U_l^- = a_l+j, U_l^+ = \frac{2m - d_l + c_l}{2}, \eta_l^- = j, \eta_l^+ = \frac{2m - d_l -c_l}{2}, M_l =m\right)\nonumber\\
+ & \Pr\left(U_l^- = a_l + j, U_l^+ = \frac{d_l + c_l}{2}, \eta_l^- = j, \eta_l^+ = \frac{d_l - c_l}{2}, M_l = m\right)\ind(m \neq d_l)\biggr\}\\
h_2(a_l, b_l, c_l, d_l) &= \ind(|a_l| = b_l)\ind(|c_l| = d_l)\nonumber\\
+ &\sum_{m=0}^{N/2}\sum_{j=0}^{m}\sum_{k=0}^m \left\{\Pr(U_l^- = a_l + k, U_l^+ = c_l + j,  \eta_l^- = k, \eta_l^+ = j, M_l = m)\right\}\\
h_3(a_l, b_l, c_l, d_l) &= \ind(|c_l| = d_l)\sum_{m=0}^{N_l/2}\sum_{j=0}^{m}\biggl\{\nonumber\\
&\Pr\left(U_l^- = \frac{2m - b_l + a_l}{2}, U_l^+ = c_l + j, \eta_l^- = \frac{2m - b_l - a_l}{2}, \eta_l^+ = j,  M_l = m\right)\nonumber\\
+ &\Pr\left(U_l^- = \frac{b_l+a_l}{2}, U_l^+ = c_l + j, \eta_l^- = \frac{b_l-a_l}{2}, \eta_l^+ = j,  M_l = m\right)\ind(m \neq b_l)\biggr\}.
\end{align}
\end{theorem}
Note that the exact form of the probabilities in $h_1, h_2, h_3$ is given by Lemma \ref{Lem:TruncatedSharp}. The distribution is essentially a product of many binomial distributions truncated to be defined only on portions of their domains that include possible ranges allowed by the constraints on $\Agood$.

\proof{Proof.}All the probability statements throughout the proof are made conditionally on $X$ and $\Hsharp$, for this reason we omit the conditional notation from these statements. Recall also that all the quantities representing counts of units in each stratum are nonnegative integers. \\
Note first that, for any stratum $l$:
\begin{align}
0 \leq U_l^+ &= U_l - \max(U_l - N_l^c, 0)&&\mbox{(By definition of $U_l^+$)}\nonumber\\
&\leq N_l^1,&&\mbox{(By definition of $U_l$)}\label{App:Eq:UlpNl1}
\end{align}
and
\begin{align}
0 \leq \eta_l^+ &= \max(\eta_l - G_l^+, 0)&&\mbox{(By definition of $\eta_l^+$)}\\
& = \max(N_l^1 - U_l - G_l^+, 0)&&\mbox{(By definition of $\eta_l$)}\\
&\leq N_l^1.\label{App:Eq:ElpNl1}
\end{align}
This implies that:
\begin{align*}
TE^+ &= \sum_{l=1}^L U_l^+ - \eta_l^+&&\qquad\mbox{(By definition of $TE^+$)}\\
&\leq \sum_{l=1}^L N_l^1 - 0, &&\mbox{(By \eqref{App:Eq:UlpNl1} and \eqref{Eq:ElpSharp})}\\
&= N^1
\intertext{and: }
TE^+ &= \sum_{l=1}^L U_l^+ - \eta_l^+&&\qquad\mbox{(By definition of $TE^+$)}\\
&\geq \sum_{l=1}^L 0 - N_l^1. &&\mbox{(By \eqref{Eq:UlpSharp} and \eqref{App:Eq:ElpNl1})}\\
&= -N^1,
\end{align*}
where recall that $N^1 = \sum_{l=1}^L N_l^1$, as defined in the statement of the theorem. By the same argument we can see that $-N^1 \leq TE^- \leq N^1$. This fact is useful to bound the domain of $TE^+$ and $TE^-$: while it is likely that not all integers between $-N^1$ and $N^1$ have positive probability for $TE^+$ and $TE^-$ under the distribution of $\chichi$, we know that no integers outside the range above will have positive mass under that distribution.

To bound the domain of $\chi = \frac{B - C - 1}{\sqrt{B + C +1}}$ note that, by definition, $B_l$ represents the number of matched pairs in stratum $l$ that have outcome 1 for the treated unit and 0 for the control unit, independently of how matches are made. Because of this we know that $B_l$ can never be greater than either the number of treated units with outcome 1 or control units with outcome 0 in stratum $l$, and, therefore: $B_l \leq \min(N_l^1, N_l^0)$, independently of how many units are assigned to treatment. For the same reason we conclude that $C_l \leq \min(N_l^1, N_l^0)$ in every stratum, independently of how many units are treated or how the matches are made. Recall that $N^m = \sum_{l=1}^L \min(N_l^1, N_l^0)$, using these facts we can see that:
\begin{align}
\chi \in \mathcal{X}_{N^{m}} = \left\{\frac{b - c - 1}{\sqrt{b + c + 1}}: b, c \in \left\{0, \dots, N^m\right\} \right\}.
\end{align}
Note finally that, since $\chi \in \mathcal{X}_{N^{m}} $ regardless of how matches are made, we know that $\chi^+ \in \mathcal{X}_{N^{m}} $ and $\chi^- \in \mathcal{X}_{N^{m}} $ because $\chi^+$ and $\chi^-$ are special cases of $\chi$ in which matches are made with Algorithms \ref{App:Alg:MaxMcN} and \ref{App:Alg:MinMcN} respectively. Because of this we conclude that $\chichi \in \mathcal{X}_{N^{m}} \times \mathcal{X}_{N^{m}}$. This set is fast to enumerate computationally and summations over its elements can be performed efficiently. While the distribution of $\chichi$ likely does not place positive probability over all values in $\mathcal{X}_{N^{m}} \times \mathcal{X}_{N^{m}}$, we shall consider values in this set and derive conditions under which they have positive probability as well as what their probability is under this distribution.\\

\noindent Begin now by writing the pmf of the range $\chichi$. As stated in Theorem \ref{Thm:Sharp}, assume that $N_l^1, N_l^0$ are fixed and known for all strata $l=1, \dots, L$. For any two values $(r, s) \in \mathcal{X}_{N^{m}} \times \mathcal{X}_{N^{m}}$ we have:
\begin{align}
\Pr(\chi^- = s, \chi^+ = r) = &\Pr\left(\frac{TE^- - 1}{\sqrt{SD^- + 1}} = s, \frac{TE^+ - 1}{\sqrt{SD^+ + 1}} = r\right)&&\mbox{(By def. of $\chichi$)}\nonumber\\
= &\Pr\left(TE^- < 1, \frac{TE^--1}{\sqrt{S^-+1}} = s, TE^+ < 1, \frac{TE^+-1}{\sqrt{R^+ + 1}} = r \right)\qquad&&\mbox{(By Claim \ref{CDef})} \nonumber\\
+ &\Pr\left(TE^- < 1, \frac{TE^--1}{\sqrt{S^-+1}} = s, TE^+ \geq 1, \frac{TE^+-1}{\sqrt{S^++1}} = r\right) \nonumber\\
+ &\Pr\left(TE^- \geq 1, \frac{TE^--1}{\sqrt{R^-+1}} = s, TE^+ < 1, \frac{TE^+-1}{\sqrt{R^++1}} = r \right) \nonumber\\
+ &\Pr\left(TE^- \geq 1, \frac{TE^--1}{\sqrt{R^-+1}} = s, TE^+ \geq 1, \frac{TE^+-1}{\sqrt{S^++1}} = r \right) \label{App:Eq:f1}\\
= &f_1(r, z) + f_2(r, z) + f_3(r, z) + f_4(r, z). \nonumber
\end{align}
The equality in \eqref{App:Eq:f1} follows by the representation of $SD^+$ and $SD^-$ in \eqref{App:Eq:DefSDp} and \eqref{App:Eq:DefSDm} of Claim \ref{CDef} respectively. Now we work with each of the four parts separately.\\

Starting with $f_3$, we now show that it must always be that $f_3 = 0$.  Consider the event set $\{TE^- \geq 1, TE^+ < 1\}$. By definition of $\chi^+$, we know that, in this case:
\begin{align*}
\chi^- &= \frac{TE^- - 1}{\sqrt{SD^- + 1}}\qquad&&\mbox{(By definition of $\chi^-$)}\\
&\geq 0&&\mbox{(Because of the event $TE^- \geq 1$)}
\intertext{and:}
\chi^+ &= \frac{TE^+ - 1}{\sqrt{SD^+ + 1}}\qquad&&\mbox{(By definition of $\chi^+$)}\\
&< 0.&&\mbox{(Because of the event $TE^+ < 1$)}
\end{align*}
Therefore, the event $\{TE^- \geq 1, TE^+ < 1\}$ implies the event $\{\chi^+ < \chi^-\}$, but this is a contradiction, since $\chi^+$ and $\chi^-$ are, respectively, the maximum and minimum of the same optimization problem (Formulation 2). Because of this it will never be the case that $\chi^+ < \chi^-$. It must be, then, that the probability of this event under the distribution of interest is 0, and that, therefore, the probability of the event $\{TE^- \geq 1, TE^+ < 1\}$ is also 0. Since $f_3$ is the probability of the event set $\{TE^- \geq 1, \frac{TE^--1}{\sqrt{R^-+1}} = s, TE^+ < 1, \frac{TE^+-1}{\sqrt{R^++1}} = r \}$, which is included in the 0-probability set above, then it must be that $f_3 = 0$.\\

This leaves us with $f_1, f_2, f_4$ to write out. For convenience we repeat the definitions of the sets introduced in the statement of the theorem:
\begin{align}
\mathcal{A}(y) &= \left\{\mathbf{a}=(a_1, \dots, a_l): \sum_{l=1}^L a_l = y,\, a_l \in\{0, \dots, y\}\right\},\\
\mathcal{B}(y,s) &= \left\{\mathbf{b}=(b_1, \dots, b_l): \sum_{l=1}^L b_l = \left(\frac{y-1}{s}\right)^2 - 1,\, b_l \in\left\{0, \dots,\left(\frac{y-1}{s}\right)^2 -1\right\}\right\},\\
\mathcal{C}(x) &= \left\{\mathbf{c}=(c_1, \dots, c_l): \sum_{l=1}^L c_l= x,\, c_l \in \{0,\dots,x\}\right\},\\
\mathcal{D}(x, r) &= \left\{\mathbf{d}=(d_1, \dots, d_l): \sum_{l=1}^L d_l = \left(\frac{x-1}{r}\right)^2 - 1,\, d_l \in \left\{0, \dots, \left(\frac{x-1}{r}\right)^2 -1\right\}\right\}\label{Eq:DefDset}.
\end{align}
In addition, let $\mathcal{H}(x, y, r, s) = \mathcal{A}(y) \times \mathcal{B}(y,s) \times \mathcal{C}(x) \times \mathcal{D}(x, r)$ be the Cartesian product of the sets above, such that each element of $\mathcal{H}(x, y, r, s)$ is a 4-tuple of vectors each of length $L$: $(\mathbf{a}, \mathbf{b}, \mathbf{c}, \mathbf{d})$.
We are now ready to derive $f_1(r,s)$:
\begin{align}
f_1(r, s) &= \Pr\left(TE^- < 1, \frac{TE^--1}{\sqrt{S^- + 1}} = s, TE^+ < 1, \frac{TE^+-1}{\sqrt{R^+ + 1}} = r \right)\label{App:Eq:f11}\\
&= \sum_{y = -N^1}^0\sum_{x = -N^1}^0 \Pr\left(TE^- = y, \frac{TE^- - 1}{\sqrt{S^- + 1}} = s,TE^+ = x, \frac{TE^+ - 1}{\sqrt{R^+ + 1}} = r\right)\label{App:Eq:f12}\\
&= \sum_{y = -N^1}^0\sum_{x = -N^1}^0 \Pr\left(TE^- = y, S^- = \left(\frac{y-1}{s}\right)^2 - 1, TE^+ = x, R^+ = \left(\frac{x-1}{r}\right)^2 - 1\right)\label{App:Eq:f13}\\
&=  \sum_{y = -N^1}^0\sum_{x = -N^1}^0\Pr\left(\sum_{l=1}^LTE_l^- = y, \sum_{l=1}^LS_l^- = \left(\frac{y-1}{s}\right)^2 - 1, \sum_{l=1}^LTE_l^+ = x, \sum_{l=1}^LR_l^+ = \left(\frac{x-1}{r}\right)^2 - 1\right)\\
&= \sum_{y = -N^1}^0 \sum_{x = -N^1}^0\sum_{(\mathbf{a}, \mathbf{b}, \mathbf{c}, \mathbf{d}) \in \mathcal{H}(x, y, r, s)}\Pr(TE^-_1 = a_1, S^-_1=b_1, TE^+_1=c_1,R^+=d_1, \dots, \label{App:Eq:f14}\\
&\qquad\qquad\qquad\qquad\qquad\qquad\qquad\qquad TE^-_L = a_L, S^-_L=b_L, TE^+_L=c_L,R^+=d_L)\label{App:Eq:f15}\\
&=  \sum_{y = -N^1}^0\sum_{x = -N^1}^0\sum_{(\mathbf{a}, \mathbf{b}, \mathbf{c}, \mathbf{d}) \in \mathcal{H}(x, y, r, s)}\prod_{l=1}^L\Pr\left(TE_l^- = a_l, S^-_l = b_l, TE_l^+ = c_l, R_l^+ = d_l\right)\label{App:Eq:f16}.
\end{align}
Line \eqref{App:Eq:f11} follows from the definition of $f_1$ given in \eqref{App:Eq:f1}. Equation \eqref{App:Eq:f13} follows from rearranging the terms in the previous line. Equation \eqref{App:Eq:f14} follows from the definitions in Claim \ref{CDef}. Equation \eqref{App:Eq:f15} follows from the definition of $\mathcal{H}(x, y, r, s)$, and equation \eqref{App:Eq:f16} follows from independence of $TE^+_l, TE^-_l, S^+_l, S^-_l, R^+_l, R_l^-$ in stratum $l$ from the same quantities in any other stratum. This independence is evident from the definitions of these quantities given in Claim \ref{CDef}.
The same exact derivation steps leads us to a similar definition for $f_2$:
\begin{align*}
f_2(r, s) &= \Pr\left(TE^- < 1, \frac{TE^- - 1}{\sqrt{S^-}} = s, TE^+ \geq 1, \frac{TE^+ - 1}{\sqrt{S^+}} = r\right)\\
&= \sum_{y=-N^1}^0\sum_{x=1}^{N^1}\Pr\left(TE^- = y , \frac{TE^- - 1}{\sqrt{S^-+1}} = s, TE^+ = x, \frac{TE^+ - 1}{\sqrt{S^++1}} = r\right)\\
&= \sum_{y=-N^1}^0\sum_{x = 1}^{N^1}\Pr\left(TE^- = y, S^- = \left(\frac{y-1}{s}\right)^2 - 1, TE^+ = x, S^+ = \left(\frac{x-1}{r}\right)^2 - 1\right)\\
&= \sum_{y=-N^1}^0\sum_{x = 1}^{N^1}\Pr\left(\sum_{l=1}^LTE_l^- = y, \sum_{l=1}^LS_l^- = \left(\frac{y-1}{s}\right)^2 - 1, \sum_{l=1}^LTE_l^+ = x, \sum_{l=1}^LS_l^+ = \left(\frac{x-1}{r}\right)^2 - 1\right)\\
&=  \sum_{y = -N^1}^0\sum_{x = 1}^{N^1}\sum_{(\mathbf{a}, \mathbf{b}, \mathbf{c}, \mathbf{d}) \in \mathcal{H}(x, y, r, s)}\Pr(TE^-_1 = a_1, S^-_1=b_1, TE^+_1=c_1,S^+=d_1, \dots\\
&\qquad\qquad\qquad\qquad\qquad\qquad\qquad\qquad TE^-_L = a_L, S^-_L=b_L, TE^+_L=c_L,S^+=d_L)\\
&= \sum_{y=-N^1}^0\sum_{x = 1}^{N^1}\sum_{(\mathbf{a}, \mathbf{b}, \mathbf{c}, \mathbf{d}) \in \mathcal{H}(x,y,r,s)}\prod_{l=1}^L\Pr\left(TE_l^- = a_l, S^-_l = b_l, TE_l^+ = c_l, S_l^+ = d_l\right).
\end{align*}
In the case of $f_4$ we can follow the same steps to derive:
\begin{align*}
f_4(r,s) &= (TE^- \geq 1, \frac{TE^- - 1}{\sqrt{R^- +1}} = s, TE^+ \geq 1, \frac{TE^+ - 1}{\sqrt{S^+ + 1}} = r)\\
&= \sum_{y=1}^{N^1}\sum_{x=1}^{N^1}\Pr\left(TE^- = y, \frac{TE^- - 1}{\sqrt{R^-+1}} = s, TE^+ = x , \frac{TE^+ - 1}{\sqrt{S^++1}} = r\right)\\
&= \sum_{y=1}^{N^1}\sum_{x = 1}^{N^1}\Pr\left(TE^- = y, R^- = \left(\frac{y-1}{s}\right)^2 - 1, TE^+ = x, S^+ = \left(\frac{x-1}{r}\right)^2 - 1\right)\\
&= \sum_{y=1}^{N^1}\sum_{x = 1}^{N^1}\Pr\left(\sum_{l=1}^LTE_l^- = y, \sum_{l=1}^LR_l^- = \left(\frac{y-1}{s}\right)^2 - 1, \sum_{l=1}^LTE_l^+ = x, \sum_{l=1}^LS_l^+ = \left(\frac{x-1}{r}\right)^2 - 1\right)\\
&=  \sum_{y = 1}^{N^1}\sum_{x = 1}^{N^1}\sum_{(\mathbf{a}, \mathbf{b}, \mathbf{c}, \mathbf{d}) \in \mathcal{H}(x, y, r, s)}\Pr(TE^-_1 = a_1, R^-_1=b_1, TE^+_1=c_1, S^+=d_1, \dots\\
&\qquad\qquad\qquad\qquad\qquad\qquad\qquad\qquad TE^-_L = a_L, R^-_L=b_L, TE^+_L=c_L,S^+=d_L)\\
 &= \sum_{y=1}^{N^1}\sum_{x = 1}^{N^1}\sum_{(\mathbf{a}, \mathbf{b}, \mathbf{c}, \mathbf{d}) \in \mathcal{H}(x,y,r,s)}\prod_{l=1}^L\Pr\left(TE_l^- = a_l, R^-_l = b_l, TE_l^+ = c_l, S_l^+ = d_l\right).
\end{align*}
Note that $f_1$, $f_2$ and $f_4$ are nonzero on disjoint parts of the sets $\{-N^1, \dots, N^1\}$, as well as the fact that the definition of $\mathcal{A}(y), \mathcal{B}(y, z), \mathcal{C}(x), \mathcal{D}(x, r)$ and $\mathcal{H}(x,y,r,s)$ doesn't change for any of the functions. For these reasons we can write the pmf of the range as in Eq~\eqref{Eq:SharpRange}.
It remains to show that the three functions $h_1$, $h_2$, $h_3$ have the form stated in the theorem. This can be accomplished by expanding the inner probabilities of $TE_l^+, TE^-_l, SD^+_l, SD^-_l$ in each stratum. Before doing this, it will be useful to note that, for any stratum $l$:
\begin{align*}
0 \leq M_l &= \min(N_l^t, N_l^c) &&\mbox{(By definition of $M_l$)}\\
&= \min(N_l^t, N_l - N_l^t) && \mbox{(By definition of $N_l^c$)}\\
&\leq N_l/2.
\end{align*}
Because of this we consider values of $M_l$ in the integer range $0, \dots, N_l/2$. Beginning with $\Pr(TE_l^- = a_l, S_l^- = b_l, TE_l^+ = c_l, R_l^+ = d_l)$:
\begin{align}
&\Pr(TE_l^- = a_l, S_l^- = b_l, TE_l^+ = c_l, R_l^+ = d_l)\\
=& \sum_{m = 0}^{N_l/2}\Pr(U_l^- - \eta_l^- = a_l, |U_l^- - \eta_l^-| = b_l, U_l^+ - \eta_l^+ = c_l, M_l - |U_l^+ + \eta_l^+ - M_l| = d_l, M_l = m)\label{App:Eq:h11}\\
=& \sum_{m = 0}^{N_l/2}\Pr(|U_l^- - \eta_l^-| = b_l, U_l^+ - \eta_l^+ = c_l, m - |U_l^+ + \eta_l^+ - m| = d_l, M_l = m|U_l^- - \eta_l^- = a_l)\nonumber\\
&\quad\times\Pr(U_l^- - \eta_l^- = a_l)\label{App:Eq:h12}\\
=& \sum_{m = 0}^{N_l/2}\Pr(|a_l| = b_l, U_l^+ - \eta_l^+ = c_l, m - |U_l^+ + \eta_l^+ - m| = d_l, M_l = m|U_l^- - \eta_l^- = a_l)\nonumber\\
&\quad\times\Pr(U_l^- - \eta_l^- = a_l)\label{App:Eq:h13}\\
=& \sum_{m = 0}^{N_l/2}\ind(|a_l| = b_l)\Pr(U_l^+ - \eta_l^+ = c_l, m - |U_l^+ + \eta_l^+ - m| = d_l, M_l = m|U_l^- - \eta_l^- = a_l)\nonumber\\
&\quad\times\Pr(U_l^- - \eta_l^- = a_l)\label{App:Eq:h14}\\
=& \ind(|a_l| = b_l)\sum_{m = 0}^{N_l/2}\Pr(U_l^- - \eta_l^- = a_l, U_l^+ - \eta_l^+ = c_l, |U_l^+ + \eta_l^+ - m| = m - d_l, M_l = m)\label{App:Eq:h15}\\
=& \ind(|a_l| = b_l)\sum_{m = 0}^{N_l/2}\Pr(U_l^- - \eta_l^- = a_l, U_l^+ - \eta_l^+ = c_l,  U_l^+ - \eta_l^+ + m = m - d_l, M_l = m)\nonumber\\
&\qquad\qquad\qquad + \Pr(U_l^- - \eta_l^- = a_l, U_l^+ - \eta_l^+ = c_l, U_l^+ - \eta_l^+ + m = d_l - m, M_l = m)\ind(m \neq d_l)\label{App:Eq:h16}\\
= &\mathbb{I}(|a_l| = b_l)\sum_{m = 0}^{N_l/2}\Pr\left(U_l^- = a_l + \eta_l^-, U_l^+ = c_l + \eta_l^+, \eta_l^+ = \frac{2m - d_l - c_l}{2}, M_l = m\right)\nonumber\\
&\qquad\qquad\qquad\;+\Pr\left(U_l^- = a_l + \eta_l^-, U_l^+ = c_l + \eta_l^+, \eta_l^+ = \frac{d_l - c_l}{2}, M_l = m\right)\ind(m \neq d_l)\label{App:Eq:h17}\\
= &\mathbb{I}(|a_l| = b_l)\sum_{m = 0}^{N_l/2}\sum_{j=0}^{m}\Pr\left(U_l^- = a_l + j, U_l^+ = \frac{2m - d_l + c_l}{2}, \eta_l^- = j, \eta_l^+ = \frac{2m - d_l - c_l}{2}, M_l = m\right)\nonumber\\
& \qquad\qquad\qquad\quad + \Pr\left(U_l^- = a_l + j, U_l^+ = \frac{d_l + c_l}{2},\eta_l^- = j, \eta_l^+ = \frac{d_l - c_l}{2}, M_l = m\right)\ind(m \neq d_l)\label{App:Eq:h18}\\
&= h_1(a_l, b_l, c_l, d_l).
\end{align}
Equation \eqref{App:Eq:h11} follows from the representations of $TE$ and $SD$ given in Claim \ref{CDef}, Equation \eqref{App:Eq:h12} from the definition of conditional probability, Equation \eqref{App:Eq:h14} from the fact that $a$ and $b$ are constants and therefore they are independent from the other quantities in the equation. Equation \eqref{App:Eq:h15} follows from the definition of conditional probability. Equations \eqref{App:Eq:h16} follows from the fact that the event $\{M_l - |U_l^+ + \eta_l^+ - M_l| = d_l\}$ is equal to the event $\{M_l - U_l^+ - \eta_l^+ + M_l = d_l\} \cup \{M_l + U_l^+ + \eta_l^+ - M_l = d_l, M_l \neq d_l\}$, and therefore the probability of its occurrence is equal to the sum of the probability of these two events. Equation \eqref{App:Eq:h17} follow by rearranging the terms in the previous line, and Equation \eqref{App:Eq:h18} from summing over values of $\eta_l^-$. The final line of the derivation is from the definition of $h_1$ given in the statement of the theorem. The following derivations for $h_2$ and $h_3$ follow exactly the same steps, starting with $h_2$:
\begin{align*}
&\Pr(TE_l^- = a_l, S_l^- = b_l, TE_l^+ = c_l, S_l^+ = d_l)\\
=&\sum_{m=0}^{N_l/2}\Pr(U_l^- - \eta_l^- = a_l, |U_l^- - \eta_l^-| = b_l, U_l^+ - \eta_l^+ = c_l, |U_l^+ - \eta_l^+| = d_l, M_l = m)\\
=&\sum_{m=0}^{N_l/2}\Pr(|U_l^- - \eta_l^-| = b_l,  |U_l^+ - \eta_l^+| = d_l, M_l = m|U_l^- - \eta_l^- = a_l, U_l^+ - \eta_l^+ = c_l)\\
&\quad\times\Pr(U_l^- - \eta_l^- = a_l, U_l^+ - \eta_l^+ = c_l)\\
=&\sum_{m=0}^{N_l/2}\Pr(|a_l| = b_l,  |c_l| = d_l, M_l = m|U_l^- - \eta_l^- = a_l, U_l^+ - \eta_l^+ = c_l)\Pr(U_l^- - \eta_l^- = a_l, U_l^+ - \eta_l^+ = c_l)\\
=&\sum_{m=0}^{N_l/2}\mathbb{I}(|a_l| = b_l)\mathbb{I}(|c_l| = d_l)\Pr(M_l = m|U_l^- - \eta_l^- = a_l, U_l^+ - \eta_l^+ = c_l)\Pr(U_l^- - \eta_l^- = a_l, U_l^+ - \eta_l^+ = c_l)\\
=&\mathbb{I}(|a_l| = b_l)\mathbb{I}(|c_l| = d_l)\sum_{m=0}^{N_l/2}\Pr(U_l^- - \eta_l^- = a_l, U_l^+ - \eta_l^+ = c_l, M_l = m)\\
=&\mathbb{I}(|a_l| = b_l|)\mathbb{I}(|c_l| = d_l)\sum_{m=0}^{N_l/2}\sum_{j=0}^{m}\sum_{k = 0}^{m}\Pr(U_l^- = a_l + k, U_l^+ = c_l + j, \eta_l^- = k, \eta_l^+ = j, M_l = m)\\
=&h_2(a_l, b_l, c_l, d_l).
\end{align*}
Finally, $h_3$ can be derived from:
\begin{align*}
&\Pr(TE_l^- = a_l, R_l^- = b_l, TE_l^+ = c_l, S_l^+ = d_l)\\
=& \sum_{m=0}^{N_l/2}\Pr(U_l^- - \eta_l^- = a_l, M_l - |U_l^- + \eta_l^- - M_l| = b_l, U_l^+ - \eta_l^+ = c_l, |U_l^+ - \eta_l^+| = d_l, M_l = m)\\
=& \sum_{m=0}^{N_l/2}\Pr(U_l^- - \eta_l^- = a_l, m - |U_l^- + \eta_l^- - m| = b_l, |U_l^+ - \eta_l^+| = d_l, M_l = m| U_l^+ - \eta_l^+ = c_l)\\
&\quad\times\Pr(U_l^+ - \eta_l^+ = c_l)\\
=& \sum_{m=0}^{N_l/2}\Pr(U_l^- - \eta_l^- = a_l, m - |U_l^- + \eta_l^- - m| = b_l, |c_l| = d_l, M_l = m| U_l^+ - \eta_l^+ = c_l)\\
&\quad\times\Pr(U_l^+ - \eta_l^+ = c_l)\\
=& \sum_{m=0}^{N_l/2}\ind(|c_l| = d_l)\Pr(U_l^- - \eta_l^- = a_l, m - |U_l^- + \eta_l^- - m| = b_l, M_l = m|U_l^+ - \eta_l^+ = c_l)\\
&\quad\times\Pr(U_l^+ - \eta_l^+ = c_l)\\
=& \ind(|c_l| = d_l)\sum_{m=0}^{N_l/2}\Pr(U_l^- - \eta_l^- = a_l, m - |U_l^- + \eta_l^- - m| = b_l, U_l^+ - \eta_l^+ = c_l, M_l = m)\\
=& \ind(|c_l| = d_l)\sum_{m=0}^{N_l/2}\Pr(U_l^- - \eta_l^- = a_l, m - U_l^- - \eta_l^- + m = b_l, U_l^+ - \eta_l^+ = c_l, M_l = m)\\
& \qquad\qquad\qquad+\Pr(U_l^- - \eta_l^- = a_l, m + U_l^- + \eta_l^- - m = b_l, U_l^+ - \eta_l^+ = c_l, M_l = m)\ind(m \neq b_l)\\
=& \mathbb{I}(|c_l| = d_l)\sum_{m=0}^{N_l/2}\Pr(U_l^- = a_l + \eta_l^-, U_l^-=m-b_l + m - \eta_l^-, U_l^+ = c_l + \eta_l^+, M_l =m)\\
& \qquad\qquad\qquad+ \Pr(U_l^- = a_l + \eta_l^-, U_l^- = b_l - \eta_l^-, U_l^+ = c_l + \eta_l^+, M_l = m)\ind(m \neq b_l)\\
=& \mathbb{I}(|c_l| = d_l)\sum_{m=0}^{N_l/2}\sum_{j=1}^{m}\Pr\left(U_l^- = \frac{2m - b_l + a_l}{2}, U_l^+ = c_l + j, \eta_l^- = \frac{2m - b_l - a_l}{2}, \eta_l^+ = j, M_l =m\right)\\
&\qquad \qquad \qquad \quad + \Pr\left(U_l^- = \frac{b_l +a_l}{2}, U_l^+ = c_l + j, \eta_l^- = \frac{b_l - a_l}{2},\eta_l^+ = j, M_l =m\right)\ind(m \neq b_l)\\
=& h_3(a_l, b_l, c_l, d_l).
\end{align*}
This concludes the derivation of all the quantities in Theorem \ref{Thm:Sharp}.  \endproof

\section{Distribution of $\chichi$ Under Exclusively Binning Constraints}\label{App:Sec:Null}

To compute the  distribution we seek, we define the quantities $p_l^t = \Pr(Y = 1|T = 1, X = x_i, i \in S_l)$ and $\Pr(Y = 1|T = 0, X = x_j, j \in S_l)$.  We can encode $\Hzero$ in this distribution by requiring: $p_l^t = p_l^c = p_l$ for all $l$. In order for the distribution of interest to correctly be estimated, Assumption~\ref{As:StratIgnorability} must hold. Note that this assumption implies directly that $\Pr(Y(t)=1|X = x_i, i \in S_l) = \Pr(Y=1|T = t, X = x_i, i \in S_l)$. With it, and by requiring $p_l^t = p_l^c$ we obtain $\Hzero$: $\Pr(Y = 1|T = 1, X = x_i, i \in S_l) = \Pr(Y(1)=1|X = x_i, i \in S_l) = \E(Y(1)|X = x_i, i \in S_l) = \E(Y(0)|X = x_i, i \in S_l) = \Pr(Y(0)=1|X = x_i, i \in S_l) = \Pr(Y=1|T= 0, X = x_i, i \in S_l)$.  As stated before, while Assumption \ref{As:StratIgnorability} is necessary to obtain the distribution of $\chichi$.

An estimate of $p_l^t$ and $p_l^c$ can be produced in an unbiased and consistent way with $\hat{p}_l^t = \frac{\sum_{i \in S_l}{y_i^t}}{N_l^t}$, or any other estimator of choice. When one wants to test the null hypothesis that $p_l^t=p_l^c$, one can use $\hat{p}_l^t=\hat{p}_l^c = \frac{\sum_{i \in S_l}y_i^t + \sum_{j \in S_l}y_j^c}{N_l}$.

Recall that exactly $M_l = \min(N_l^t, N_l^c)$ matches are made, with $M = \sum_{l=1}^L M_l$. Now denote with $N_l^t$ the number of treated units in stratum $S_l$ and with $N_l^c$ the number of control units in that same stratum. For each stratum, $l=1, \dots, L$, the data are generated as follows:
\begin{alignat}{1}
U_l &\overset{iid}{\sim} Bin(p_l^t, N_l^t)\label{Eq:ConditionalDGP1} \\
V_l &= N_l^t - U_l \\
\eta_l &\overset{iid}{\sim} Bin(p_l^c, N_l^c) \\
\nu_l &= N_l^c - \eta_l.\label{Eq:ConditionalDGP2}
\end{alignat}

The joint distribution of the truncated variables introduced in Equations \eqref{Eq:Trunc1}-\eqref{Eq:Trunc4} under this DGP is given in the following lemma:
\begin{lemma}{(Distribution of Truncated Variables)}\label{Lem:TruncatedConditional} For all $l = 1, \dots, L$, let $(N_l^t, N_l^c, p_l^t, p_l^c)$ be fixed and known, and let $G_l^+ = \max(N_l^c - N_l^t, 0)$, $G_l^- = \max(N_l^t - N_l^c, 0)$. Let data $\mathbb{D} = \{(U_l, V_l, \eta_l, \nu_l)\}_{l=1}^L$ be drawn according to Equations \eqref{Eq:ConditionalDGP1}-\eqref{Eq:ConditionalDGP2}. The truncated variables have the following joint distributions:
\begin{align}
\Pr(U_l^- = a, U_l^+ = b) &= \begin{cases}
\Pr(U_l = b) &\mbox{if } b < \min(N_l^c, G_l^- + 1),\; a = 0\\
\Pr(U_l = b) &\mbox{if } b = a + G_l^-,\; 0 < a <N_l^c - G_l^-\\
\Pr(N_l^c \leq U_l \leq G_l^-) &\mbox{if } b = N_l^c,\; a = 0\\
\Pr(U_l = a + G_l^-) &\mbox{if } b = N_l^c,\; a \geq \max(N_l^c - G_l^-,1)\\
0 &\mbox{otherwise.}\label{Eq:TruncatedATE}
\end{cases}\\
\Pr(\eta_l^- = a, \eta_l^+ = b) &=
\begin{cases}
\Pr(\eta_l = a) &\mbox{if } a < \min(N_l^t, G_l^+ + 1),\; b = 0\\
\Pr(\eta_l = a) &\mbox{if } a = b + G_l^+,\; 0 < b <N_l^t - G_l^+\\
\Pr(N_l^c \leq \eta_l \leq G_l^+) &\mbox{if } a = N_l^t,\; b = 0\\
\Pr(\eta_l = b + G_l^+) &\mbox{if } a = N_l^t,\; b \geq \max(N_l^t - G_l^+,1)\\
0 &\mbox{otherwise.}\\
\end{cases}
\end{align}
where $\Pr(U_l = x) = Bin(x, p_l^t, N_l^t)$, $\Pr(x \leq U_l \leq y) = \sum_{z = x}^y Bin(z, p_l^t, N_l^t)$, $\Pr(\eta_l = x) = Bin(x, p_l^c, N_l^c)$, and $\Pr(x \leq \eta_l \leq y) = \sum_{z = x}^y Bin(z, p_l^c, N_l^c)$.
\end{lemma}

\proof{Proof} We will prove the result for $\Pr(U_l^- = a, U_l^+ = b)$ only, as the proof for $\Pr(\eta_l^- = a, \eta_l^+ = b)$ is symmetrical. Recall first that we are now in a case in which the number of treated and control units in each stratum, $N_l^t$ and $N_l^c$ respectively, are fixed.  We use this together with Assumption \ref{As:StratIgnorability} to write the marginal distributions of $U_l^+$ and $U_l^-$:
\begin{align*}
\Pr(U_l^+ = b) &= \Pr(U_l - \max(U_l  - N_l^c, 0) = b) = \begin{cases} \Pr(U_l = b) \quad \mbox{if } b < N_l^c \\ \Pr(U_l \geq N_l^c) \quad \mbox{if } b=N_l^c\end{cases}\\
\Pr(U_l^- = a) &= \Pr(\max(U_l - G_l^-, 0) = a) = \begin{cases}\Pr(U_l \leq G_l^-) \quad \mbox{if } a = 0\\\Pr(U_l = a + G_l^-) \quad \mbox{if } a > 0.\end{cases}
\end{align*}
In the above, $G_l^- = \max(N_l^t - N_l^c, 0)$, a fixed quantity. The first equality in both formulas follows from the definition of $U_l^+$ and $U_l^-$ given in Equations \eqref{Eq:Trunc1} and \eqref{Eq:Trunc2} respectively. The two cases of each definition follow directly from these definitions. Given the marginal distributions above, we can easily find distributions for $U_l^+$ and $U_l^-$ conditional on $U_l = x$:
\begin{align}
\ind(U_l^+ = b|U_l = x) &= \begin{cases} 1 \quad \mbox{if } b < N_l^c, x = b\\ 1 \quad \mbox{if } b=N_l^c, x \geq N_l^c\\ 0 \quad \mbox{otherwise.}\end{cases}\label{App:Eq:UlpMarg}\\
\ind(U_l^- = a|U_l = x) &= \begin{cases}1 \quad \mbox{if } a = 0, x \leq G_l^-\\ 1 \quad \mbox{if } a > 0, x = a + G_l^-\\ 0 \quad \mbox{otherwise.}\end{cases}\label{App:Eq:UlmMarg}
\end{align}
This is because, if the number of treated and control units in each stratum is fixed, randomness in $U_l^+$ and $U_l^-$ comes only from $U_l$, as is evident from the definitions of $U_l^+$ and $U_l^-$ given in Eqns. \eqref{Eq:UlpSharp} and  \eqref{Eq:UlmSharp}. Therefore once that is fixed the two quantities become constant and, as such, independent of each other. Given this fact, it is evident that $\Pr(U_l^- = a, U_l^+ = b|U_l =x) = \ind(U_l^- = a, U_l^+ = b|U_l =x) = \ind(U_l^- = a|U_l=x)\ind(U_l^+ = b|U_l=x)$. This product of indicator functions can also be written as four conditions, each one representing the intersection of the event sets on which both distributions place nonzero probability:
\begin{align*}
\ind(U_l^- = a|U_l=x)\ind(U_l^+ = b|U_l = x) &= \begin{cases}
1 &\mbox{if } b < N_l^c, x=b \mbox{ and }a = 0, x \leq G_l^-\\
1 &\mbox{if } b < N_l^c, x=b \mbox{ and }a > 0, x = a + G_l^-\\
1 &\mbox{if } b = N_l^c, x \geq N_l^c \mbox{ and }a = 0, x \leq G_l^-\\
1 &\mbox{if } b = N_l^c, x \geq N_l^c \mbox{ and }a > 0, x = a + G_l^-\\
0 &\mbox{otherwise.}
\end{cases}
\end{align*}
This is evident from Equations \eqref{App:Eq:UlpMarg} and \eqref{App:Eq:UlmMarg}. The event sets in the equation can be simplified further as some of their elements are redundant, we show this in the following.\\

\textbf{Case 1}: $b < N_l^c, x=b$ and $a = 0, x \leq G_l^-$.\\
First, using the second and fourth condition we can write $b \leq G_l^-$. Then, since $b$ has to be less than both $G_l^-$ and $N_l^c$ we can rewrite the condition as: $b < \min(N_l^c, G_l^- + 1)$, where we replace the equality on $G_l^-$ with $b$ being strictly less than $G_l^- + 1$. This is possible because $G_l^+$ is, by definition, a nonnegative integer. The final event here becomes: $x = b,\; b < \min(N_l^c, G_l^- + 1),\; a = 0$. \\

\textbf{Case 2}: $b < N_l^c, x=b$ and $a > 0, x = a + G_l^-$.\\
First, using the first, second and fourth condition we have: $N_l^c > b = x = a + G_l^-$, which we rewrite as the event: $x = b, b = a + G_l^-, a < N_l^c - G_l^-$. With this the final event becomes: $x = b,\; b = a + G_l^-,\; 0 < a < N_l^c - G_l^-$.\\

\textbf{Case 3}: $b = N_l^c, x \geq N_l^c$ and $a = 0, x \leq G_l^-$.\\
Here we can combine the second and fourth conditions to obtain: $N_l^c \leq x \leq G_l^-$. This leads to the event: $b = N_l^c,\; N_l^c \leq x \leq G_l^-,\; a = 0$. \\

\textbf{Case 4}: $b = N_l^c, x \geq N_l^c$ and $a > 0, x = a + G_l^-$.\\
First from the second and fourth events we have: $x = a + G_l^- \geq N_l^c$, which can be used to obtain the event: $a \geq N_l^c - G_l^-$. We also rewrite the third condition as $a \geq 1$, leading to the following final representation for the event of this case: $b = N_l^c,\; a \geq \max(N_l^c - G_l^-,1),\; x = a + G_l^-$.\\

Finally, we can put all these events together to obtain a simplified formulation for the joint conditional distribution of $U_l^+, U_l^-$:
\begin{align*}
\ind(U_l^- = a, U_l^+ = b|U_l = x) &= \begin{cases}
1 &\mbox{if } x = b,\; b < \min(N_l^c, G_l^- + 1),\; a = 0\\
1 &\mbox{if } x = b,\; b = a + G_l^-,\; 0 < a < N_l^c - G_l^-\\
1 &\mbox{if } b = N_l^c,\; N_l^c \leq x \leq G_l^-,\; a = 0\\
1 &\mbox{if } b = N_l^c,\; a \geq \max(N_l^c - G_l^-,1),\; x = a + G_l^-\\
0 &\mbox{otherwise.}
\end{cases}
\end{align*}
The marginal distribution of $U_l^+, U_l^-$ can now be found by using the law of total probability to sum over $U_l^+$:
\begin{align*}
\Pr(U_l^- = a, U_l^+ = b) &= \sum_{x = 0}^{N_l^t}\Pr(U_l = x)\ind(U_l^+ = a, U_l^- = b|U_l = x)\\
&= \sum_{x = 0}^{N_l^t} \Pr(U_l = x)\begin{cases}
1 &\mbox{if } x = b,\; b < \min(N_l^c, G_l^- + 1),\; a = 0\\
1 &\mbox{if } x = b,\; b = a + G_l^-,\; 0 < a < N_l^c - G_l^-\\
1 &\mbox{if } b = N_l^c,\; N_l^c \leq x \leq G_l^-,\; a = 0\\
1 &\mbox{if } b = N_l^c,\; a \geq \max(N_l^c - G_l^-,1),\; x = a + G_l^-\\
0 &\mbox{otherwise.}
\end{cases}\\
&= \sum_{x = 0}^{N_l^t}\begin{cases}
\Pr(U_l = x) &\mbox{if } x = b,\; b < \min(N_l^c, G_l^- + 1),\; a = 0\\
\Pr(U_l = x) &\mbox{if } x = b,\; b = a + G_l^-,\; 0 < a < N_l^c - G_l^-\\
\Pr(U_l = x) &\mbox{if } b = N_l^c,\; N_l^c \leq x \leq G_l^-,\; a = 0\\
\Pr(U_l = x) &\mbox{if } b = N_l^c,\; a \geq \max(N_l^c - G_l^-,1),\; x = a + G_l^-\\
0 &\mbox{otherwise}
\end{cases}\\
&= \begin{cases}
\Pr(U_l = b) &\mbox{if } b < \min(N_l^c, G_l^- + 1),\; a = 0\\
\Pr(U_l = b) &\mbox{if } b = a + G_l^-,\; 0 < a < N_l^c - G_l^-\\
\Pr(N_l^c \leq U_l \leq G_l^-) &\mbox{if } b = N_l^c,\; a = 0\\
\Pr(U_l = a + G_l^-) &\mbox{if } b = N_l^c,\; a \geq \max(N_l^c - G_l^-,1)\\
0 &\mbox{otherwise.}
\end{cases}
\end{align*}
To obtain the final line the condition on $x$ in the event is substituted in the probability. This concludes the derivation of the pmf of $U_l^+, U_l^-$ and the proof of this lemma.
 \\ \endproof

Using Lemma \ref{Lem:TruncatedConditional}, these distributions are easy to enumerate and to construct a lookup table for. They are also clearly symmetrical and their form implies that $\Pr(U_l^- = a, U_l^+ = b, \eta_l^- = c, \eta_l^+ = d) = \Pr(U_l^- = a, U_l^+ = b)\Pr(\eta_l^- = c, \eta_l^+ = d)$. This fact is going to allow us to derive an expression for the distribution of $\chichi$, given in the following theorem.
\begin{theorem}{(Distribution of $\chichi$)}\label{Thm:Null}
For all $l = 1, \dots, L$, let $(N_l^t, N_l^c, p_l^t, p_l^c)$ be fixed and known, and let $M_l = \min(N_l^t, N_l^c)$, $M = \sum_{l=1}^L M_l$. Let data $\mathbb{D} = \{(U_l, V_l, \eta_l, \nu_l)\}_{l=1}^L$ be drawn according to Equations \eqref{Eq:ConditionalDGP1}-\eqref{Eq:ConditionalDGP2}. Let $\chi^+$ be the maximum of Formulation 2 on $\mathbb{D}$ and let $\chi^-$ be the minimum of Formulation 2 also on $\mathbb{D}$. Let $\mathcal{X}_{M} := \left\{\frac{b - c - 1}{\sqrt{b + c + 1}}: b, c \in \{0, \dots, M\}\right\}$. Additionally, let $\mathcal{A}(y), \mathcal{B}(y,s), \mathcal{C}(x), \mathcal{D}(x, r), \mathcal{H}(x,y,r,s)$ be defined as in Theorem~\ref{Thm:Sharp}. The pmf of $\chichi$, for two values $s, r \in \mathcal{X}_M$ is given by:
\begin{align}
\Pr(\chi^- = s, \chi^+ = r|X) = \sum_{x=-M}^{M}\sum_{y=-M}^{M}\sum_{(\mathbf{a}, \mathbf{b}, \mathbf{c}, \mathbf{d}) \in \mathcal{H}(x,y,r,z)}\prod_{l=1}^L\begin{cases}
g_1(a_l,b_l,c_l,d_l) \quad & \mbox{if } x < 1, y < 1\\
g_2(a_l,b_l,c_l,d_l) \quad & \mbox{if } x \geq 1, y < 1 \\
g_3(a_l,b_l,c_l,d_l) \quad & \mbox{if } x \geq 1, y \geq 1\\
0 \quad & \mbox{otherwise,}\end{cases}\label{Eq:RangeDist}
\end{align}
where:
\begin{alignat}{1}
&g_1(a_l, b_l, c_l,d_l) =\nonumber\\
& \ind(|a_l| = b_l) \sum_{j = 0}^{M_l}\biggl[\Pr\left(U_l^- = a_l + j, U_l^+ = \frac{2M_l - d_l + c_l}{2}\right)\Pr\left(\eta_l^- = j, \eta_l^+ = \frac{2M_l - d_l - c_l}{2}\right)\nonumber\\
&\qquad\qquad\qquad\quad+ \Pr\left(U_l^- = a_l + j, U_l^+ = \frac{d_l  + c_l}{2}\right)\Pr\left(\eta_l^- = j,\eta_l^+ = \frac{d_l - c_l}{2}\right)\ind(M_l \neq d_l)\biggr] \label{Eq:g1}\\
&g_2(a_l, b_l, c_l, d_l) =\nonumber\\
& \ind(|a_l| = b_l)\ind(|c_l| = d_l)\sum_{j = 0}^{M_l}\sum_{k=0}^{M_l}\Pr(U_l^- = a_l + j, U_l^+ = c_l + k)\Pr(\eta_l^- = j, \eta_l^+ = k)\label{Eq:g2}\\
&g_3(a_l, b_l, c_l,d_l) =\nonumber\\
& \mathbb{I}(|c_l| = d_l) \sum_{k = 0}^{M_l}\biggl[\Pr\left(U_l^- = \frac{2M_l - b_l + a_l}{2}, U_l^+ = c_l + k\right)\Pr\left(\eta_l^- = \frac{2M_l - b_l - a_l}{2}, \eta_l^+ = k\right)\nonumber\\
&\qquad\qquad\qquad\quad+ \Pr\left(U_l^- = \frac{b_l  + a_l}{2}, U_l^+ = c_l + k\right)\Pr\left(\eta_l^- = \frac{b_l - a_l}{2}, \eta_l^+ = k\right)\ind(M_l \neq b_l)\biggr] \label{Eq:g3}.
\end{alignat}
The probabilities $\Pr(U_l^- = a, U_l^+ = b)$ and $\Pr(\eta_l^- = a, U_l^+ = b)$ are given in Lemma \ref{Lem:TruncatedConditional} and depend on $N_l^t, N_l^c, p_l^t, p_l^c$.
\end{theorem}

\proof{Proof} The proof follows closely the template of the proof of Theorem \ref{Thm:Sharp}.
Note first that in this case $N_l^t$ and $N_l^c$ are fixed and known for all strata $l$.
We first bound the values of $U_l^+$ and $\eta_l^+$:
\begin{align*}
0 \leq U_l^+ &= M_l - V_l^+ &&\mbox{(By definition of $V_l^+$)}\\
&\leq M_l, &&\mbox{(Both $M_l$ and $V_l^+$ are nonnegative integers)}
\intertext{and:}
0 \leq \eta_l^+ &= M_l - \nu_l^+&&\mbox{(By definition of $\nu_l^+$)}\\
&\leq M_l.&&\mbox{(Both $M_l$ and $\nu_l^+$ are nonnegative integers)}
\end{align*}
From this it follows that $ -M_l \leq TE_l^+ = U_l^+ - \eta_l^+ \leq M_l$. Recall that $M = \sum_{l=1}^M M_l$ and that $TE^+ = \sum_{l=1}^L TE^+_l$: it must be that $ - M\leq TE^+ \leq M$. An exactly symmetric argument shows that $-M \leq TE^- \leq M$. Because of the above the values that we will consider for $TE^+$ and $TE^-$ in this proof are the integers ranging between $-M$ and $M$.
Finally, as detailed in the proof of Theorem \ref{Thm:Sharp}, it must be that $B$ and $C$ are always less than $M$, the total number of matches made, as they represent counts of matched pairs. This allows us to bound the set of values that have positive probability under the distribution of $\chichi$:
\begin{align*}
\mathcal{X}_{M} &= \left\{\frac{b - c - 1}{\sqrt{b + c + 1}}: b, c \in \{0, \dots, M\}\right\}.
\end{align*}
Given the set above, we will consider values of $\chichi$ in the set $\mathcal{X}_{M}$, even though not all of the set's elements will have positive probability under the distribution of interest. As done in Theorem \ref{Thm:Sharp} we derive a set of conditions on values within $\mathcal{X}_{M}$ that determine whether or not those values have positive probability and, if yes, what is their probability under the distribution of $\chichi$ under Assumption \ref{As:StratIgnorability}. \\

The steps leading to the form of \eqref{Eq:RangeDist} in the main statement of the theorem are exactly the same as those in the proof of Theorem \ref{Thm:Sharp}, with the sole difference that the possible values of $TE^+$ and $TE^-$ now are between $-M$ and $M$. Because of this, the summations over $x$ and $y$ both in the definition of the joint probability of interest and in the functions, $f_1, f_2, f_3, f_4$, introduced in the proof of Theorem \ref{Thm:Sharp}, range from $-M$ to $M$ in this case. In particular, the summation over $x$ and $y$ in $f_1$ ranges from $-M$ to 0 in this case, the summation over $x$ in $f_2$ ranges from $1$ to $M$, the summation over $y$ in $f_2$ now ranges from $-M$ to 0, and the summations over $x$ and $y$ in $f_4$ range from 1 tom $M$ in this case. The definitions of $f_1, f_2, f_4$ are also exactly the same as those in the proof of Theorem \ref{Thm:Sharp}, including the fact that $f_3 = 0$. The argument made for this in the proof of Theorem \ref{Thm:Sharp} applies exactly in the same way for this case. \\

The only other difference is in the inner probabilities for the joint distribution of the statistics in each stratum. These inner probabilities result in equations \eqref{Eq:g1}-\eqref{Eq:g3}, and can be derived as follows:
\begin{align}
&\Pr(TE_l^- = a_l, S_l^- = b_l, TE_l^+ = c_l, R_l^+ = d_l)\nonumber\\
&= \Pr(U_l^- - \eta_l^- = a_l, |U_l^- - \eta_l^-| = b_l, U_l^+ - \eta_l^+ = c_l, M_l - |U_l^+ + \eta_l^+ - M_l| = d_l)\label{App:Eq:g11}\\
&= \Pr(U_l^- - \eta_l^- = a_l, |a_l| = b_l, U_l^+ - \eta_l^+ = c_l, |U_l^+ + \eta_l^+ - M_l| = M_l - d_l)\label{App:Eq:g12}\\
&= \ind(|a_l| = b_l)\Pr(U_l^- - \eta_l^- = a_l, U_l^+ - \eta_l^+ = c_l, |U_l^+ + \eta_l^+ - M_l| = M_l - d_l)\label{App:Eq:g13}\\
&= \ind(|a_l| = b_l)\biggl[\Pr(U_l^- - \eta_l^- = a_l, U_l^+ - \eta_l^+ = c_l, U_l^+ + \eta_l^+ - M_l = M_l - d_l)\nonumber\\
&\qquad\qquad\qquad+ \Pr(U_l^- -\eta_l^- = a_l, U_l^+ - \eta_l^+ = c_l, U_l^+ + \eta_l^+ - M_l = d_l - M_l)\ind(M_l \neq d_l)\biggr]\label{App:Eq:g14}\\
&= \mathbb{I}(|a_l| = b_l)\biggl[\Pr\left(U_l^- = a_l + \eta_l^-, U_l^+ = c_l + \eta_l^+, \eta_l^+ = \frac{2M_l - d_l - c_l}{2}\right)\nonumber\\
&\qquad\qquad\qquad+\Pr\left(U_l^- = a_l + \eta_l^-, U_l^+ = c_l + \eta_l^+, \eta_l^+ = \frac{d_l - c_l}{2}\right)\ind(M_l \neq d_l)\biggr]\label{App:Eq:g15}\\
&= \mathbb{I}(|a_l| = b_l)\sum_{j=0}^{M_l}\Pr\left(U_l^- = a_l + j, U_l^+ = \frac{d_l + c_l}{2}\right)\Pr\left(\eta_l^- = j, \eta_l^+ = \frac{d_l - c_l}{2}\right)\ind(M_l \neq d_l)\nonumber\\
& \qquad\qquad\qquad\quad + \Pr\left(U_l^- = a_l + j, U_l^+ = \frac{2M_l - d_l + c_l}{2}\right)\Pr\left(\eta_l^- = j, \eta_l^+ = \frac{2M_l - d_l - c_l}{2}\right)\label{App:Eq:g16}\\
&= g_1(a_l, b_l, c_l, d_l).\label{App:Eq:g17}
\end{align}
In the above, Equation \eqref{App:Eq:g11} follows from the definitions of $TE_l^-, S_l^-, TE_l^+, R_l^+$ given in Claim \ref{CDef}. Equation \eqref{App:Eq:g12} follows from rearranging the terms in the last equality in the parentheses, and from plugging in the equality for $U_l^- - \eta_l^-$, \eqref{App:Eq:g13} follows from the fact that $a_l$ and $b_l$ are both constants and so the event $|a_l| = b_l$ is independent from all the others inside the parentheses. Equation \ref{App:Eq:g14} follows from the fact that the event $\{|U_l^+ + \eta_l^+ - M_l| = M_l - d_l\} = \{U_l^+ + \eta_l^+ - M_l = M_l - d_l\} \cup \{U_l^+ + \eta_l^+ - M_l = d_l - M_l, M_l\neq d_l\}$, This follows from the definition of the absolute value function applied to positive integers. Note that the two sets are disjoint, so the probability of their events can be added and that $M_l$ and $d_l$ are fixed quantities, so they are independent of all the other events in the statement. Equation \eqref{App:Eq:g15} follows from rearranging the terms in the equalities for $\eta_l^+$. Equation \eqref{App:Eq:g16} follows from summing over all values of $\eta_l^-$ and from plugging in equalities for $\eta_l^+$ into the equation for $U_l^+$. Finally, \eqref{App:Eq:g17} follows from the definition of $g_1(a_l, b_l, c_l, d_l)$ given in the statement of the theorem. The derivations for $g_2$ and $g_3$ follow exactly the same steps. For $g_2$ we have:
\begin{align*}
&\Pr(TE^- = a_l, S_l^- = b_l, TE^+ = c_l, S_l^+ = d_l)\\
&=\Pr(U_l^- - \eta_l^- = a_l, |U_l^- - \eta_l^-| = b_l, U_l^+ - \eta_l^+ = c_l, |U_l^+ - \eta_l^+| = d_l )\\
&=\Pr(U_l^- - \eta_l^- = a_l, |a_l| = b_l, U_l^+ - \eta_l^+ = c_l, |c_l| = d_l )\\
&=\mathbb{I}(|a_l| = b_l)\mathbb{I}(|c_l| = d_l)\Pr(U_l^- - \eta_l^- = a_l, U_l^+ - \eta_l^+ = c_l)\\
&=\mathbb{I}(|a_l| = b_l)\mathbb{I}(|c_l| = d_l)\Pr(U_l^- = a_l + \eta_l^-, U_l^+ = c_l+ \eta_l^+)\\
&=\mathbb{I}(|a_l| = b_l)\mathbb{I}(|c_l| = d_l)\sum_{j=0}^{M_l}\sum_{k = 0}^{M_l}\Pr(U_l^- = a_l + k, U_l^+ = c_l + j)\Pr(\eta_l^- = k, \eta_l^+ = j)\\
&=g_2(a_l, b_l, c_l, d_l).
\end{align*}
Finally, $g_3$ can be derived with the same steps as $g_1$:
\begin{align*}
&\Pr(TE^- = a_l, R_l^- = b_l, TE_l^+ = c_l, S_l^+ = d_l)\\
&= \Pr(U_l^- - \eta_l^- = a_l, M_l - |U_l^- + \eta_l^- - M_l| = b_l, U_l^+ - \eta_l^+ = c_l, |U_l^+ - \eta_l^+| = d_l)\\
&= \Pr(U_l^- - \eta_l^- = a_l, |U_l^- + \eta_l^- - M_l| = M_l - b_l, U_l^+ - \eta_l^+ = c_l, |c_l| = d_l)\\
&= \ind(|c_l| = d_l)\Pr(U_l^- - \eta_l^- = a_l, |U_l^- + \eta_l^- - M_l| = M_l - b_l, U_l^+ - \eta_l^+ = c_l)\\
&=\mathbb{I}(|c_l| = d_l)\biggl[\Pr(U_l^- = a_l + \eta_l^-, U_l^- + \eta_l^- - M_l = M_l - b_l, U_l^+ = c_l + \eta_l^+)\\
&\qquad\qquad\qquad + \Pr(U_l^- = a_l + \eta_l^-, U_l^- +\eta_l^- - M_l = b_l - M_l, U_l^+ = c_l + \eta_l^+)\mathbb{I}(M_l \neq b_l)\biggr]\\
&=\mathbb{I}(|c_l| = d_l)\biggl[\Pr(U_l^- = a_l + \eta_l^-, \eta_l^- = \frac{2M_l - a_l - b_l}{2}, U_l^+ = c_l + \eta_l^+)\\
&\qquad\qquad\qquad + \Pr(U_l^- = a_l + \eta_l^-,\eta_l^- = \frac{b_l - a_l}{2} , U_l^+ = c_l + \eta_l^+)\mathbb{I}(M_l \neq b_l)\biggr]\\
=& \mathbb{I}(|c_l| = d_l)\sum_{k=0}^{M_l}\Pr\left(U_l^- = \frac{b_l + a_l}{2}, U_l^+ = c_l + k\right)\Pr\left(\eta_l^- = \frac{b_l - a_l}{2},\eta_l^+ = k\right)\\
&\qquad \qquad \qquad + \Pr\left(U_l^- = \frac{2M_l - b_l + a_l}{2}, U_l^+ = c_l + k\right)\Pr\left(\eta_l^- = \frac{2M_l - b_l - a_l}{2}, \eta_l^+ = k\right)\ind(M_l \neq b_l)\\
=& g_3(a_l, b_l, c_l, d_l).
\end{align*}
This concludes the proof of Theorem \ref{Thm:Null}.
 \endproof

\section{Proof of Theorem \ref{Thm:DFT}}
Recall that, by Theorems \ref{Thm:Null} and \ref{Thm:Sharp}, the distributions of matched pairs in each stratum are independent of each other because of SUTVA, and because optimization of $\chi$ is performed almost independently in each stratum. Since the calculation of the full joint distribution of McNemar's statistic is a convolution of independent discrete probabilities, it can be sped up with multidimensional Fast Fourier Transforms. The basic row-column mDFT algorithm reduces the complexity of having to generate $\mathcal{H}(x,y,r,z)$ to $O(4m N^{4m}\log N)$ \cite[see, e.g.,][]{dudgeon1983multidimensional}. Here $N$ is the total number of units, $m$ is the largest number of matches made in any of the strata. This last quantity is repeated four times, as there are four variables for which probabilities must be computed. These are the densities being convolved together in the summation. This, coupled with simple enumeration to create lookup tables for distributions of $g_1$, $g_2$, $g_3$ in each stratum (See Theorems \ref{Thm:Sharp} and \ref{Thm:Null} for the definition of these quantities), as well as the final range distribution yields an algorithm that creates a probability table for the null distribution of $\chichi$ in $O(4m N^{4m}\log N)$. This implies that the worst-case time of computing the distribution in Theorem \ref{Thm:Sharp} is polynomial.

\section{Plots of exact distributions of $\chichi$ on a simulated dataset}\label{App:Sec:SimDist}
\begin{figure}[!htbp]
\centering
  \begin{subfigure}[t]{0.43\textwidth}
    \centering
    \fbox{\includegraphics[width=\linewidth]{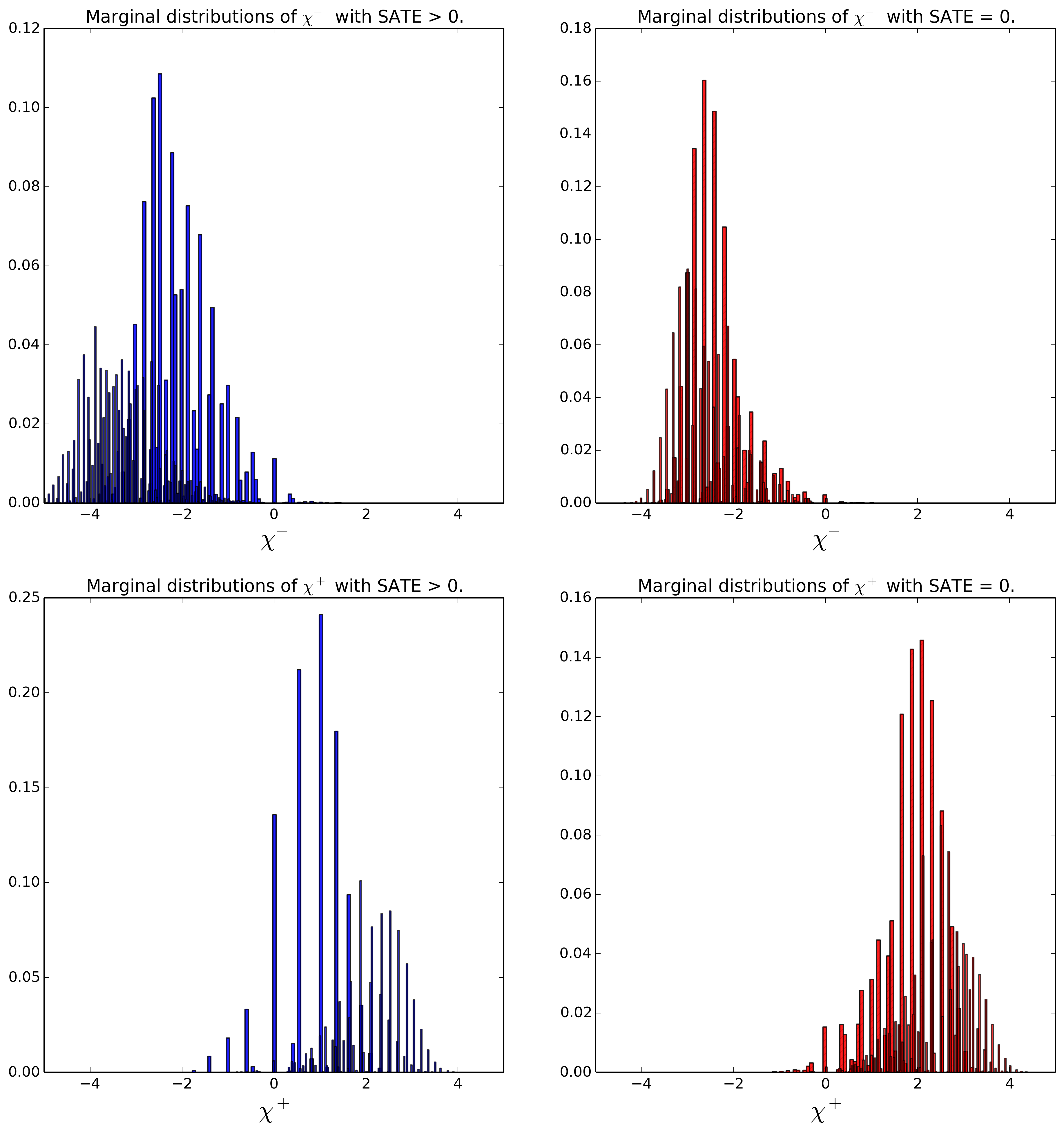}}
    \caption{}
  \end{subfigure}%
  ~
  \begin{subfigure}[t]{0.43\textwidth}
      \centering
      \fbox{\includegraphics[width=\linewidth]{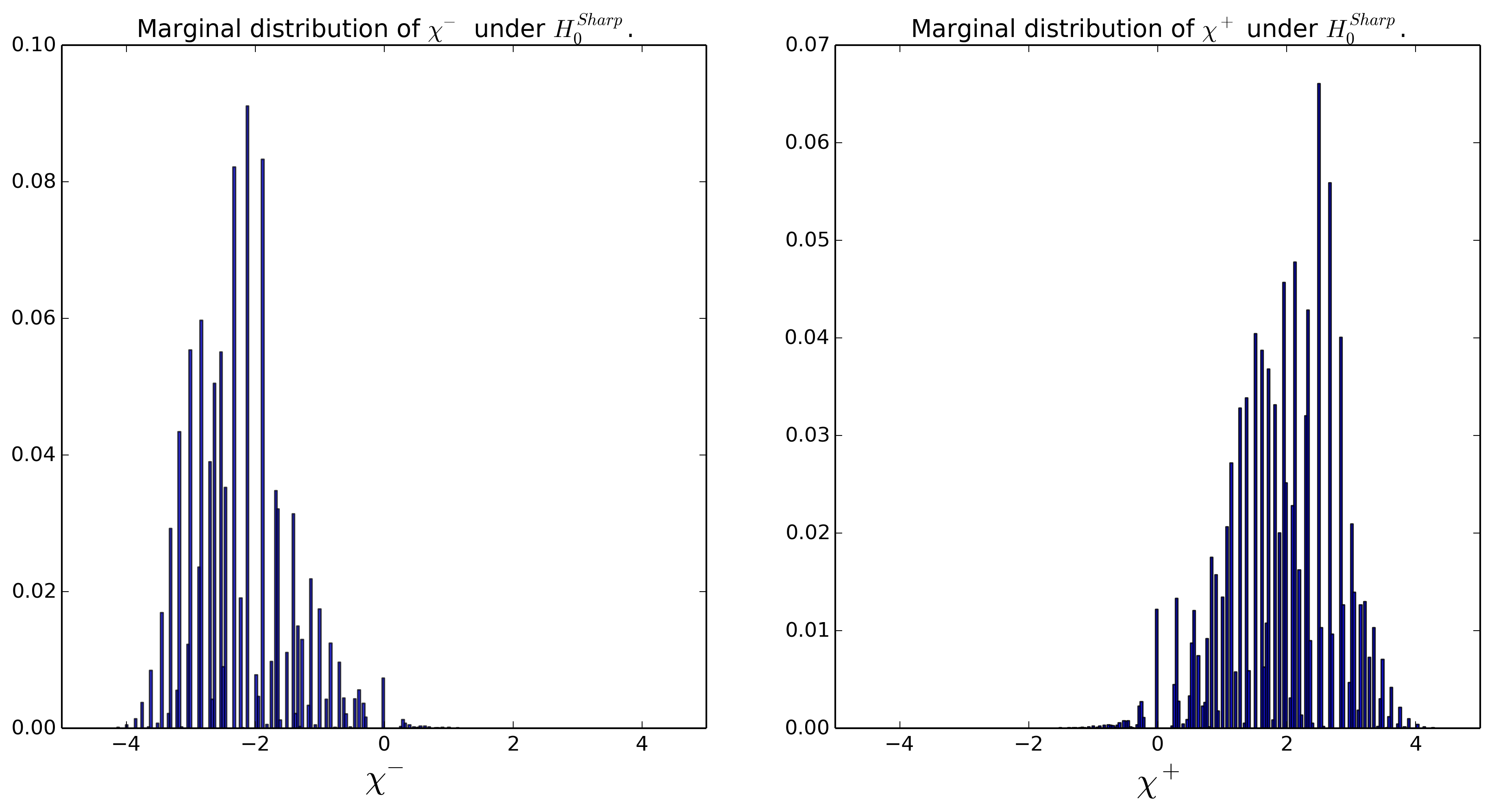}}
      \caption{}
  \end{subfigure}
  \caption{Marginal distributions of $\chichi$ on a simulated dataset with 150 units divided in 15 strata, each with more control than treated units. (a) General marginal distributions with random potential outcomes. 
    The left column shows the distributions when a SATE is present in the whole sample, and the right column when the SATE is 0. (b) Null marginal distributions under $\Hsharp$}
  \label{Fig:Distributions}
\end{figure}

Figure \ref{Fig:Distributions} shows marginal distributions of $\chi^+$ and $\chi^-$ derived from the joint distributions in Theorems \ref{Thm:Sharp} and \ref{Thm:Null}, respectively. We simulated datasets from the DGPs described in those sections and estimated the distributions with the formulas in Theorems \ref{Thm:Sharpa} and \ref{Thm:Nulla}, and the polynomial time algorithm suggested in Theorem \ref{Thm:DFT}. Panel (a) shows a location and scale difference in the distribution of $\chichi$ between when the null is true and when it is not. This demonstrates our tests' capacity to detect full-sample treatment effects.

\section{Additional Simulations}\label{App:Sec:Simulations}

\subsection{Performance of robust tests with increasing number of covariates}
We present results of simulation studies under the same settings as those in Section \ref{Sec:Simulations}, but we instead keep the number of units fixed at 200, and let the number of covariates grow. 

\begin{figure}[!htbp]
    \centering
    \includegraphics[width=\textwidth]{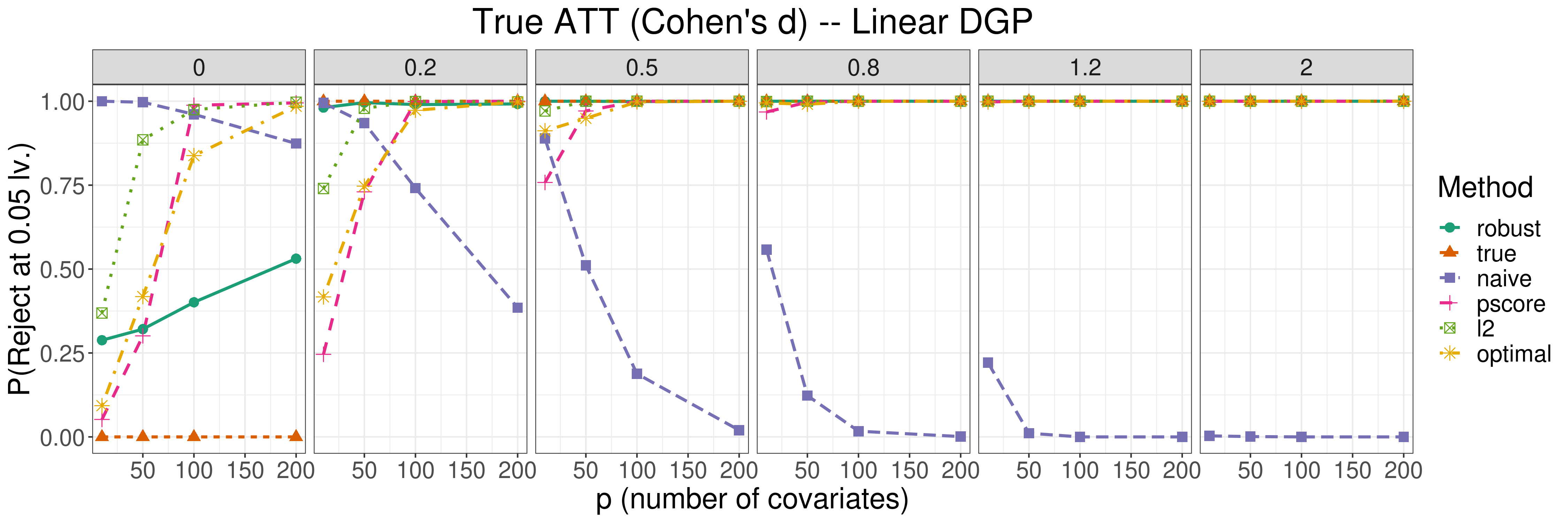}
    \includegraphics[width=\textwidth]{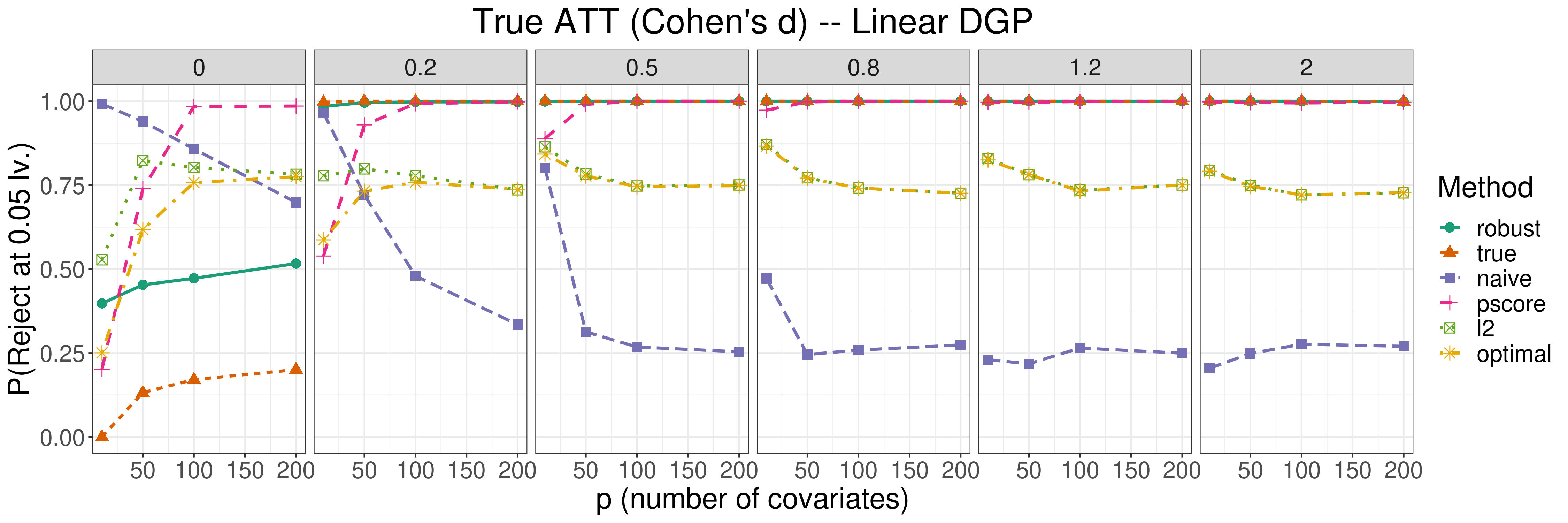}
    \caption{Performance of the robust McNemar's test on a sample size of $N=200$ units with an increasing number of covariates. Top row: simple DGP, bottom row: complex DGP. The ideal method is labelled ``true'' and represented by the dark orange dashed line in the figure. }
    \label{fig:mcn_largep}
\end{figure}

\begin{figure}[!htbp]
    \centering
    \includegraphics[width=\textwidth]{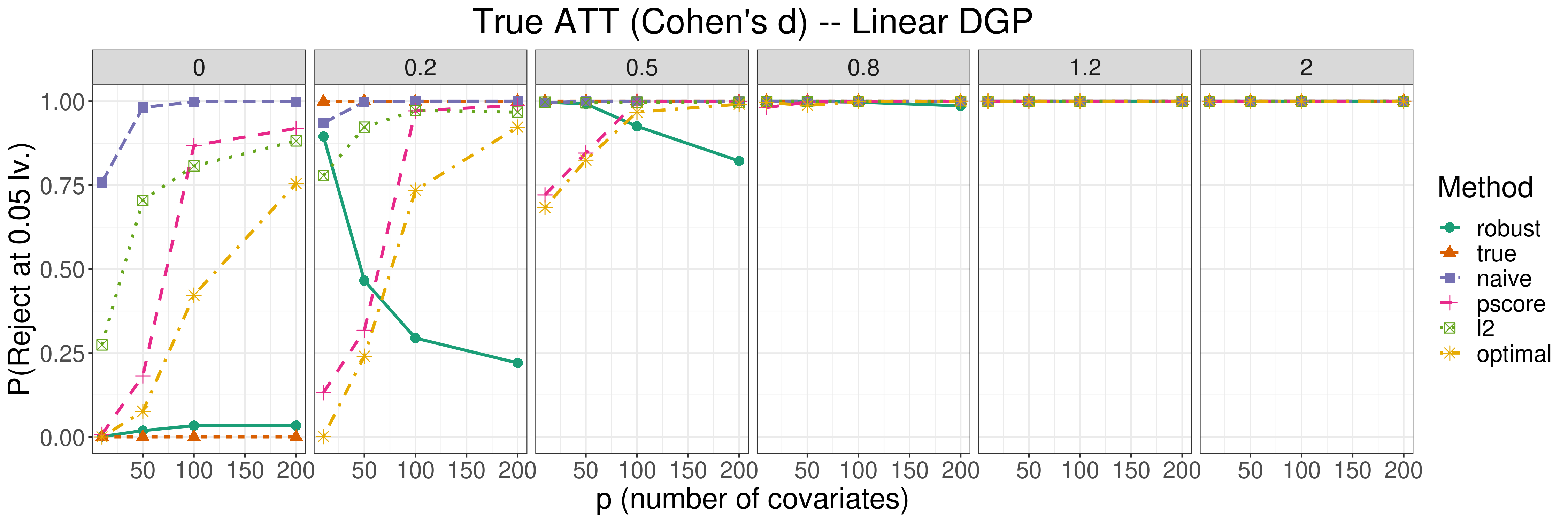}
    \includegraphics[width=\textwidth]{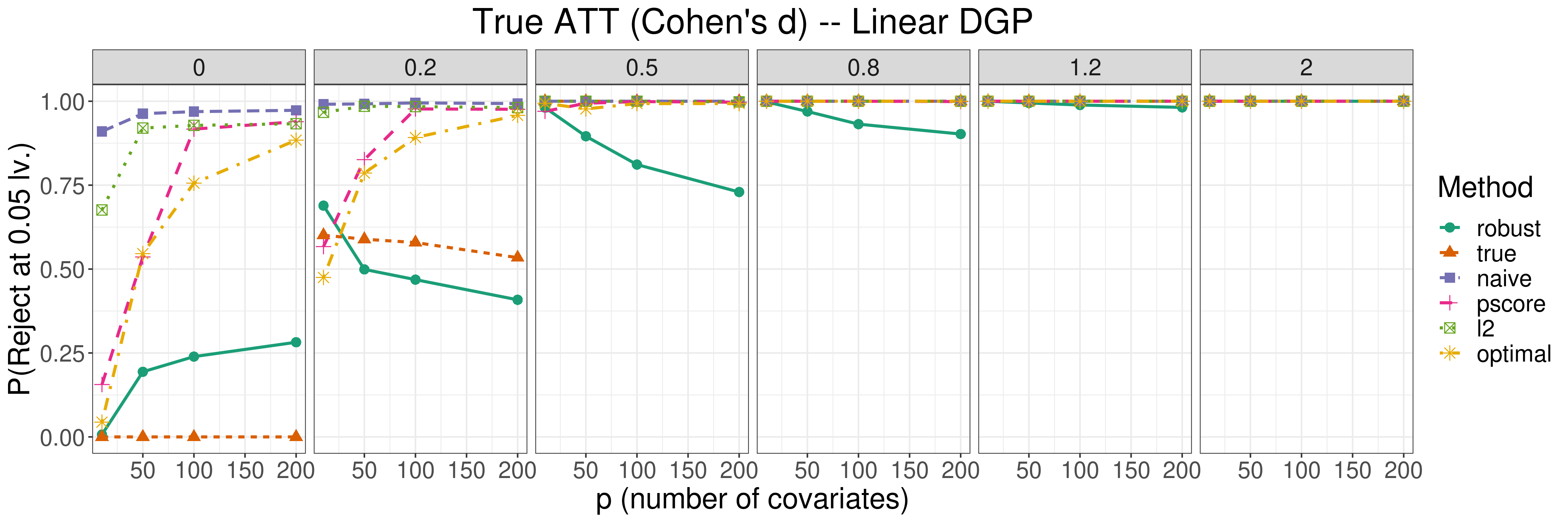}
    \caption{Performance of the robust z-test on a sample size of $N=200$ units with an increasing number of covariates. Top row: simple DGP, bottom row: complex DGP. }
    \label{fig:z_largep}
\end{figure}

Results for McNemar's test are reported in Figure \ref{fig:mcn_largep}, and results for the z-test are reported in Figure \ref{fig:z_largep}. We see that, largely, our tests tend to perform well even in the increasing $P$ regime: both tests achieve the overall lowest error rate when the ATT is 0 (closest to the error rate of the idealized test that includes both true potential outcomes), and display good statistical power as the ATT increases in relative strength. The z-test is somewhat conservative when the true ATT is weak (Cohen's d=0.2), and when there are many covariates: this is a desirable pattern since the larger number of covariates makes the matching problem harder. Notably, the benchmark test performed on the true potential outcomes displays similar behavior to our robust z-test in Figure \ref{fig:z_largep} when $d=0.2$. 

\subsection{Performance of robust tests with Exclusively Binning Constraints}
We present performance results of our robust tests when constraints are exclusively binning, i.e., when data is perfectly stratified. 
In this setting we partition the $N$ units into strata such that each stratum contains exactly $25$ units, for example, when $N=100$ there will be 4 strata. We denote the number of strata for a given $N$ with $L_N$, and use the variable $s_i = 1, \dots, L_N$ to denote which stratum unit $i$ is assigned to. For a set of fixed coefficients $\beta_1, \dots, \beta_{L_N}$ that are calibrated as described prior, we generate:
\begin{align*}
    e_i = \frac{1}{1 + \exp(- \beta_{s_i})},\quad \epsilon_i \sim \text{Normal}(0,1),\quad y_i^* = t_i \tau + \beta_{s_i} + \epsilon_i, 
\end{align*}
with treatments $t_i$ assigned to exactly $N^t$ units but each with probability $e_i$. For binary data, we set $y_i = \ind[y_i^* > 0]$, and for continuous data $y_i = y_i^*$, as done in all the other simulations. All other settings are as in the simulations presented in Section \ref{Sec:Simulations} of the main paper. 

\begin{figure}[!htbp]
    \centering
    \includegraphics[width=\textwidth]{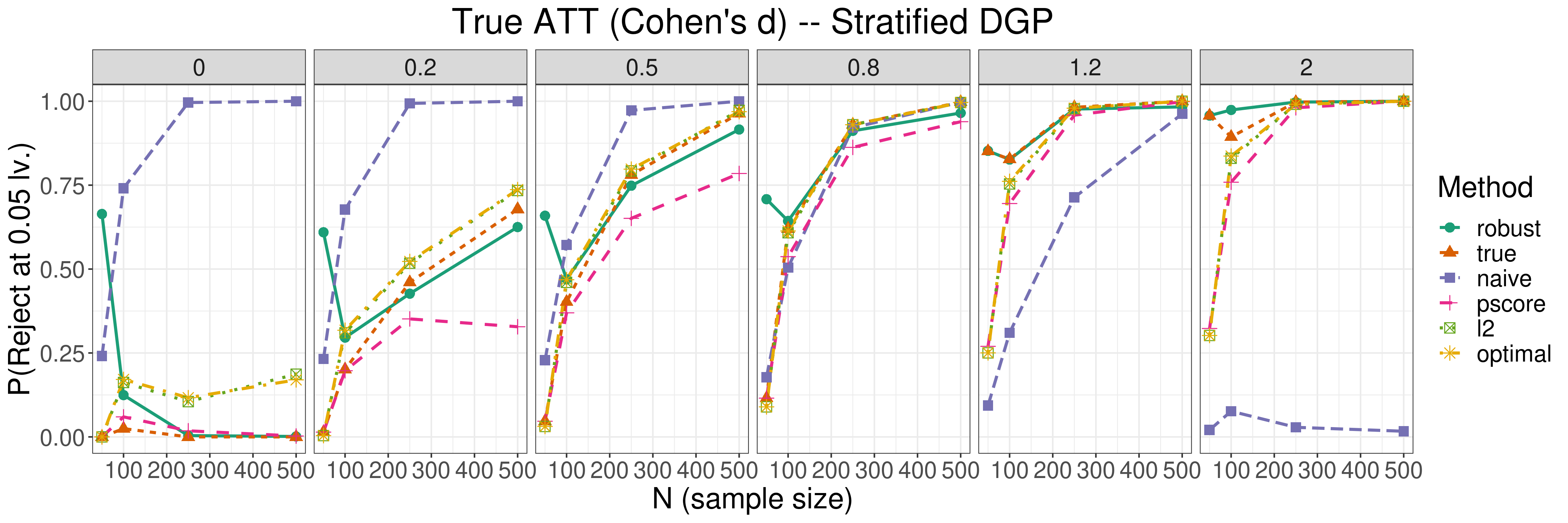}
    \caption{Performance of robust McNemar's test under exclusively binning constraints. }
    \label{fig:mcn_stratified}
\end{figure}

Results for the robust McNemar's test are presented in Figure \ref{fig:mcn_stratified}. 

\begin{figure}[!htbp]
    \centering
    \includegraphics[width=\textwidth]{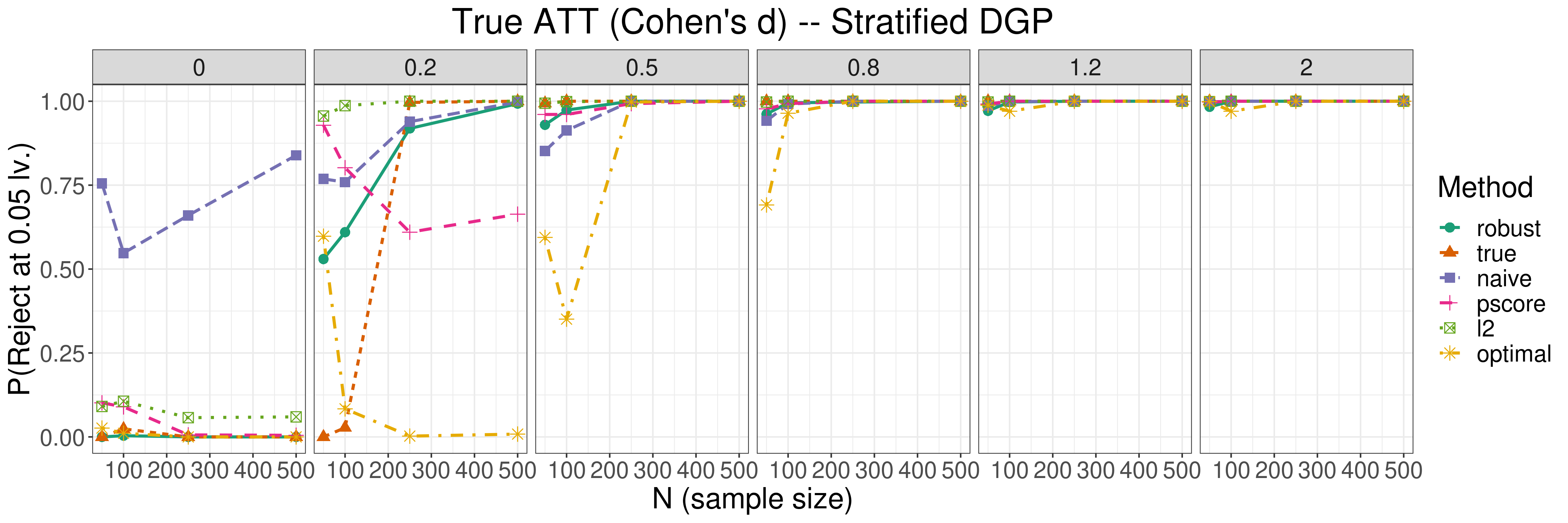}
    \caption{Performance of robust Z test under exclusively binning constraints. }
    \label{fig:z_stratified}
\end{figure}

Results for the robust z-test are presented in Figure \ref{fig:z_stratified}

We see that both test perform well under these types of constraints: error rate for the z-test is almost 0 when the ATT is 0, and the test achieves good statistical power as the ATT grows in strength. A similar patterns is visible for McNemar's test, though this test requires a slightly larger sample size ($N \geq 100$) in order to achieve a good false rejection rate.

\subsection{Performance of robust z-test with different types of constraints}

We study whether the type of match quality constraints used in our robust test affects both the error rate of the robust tests as well as their runtime. All simulation settings are kept same as before, except that our robust test is ran 3 times per iteration, each time with a different type of constraint. The types of constraints we compare are as follows:
\begin{itemize}
    \item \textbf{Moment}: We require balance between the first two moments of the empirical distributions of each covariate in the treated/control sets after matching. This is formulated as: $|\frac{1}{M}\sum_{i=1}^{N^t}\sum_{j=1}^{N^c}a_{ij}(x^t_{ip} - x^c_{jp})| \leq \epsilon_{p}^{mean} \times (\sigma(x_p^t)/2 + \sigma(x_p^c)/2)$, for $p = 1, \dots, P$, and $|\frac{1}{M}\sum_{i=1}^{N^t}\sum_{j=1}^{N^c}a_{ij}(x^t_{ip} - x^c_{jp})^2| \leq \epsilon_{p}^{var} \times (\sigma^2(x_p^t)/2 + \sigma^2(x_p^c)/2)$, for $p = 1, \dots, P$ ($2P$ constraints total).
    \item \textbf{Caliper}: We only allow units to be matched if their propensity scores are close enough. This is formulated as: $a_{ij}|e^t_i - e^c_j| \leq \epsilon^{cal}$ for $i = 1, \dots N^t$, $j = 1, \dots, N^c$ ($N^t \times N^c$ constraints total).
    \item \textbf{Quantile}: We require the empirical distributions of each covariate in the treated/control sets after matching to be similar in terms of their quantiles. For each covariate $p = 1, \dots, P$, and a pre-specified sequence of quantiles $\gamma_{p1}, \dots, \gamma_{pL}$, such that $\gamma_{p1} \leq \gamma_{p2} \leq \dots \gamma_{pL}$, we require: $|\frac{1}{M}\sum_{i = 1}^{N^t}a_{ij}\ind[x^t_{ip} \leq \gamma_p] - \sum_{j=1}^{N^c}\sum_{i=1}^{N^t} a_{ij}\ind[x^c_{jp} \leq \gamma_p]| \leq \epsilon^{qua}_p$ ($P \times L$ constraints total).
\end{itemize}
as previously done, tolerance values ($\epsilon^{mom}_{p}$, $\epsilon^{cal}$, $\epsilon_{p}^{qua}$) are chosen as the smallest feasible values on an evenly spaced grid between 0.1 and 1, and $\sigma_p^2(x_{p}^t)$, $\sigma_p(x_{p}^t)$, $\sigma_p^2(x_{p}^c)$, $\sigma_p(x_{p}^c)$ are the variance and standard deviation of covariate $x_p$ in the treated and control set respectively. 

\begin{figure}
    \centering
    \includegraphics[width=\textwidth]{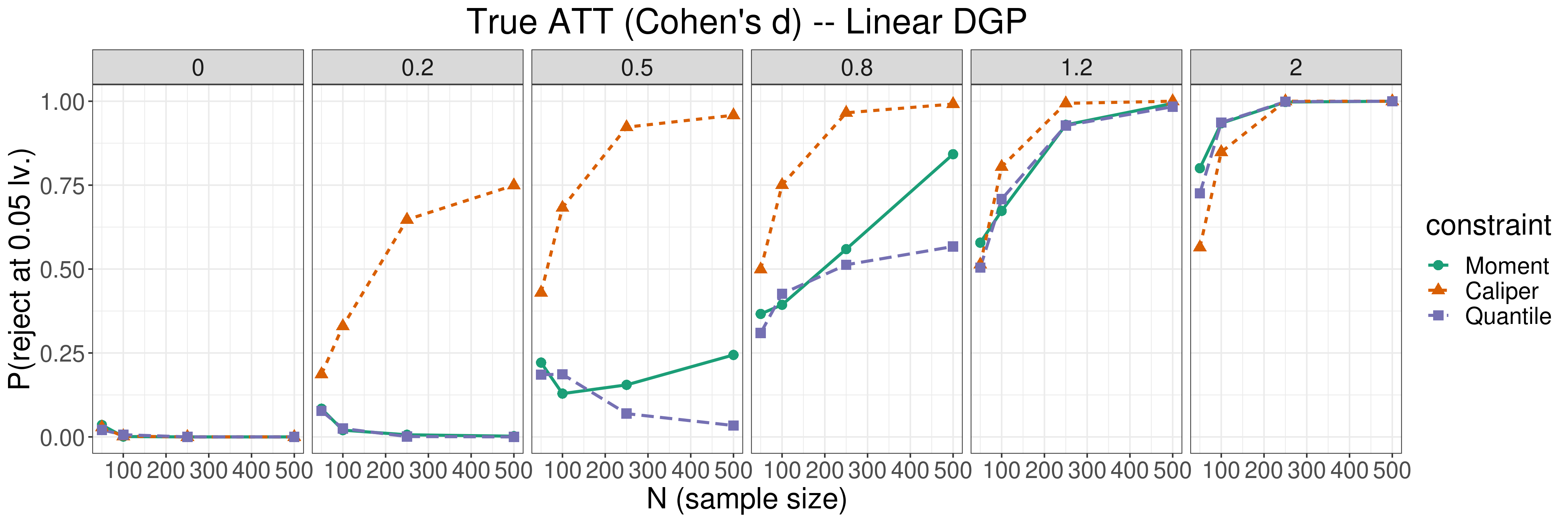}
    \includegraphics[width=\textwidth]{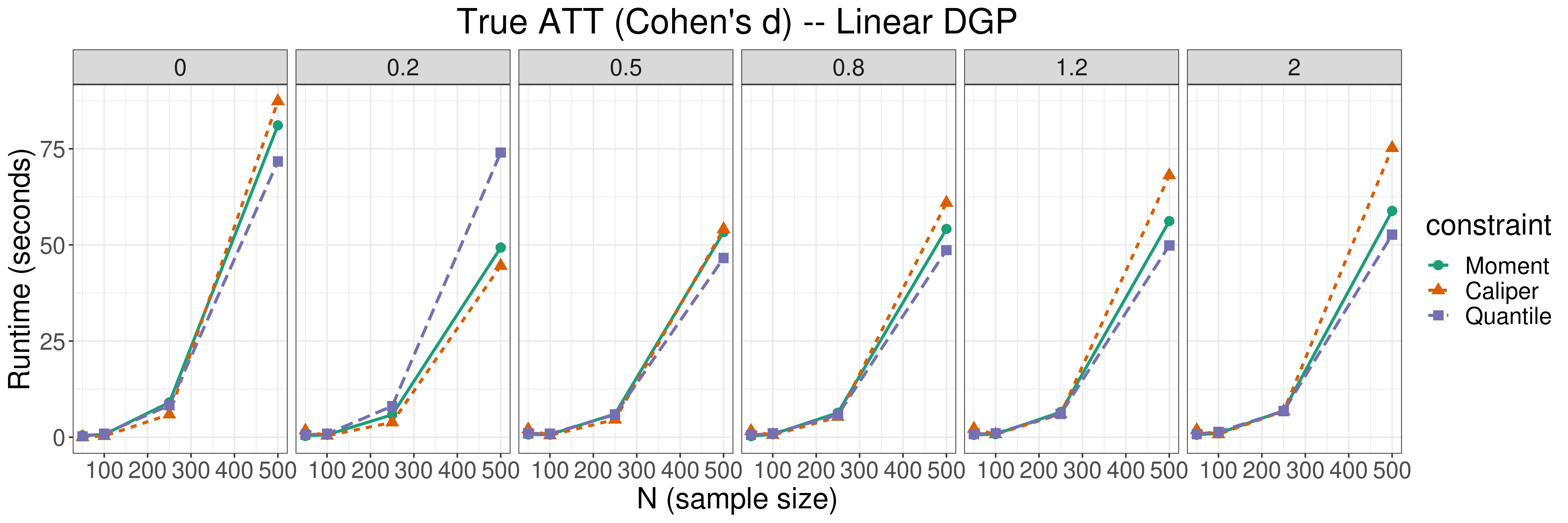}
    \includegraphics[width=\textwidth]{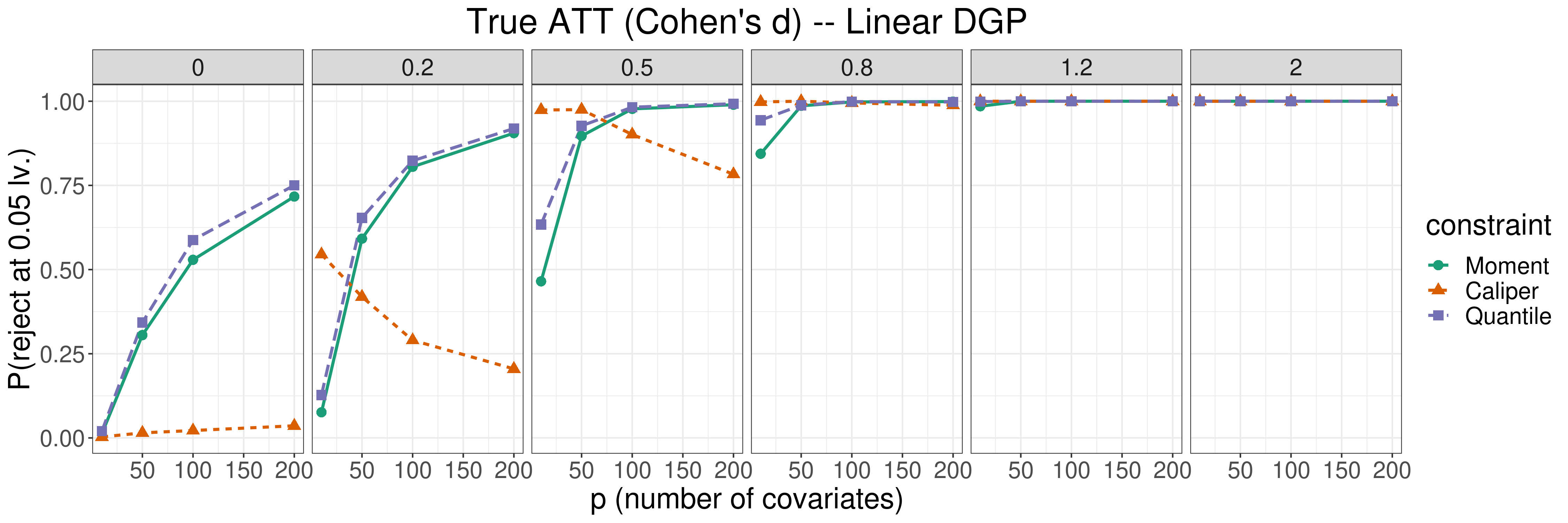}
    \includegraphics[width=\textwidth]{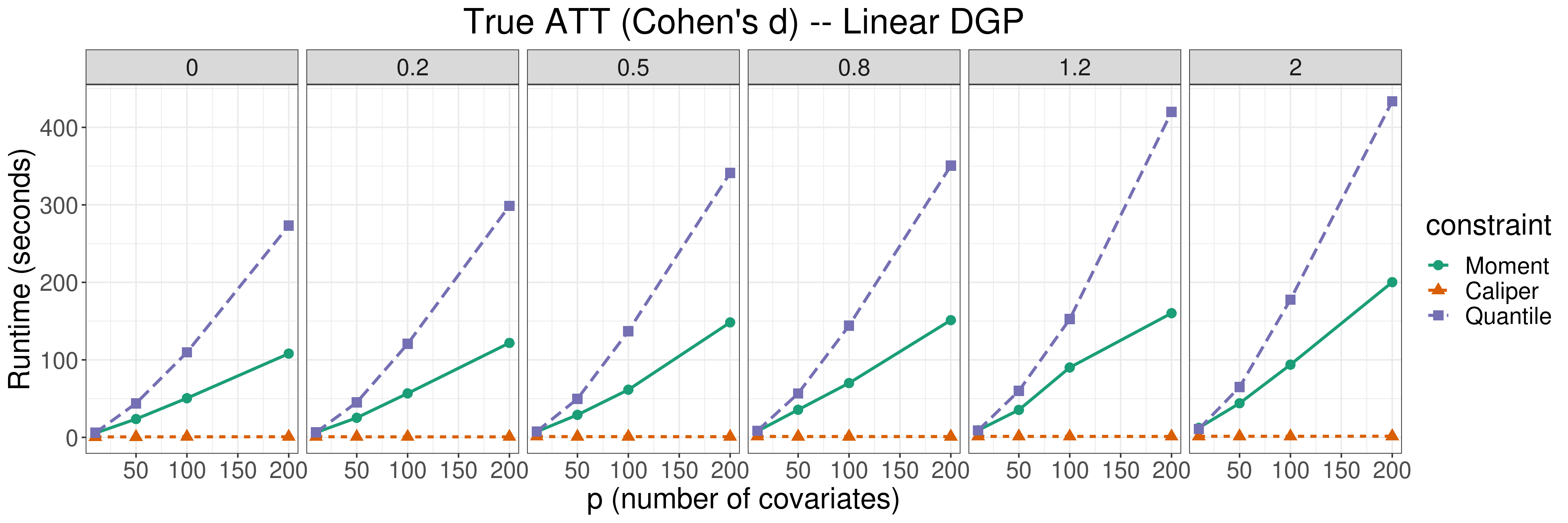}
    \caption{Comparison of rejection rate and runtime of the robust z-test with different types of constraints. }
    \label{fig:constraints}
\end{figure}

Results are reported in Figure \ref{fig:constraints}. In both fixed $P$ and fixed $N$ regimes we see that the pscore caliper constraint seems to display the lowest rejection rate when the true ATT is 0, indicating that this type of constraint is particularly suitable for the settings we studied. Interestingly, the caliper constraint seems to also be the one able to achieve better statistical power as the true ATT increases in size in the fixed $P$ regime, while it displays worse statistical power than the other two constraints in the fixed-$N$ regime. In terms of runtime, there appears to not be virtually any difference between constraint types in the fixed-$P$ regime, however the caliper constraints is much faster and unaffected by the number of covariates in the fixed-$N$ regime, as the number of covariate grows, which is expected due to the propensity score being a 1-dimensional quantity for any amount of covariates. Also expectedly, the quantile constraint requires longer execution time as the number of covariates grows compared to the moment constraint.

The main takeaway from this excercise is that it is a good idea for analyst to include several constraints at once in their application of the robust tests, requiring several different metrics of balance to all be below a certain tolerance to admit a match assignment as good. 

\subsection{Performance of robust z-test with different MIP approximation methods}

\begin{figure}[!htbp]
    \centering
    \includegraphics[width=\textwidth]{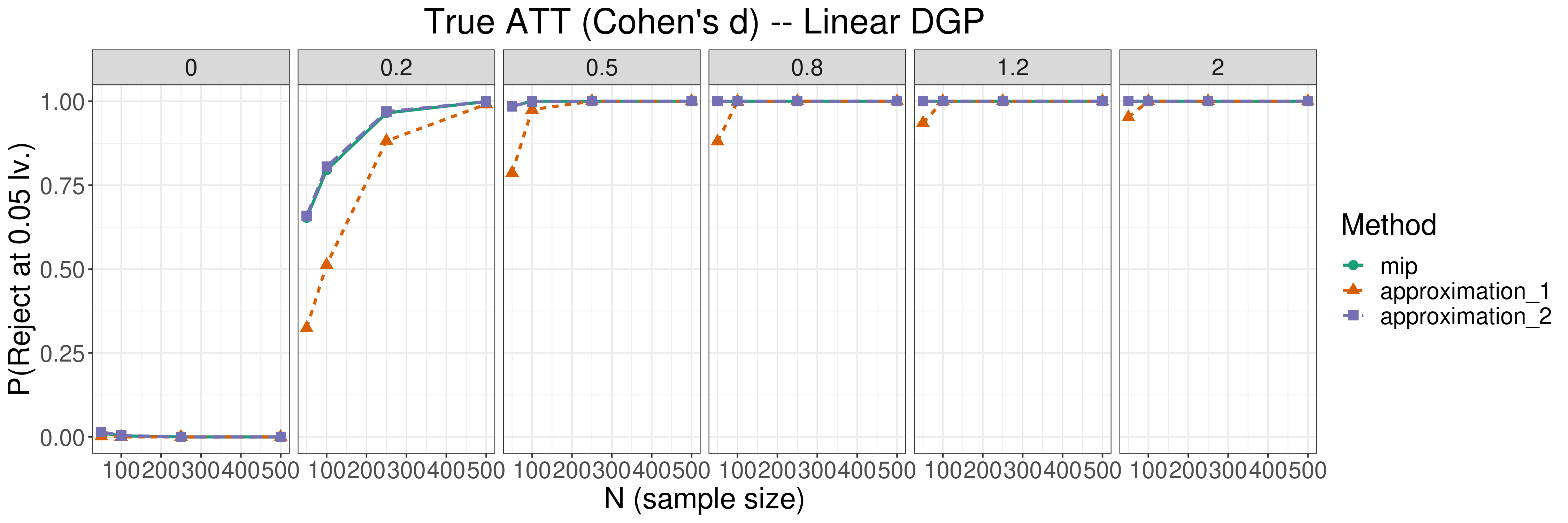}
    \includegraphics[width=\textwidth]{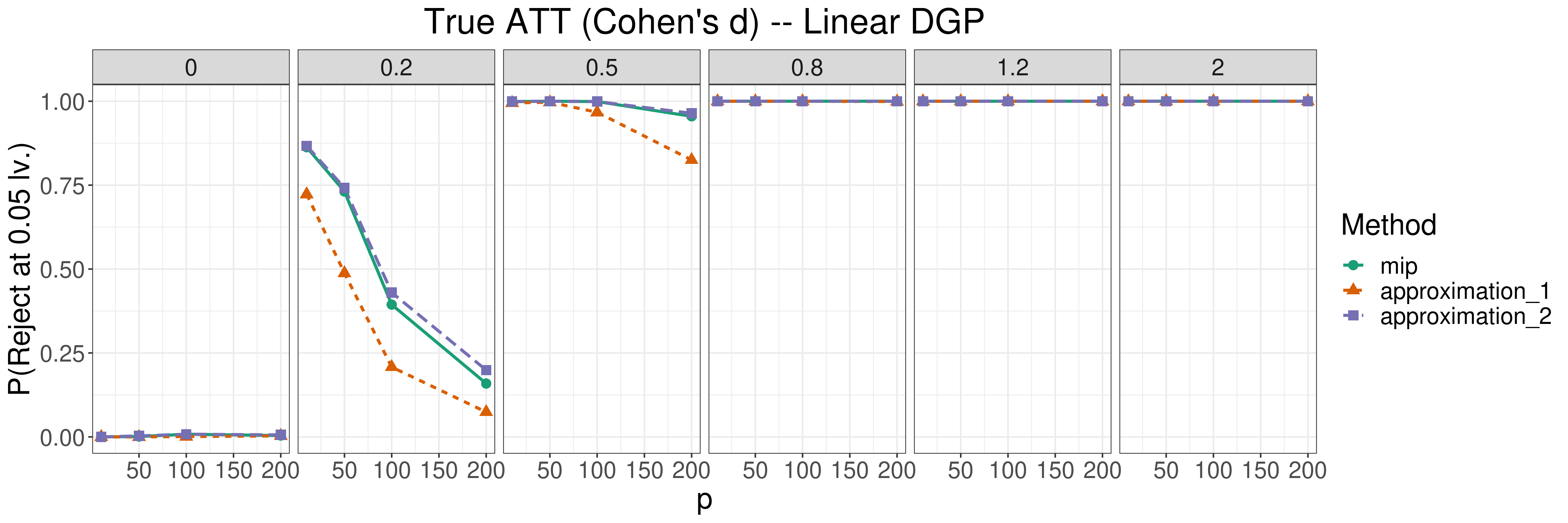}
    \caption{Performance of robust z-test with different MIP approximation methods.}
    \label{fig:approx}
\end{figure}

Since MIP solution algorithms have a runtime that is exponential in the number of decision variables (units to match, in our case), we have implemented and tested two different approximation methods to speed up computation for our robust tests. Both methods are based on first solving a linear relaxation of the original MIP, and then solving an integer program that has a solution closest to that of the relaxation. The two methods are described as follows:

\paragraph{Approximation 1}: This is the same approximation method used in \cite{Zub2012}. We first solve a linear relaxation of all the MIPs used for the robust z-test, (Formulation 2) and once an optimal solution is found via our proposed algorithm, we approximate a MIP solution by solving the following IP. Let $b_{11}, \dots, b_{N^tN^c}$ be the continuous matching indicators output by solving the linear relaxation of Formulation 2. The approximation can be obtained by solving:
\begin{align*}
    \max/\min &\sum_{i=1}^{N^t}\sum_{j=1}^{N^c} a_{ij} b_{ij}\\
    \text{Subject to:} & \\
    \sum_{j=1}^{N^c} a_{ij} &\leq 1,\quad i =1,\dots, N^t\\
    \sum_{i=1}^{N^t} a_{ij} &\leq 1,\quad j=1, \dots, N^c.
\end{align*}
The approximation tries match as many units entirely as the linear relaxation matches fractionally, while not matching any unit more than once. 
This approximation provides a substantial speedup over the MIP, since it removes all the additional constraints, however the solution could be sub-optimal and some constraints may be violated in practice. Results from comparisons on simulated data in Figure \ref{fig:approx} show that, in practice, this approximation is almost as reliable as solving the exact MIP, making it a viable option in practice. 

\paragraph{Approximation 2}: Approximation 2 solves the same problem as Approximation 1, except that all the additional constraints (including user-defined balance constraints) in Formulation 2 are kept. This ensures that the solution found satisfies the constraints specified, but is more expensive in terms of computation time. As Figure \ref{fig:approx} shows, this approximation performs even more closely to the MIP: gains in computation time are offset by almost no loss in performance. 

\subsection{Comparison To Matching Bounds \citep{Prelim}}

We perform a comparison of our robust tests to Matching Bounds \citep{Prelim}, a method for a different facet of the issue of multiple match assignment disagreeing. There are several substantial differences between the Matching Bounds approach of \cite{Prelim} and the robust testing approach proposed in this work. Namely that Matching Bounds is a tool for finding ranges of \textbf{treatment effects}, our robust tests are a tool for finding ranges of \textbf{test statistics} (i.e., we consider not just treatment effects but the uncertainty in their estimates). An analyst interested in knowing what possible treatment effects could be obtained by changing matched samples should employ Matching Bounds, while an analyst interested in obtaining the range of test statistics (and p-values) that could be obtained by matching in different ways should use the robust tests. Specifically \citep{Prelim} finds bounds on the raw value of the treatment effect of interest, our robust testing approach targets a test statistic for the treatment effect: this latter quantity depends both on TE value and on TE variance. This latter dependence implies that the optimum Z-value for a robust test will likely not be the same Z-value obtained by, for example, performing a Z-test via normal approximation on the set of units that maximizes/minimizes the TE as selected by solving the optimization problem in \citep{Prelim}. Specifically, a Z-test conducted on matches selected as in \citep{Prelim} will find a sub-optimal solution to the Z-statistic optimization problem, potentially leading to increased chances of type-I error (incorrect rejection of the null.).

To further demonstrate the differences between the two methods and their purposes, we have performed a set of simulations whose settings are similar to those in Section 6 of our paper and results are reported in Figure \ref{fig:mbcomparison}. One can see the benefit of finding the range of test statistics, as compared to just the range of treatment effects: P-values output by MBs are not as conservative as those from the robust test when the ATT is 0, and the bounds on the ATT output by MBs are better than those output by the robust test when the ATT is 0. 

\begin{figure}[!htbp]
    \centering
    \includegraphics[width=\textwidth]{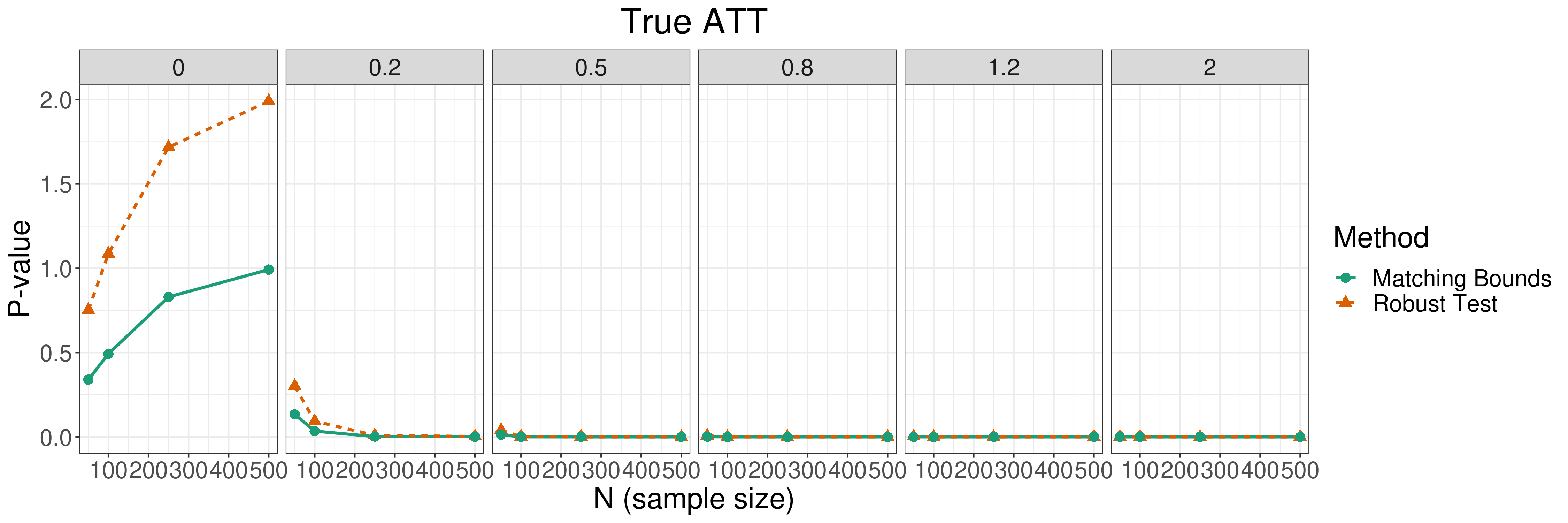}
    \includegraphics[width=\textwidth]{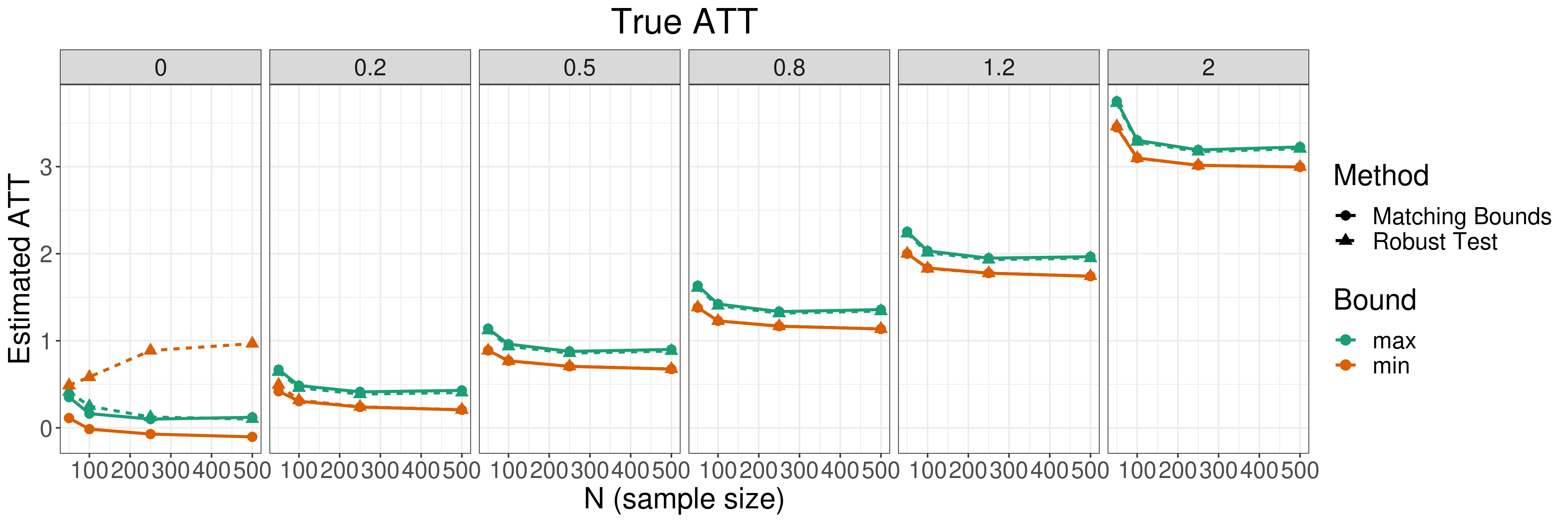}
    \includegraphics[width=\textwidth]{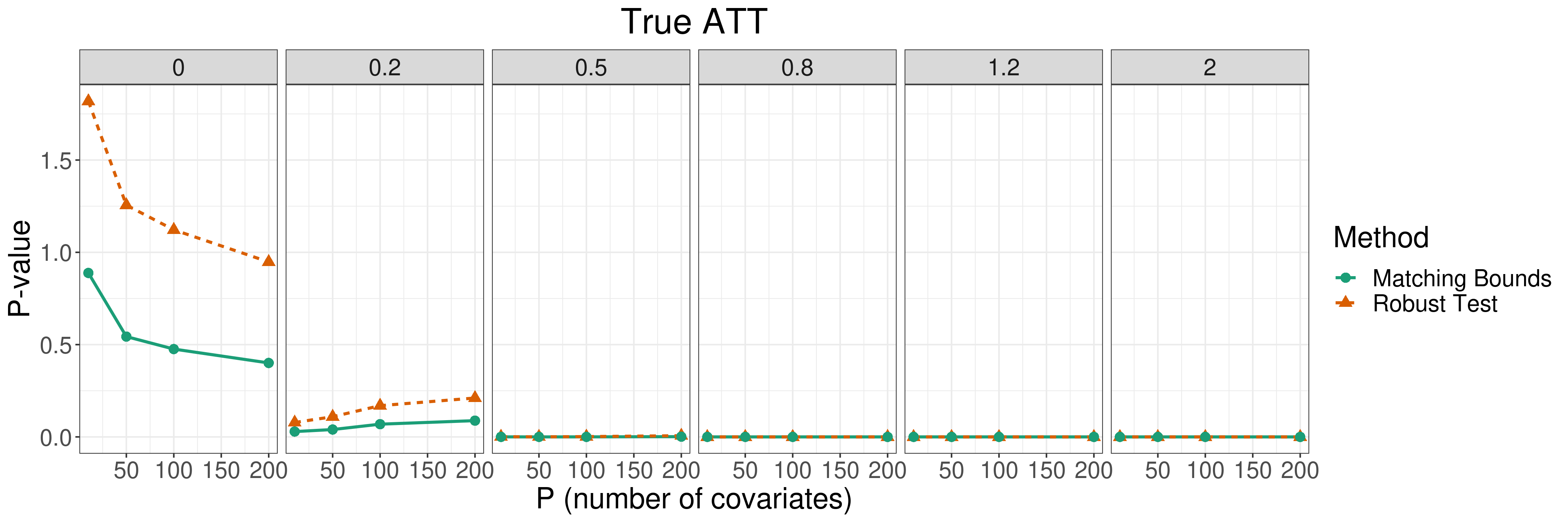}
    \includegraphics[width=\textwidth]{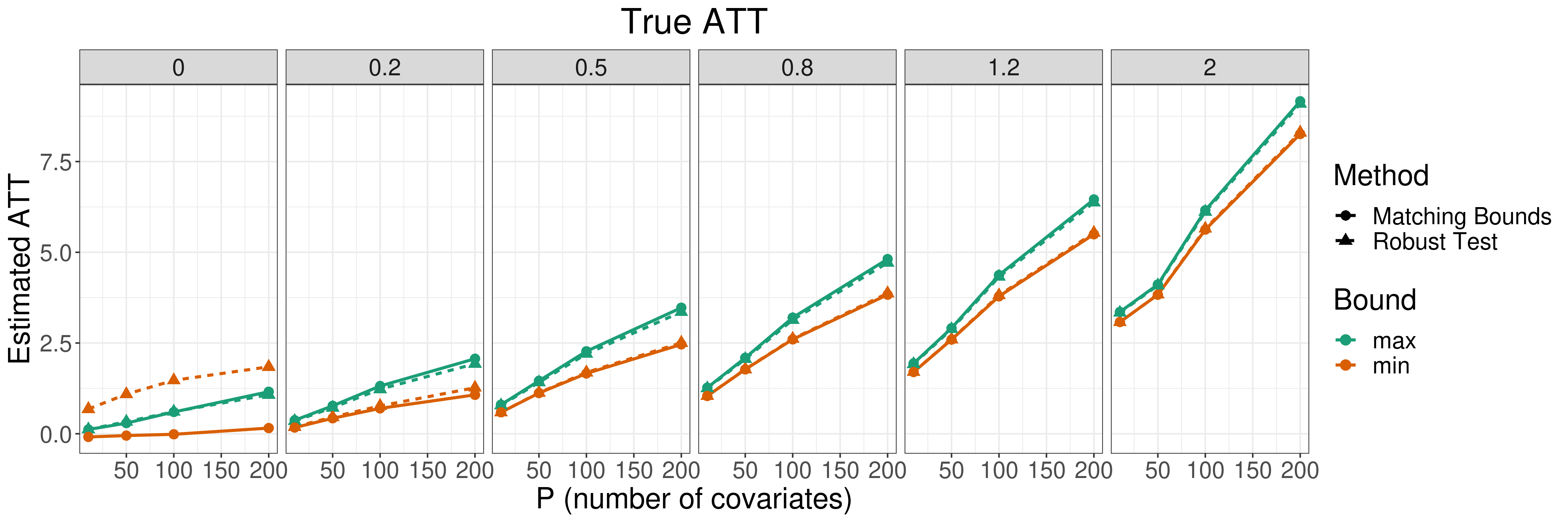}
    \caption{Comparison between Matching Bounds \citep{Prelim} and the Robust Z-test. \textit{Top Row:} The figure shows that if the true ATT is 0, our robust test have a high p-value (which is correct), and if the ATT is 0.2 or more, our robust test starts out conservatively and goes to 0 as $N$ or $P$ increases. \textit{Bottom Row}: Our robust tests give uncertainty bounds on the ATT that are narrower than Matching Bounds because they optimize test statistics that depend on both the estimated mean and variance of the ATT. }
    \label{fig:mblargep}
\end{figure}

\section{Agreement of traditional matching methods in case studies 1 and 2}

As a supplement to our case studies, we present p-values obtained after applying traditional matching algorithms to the datasets of Case Studies 1 and 2. We do so to gauge whether there is disagreement in rejection decisions between matching methods on these datasets. 

We have constrained all the matching methods used to respect the caliper constraint we employ in constructing our robust test, e.g., units whose absolute distance in the covariates exceeds the thresholds defined in Section 7 are disallowed from being matched. This ensures that all three methods compared produce at the very least balance on the covariates up to that threshold. In this sense, all these methods produce similar balance. 

Results of this comparison are presented in Table \ref{tab:csagree}. We see that, generally methods tend to fail to reject on the GLOW dataset: this is because all methods fail to maximize the effective sample size of McNemar's test. They match units with the same outcome, and therefore all produce low-powered tests. Instead, we constrain our tests to output only matches that have a good effective sample size (almost all the treated units) -- in this case we do indeed demonstrate that an effect is present. 

For the Bike Sharing case study, we see that propensity score matching and L2 distance matching fail to reject the null hypothesis, while optimal constrained matching does in fact reject it. The disagreement here is likely due to the fact that several matched sets that meet the balance requirements are present, as we argue in our paper. Our application of the robust z-test to this dataset further corroborates this hypothesis by showing that, indeed, it cannot reject the null hypothesis over all possible good matches. 

\begin{table}[!ht]
\centering
\caption{P-values for McNemar and Z-test conducted on datasets for Case Studies 1 and 2 after matching with different traditional methods}
\label{tab:csagree}
\begin{tabular}{r|rrrr}
  \hline
 & No Matching & Pscore & L2 & Optimal \\ 
  \hline
    GLOW & 0.00 & 1.00 & 1.00 & 0.42 \\ 
    Bike Share & 0.00 & 0.20 & 0.32 & 0.00 \\ 
   \hline
\end{tabular}
\end{table}

\section{Additional Information about Case Study 2}\label{App:Sec:CS2}
In our second case study, we used 2 years (2011-2012) of bike sharing data from the Capital Bike Sharing (CBS) system \citep[see][]{cbs} from Washington DC comprised of 3,807,587 records over 731 days, from which we chose 247 treatment days and 463 control days according to the weather as follows: The control group consists of days with Weather 1: Clear, Few clouds, Partly cloudy, and Partly cloudy. The treatment group consists of days with Weather 2: Mist + Cloudy, Mist + Broken clouds, Mist + Few clouds, Mist. The covariates for matching are as follows: Season (Spring; Summer; Fall; Winter), Year (2011; 2012), Workday (No; Yes), Temperature (maximum 41 degree Celsius), Humidity (maximum 100 percent), Wind speed (maximum 67). The outcome is the total number of rental bikes. We computed distance between days as follows: $\textrm{dist}_{ij}$  =1 if covariates season, year and workday were the same, and the differences in temperature, humidity and wind speed are less or equal to 2, 6 and 6, respectively for treated unit \emph{i} and control unit \emph{j}, 0 otherwise. Since the outcome variable is continuous, we focus on testing $\Hatt$ by producing ranges of p-values with the method introduced in Section \ref{Sec:approx}. Optimization models for this case study have been implemented in AMPL \citep{ampl}, and solved with the solver CPLEX \citep{cplex}. The bike sharing and GLOW datasets are publicly available. The reported solution time  with a X64-based PC with Intel(R) Core(TM) i7-4790 CPU running at 3.60 GHz with 16 GB memory and mip gap of 0.001 to solve a single instance is less than 1 second for all the tests.

\section{Additonal Case Studies}


\subsection{Case Study 3: Training Program Evaluation}
In this case study, we used training program evaluation data described in \cite{RD}, and \cite{RD1}, which were drawn from \cite{RL}. This data set contains 15,992 control units and 297 treatment units. The covariates for matching are as follows: age, education, Black (1 if black, 0 otherwise), Hispanic (1 if Hispanic, 0 otherwise), married (1 if married, 0 otherwise), nodegree (1 if no degree, 0 otherwise), and earnings in 1975. The outcome is the earnings in 1978. The treatment variable is whether an individual receives job training. We computed distance between units as follows:  $\textrm{dist}_{ij}$ =1 if covariates Black, Hispanic, married and nondegree were the same, and the differences in age, education and earnings in 1975 were less or equal to 5, 4 and 4000, respectively, for treated unit \emph{i} and control unit \emph{j}, 0 otherwise.

\begin{figure}
\begin{center}
\includegraphics[height=3in]{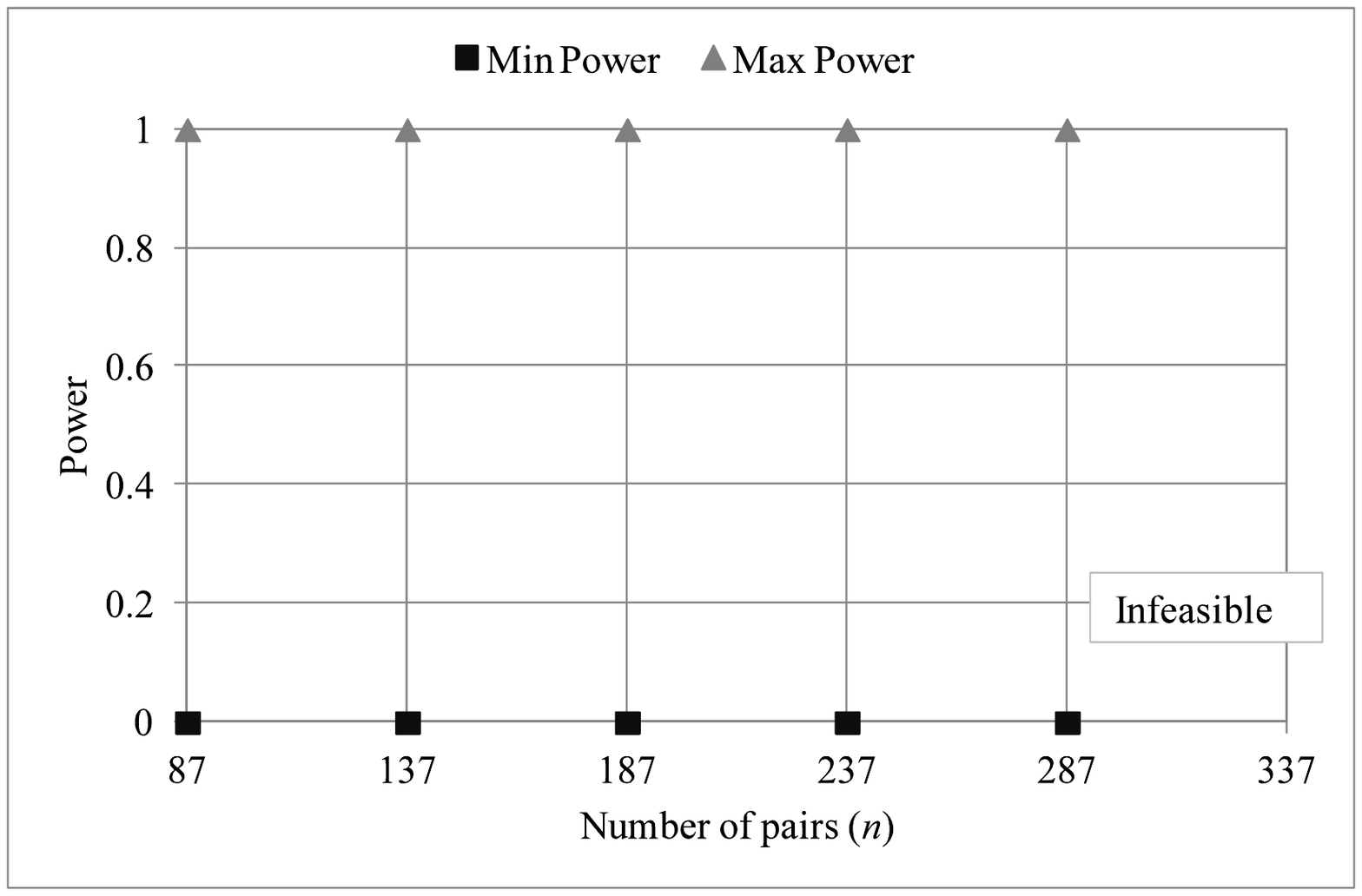}
\caption{Variation of $z$-test optimum \emph{P}-values for different \emph{n}. (Case Study 3: Training Program) \label{Figure11}}
\end{center}
\end{figure}

In the Figure \ref{Figure11}, the \emph{P}-value upper bounds are 1 and the \emph{P}-value lower bounds are  0, illustrating that there is a lot of uncertainty associated with the choice of experimenter -- one experimenter choosing 287 matched pairs can find a \emph{P}-value of $\sim$ 0 and declare a statistically significant difference while another experimenter can find a \emph{P}-value of $\sim$ 1 and declare the opposite. In this case it is truly unclear whether or not training has an effect on the earnings. To sanity check whether a reasonably sized effect would have been detected had one been present, we injected synthetic random noise (with normal distribution of mean $\simeq$ $\$$10,000 and standard deviation $\simeq$ $\$$100) on the treatment outcome, and the $z$-test robustly detects the treatment effect before the solutions become infeasible.

\subsection{Case Study 4: Crime and Transition into Adulthood}
In this case study we have used the data from a U.S. Department of Justice study regarding crime during the transition to adulthood, for youths raised in out-of-home care \citep[see][]{Courtney}. Each observation represents a youth, and the outcome is whether he or she committed a violent crime over the 3 waves of the study.

The (binary) covariates are as follows: hispanic, white, black, other race, alcohol or drug dependency, mental health diagnosis, history of neglect, entry over age of 12, entry age under 12, in school or employed, prior violent crime, any prior delinquency. The ``treatment'' variable is whether or not the individual is female; in particular we want to determine whether being female (controlling for race, criminal history, school/employment and relationship with parents) influences the probability of committing a violent crime. Here  $\textrm{dist}_{ij}$  =1 whenever all covariates of treated unit \emph{i} are the same as those of the control unit  \emph{j}, 0 otherwise.

\begin{figure}
\begin{center}
\includegraphics[height=3in]{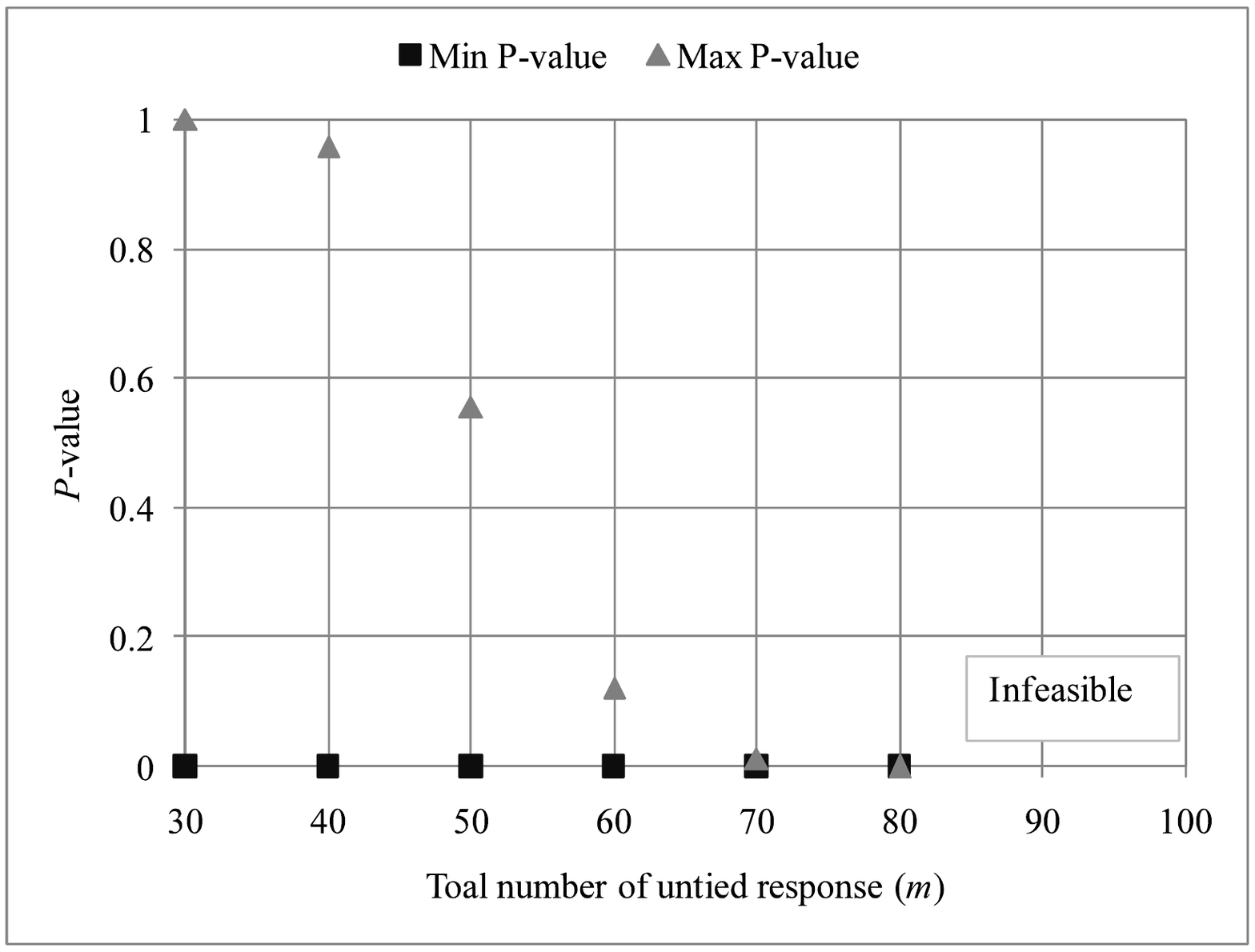}
\caption{Variation of McNemar's test optimum \emph{P}-value for different \emph{m}. (Case Study 4: Crime and Transition) \label{Figure8}}
\end{center}
\end{figure}

Figure \ref{Figure7} is constructed in an analogous way to Figure \ref{Figure8} (using McNemar's test rather the $z$-test) showing the total number of discordant pairs along the x axis. Here, any matched pairs assignment would show a significant difference for the risks of violence between males and females. This difference becomes more pronounced as the number of pairs increases. Thus, the outcome is robust to the choice of experimenter.

\section{Robust Wilcoxon Signed Rank Test}
In this section, we present an extension of our robust framework to the Wilcoxon Signed Rank Test. This extension is intended to show that our framework is general enough to be extendable to other test statistics, as well as provide one more robust test for practitioners. 

The Wilcoxon Signed Rank Test is a more powerful test on the median than the sign test, but with a stronger assumption: it assumes that the population distribution is symmetric.
We define the following parameters and decision variables to formulate the model: 

\begin{itemize}
    \item $T_{i}$ \quad is the the outcome of a treated observation $i$ in the treatment group
    \item $C_{j}$ \quad is the the outcome of a control observation $j$ in the control group
    \item $s_{ij}$ \quad is the $ij$th element of a  matrix. It takes value 1 if the covariates of treated observation $i$ \newline\indent\quad \indent \quad 
    and control observation $j$ are similar enough to be a possible matched pair, otherwise 0
    \item $\delta_{ij}$ \quad is a binary precomputed parameter, which is equal to 1 if $T_{i}$ is greater than $C_{j}$, 0 otherwise
    \item $g_{ijkl}$ \quad is a binary precomputed parameter, which is equal to 1 if $\eta_{ij}$ is greater than $\eta_{kl}$, where $\eta_{ij}=\left|T_{i}-C_{j}\right|$, 0 otherwise
\end{itemize}
There are three matrices of decision variables, namely:
\begin{itemize}
 \item $a_{ij}$ \quad is a binary variable that is 1 if \emph{i} and \emph{j} are in the same pair, otherwise 0
 \item $z_{ijkl}$ \quad is a binary variable that is 1 if $g_{ijkl}$, $a_{ij}$ and $a_{kl}$ all are equal to 1, 0 otherwise. Intuitively $z_{ijkl}$ is 1 only when $ij$ and $kl$ are both being used as pairs and when $ij$'s absolute difference is larger than $kl$'s absolute difference.
 \item $h_{ij}$ \quad is an integer variable whose value is the rank of pair $ij$. It is the count of $kl$ pairs ranked beneath pair $ij$ (plus one, so that the lowest rank is 1 rather than 0).
\end{itemize}
To compute the test statistic in the traditional way, one would compute absolute differences $\eta_{ij}$ and rank order them. The test statistic is the sum of ranks of the positive differences. 

\noindent The formulation is as follows :
\begin{equation*}
          \text{Maximize/Minimize}_{\mathbf{a,z,h}}\quad w_{+}(\mathbf{a,z,h})=\sum_{ i \in Q}\sum_{j \in R}h_{ij}\delta_{ij}
      \end{equation*}
\noindent subject to:
\begin{eqnarray}
        &g_{ijkl}+a_{ij}+a_{kl}-2 \leq z_{ijkl} \quad \forall i,j,k,l  \quad &\textrm{($z_{ijkl}$ is 1 only if $a_{ij}$, $a_{kl}$ and $g_{ijkl}$ are 1)} \label{constraint2.1}\\
        &z_{ijkl} \leq a_{ij} \quad\forall i,j,k,l \quad &\textrm {($z_{ijkl}$ is 1 only if $a_{ij}$ is 1)} \label{constraint2.2}\\
        &z_{ijkl} \leq a_{kl}  \quad  \forall i,j,k,l \quad &\textrm {($z_{ijkl}$ is 1 only if $a_{kl}$ is 1)} \label{constraint2.3}\\
        &z_{ijkl} \leq g_{ijkl} \quad \forall i,j,k,l \quad &\textrm {($z_{ijkl}$ is  1 only if $g_{ijkl}$ is 1)} \label{constraint2.4}\\
        &a_{ij}  \leq  s_{ij} \quad \forall i,j &\quad \textrm{(Choose only pairs that are allowed)}\label{constraint2.5}\\
        &\sum_{ i \in Q}\sum_{j \in R}{a}_{ij}= n  & \quad \textrm{(Choose $n$ pairs)} \label{constraint2.10}\\
        &\sum_{ i \in Q}a_{ij} \leq  1  \quad   \forall j  &\quad \textrm{(Choose at most one treatment )} \label{constraint2.11}\\
        &\sum_{ j \in R}a_{ij}  \leq  1  \quad     \forall i  &\quad \textrm{(Choose at most one control )}  \label{constraint2.12}\\
        &a_{ij} \in \{0,1\}  \quad   \forall i,j &\quad \textrm{(Defines binary variable $a_{ij}$)}\label{constraint2.6}\\ 
        &h_{ij}  = \sum_{ k \in Q}\sum_{l \in R}z_{ijkl}+a_{ij} \quad \forall i,j   &\quad \textrm {(Defines $h_{ij}$).} \label{constraint2.7} \\
        \lefteqn{\textrm{(optional covariate balance constraints).}}\nonumber     
      \end{eqnarray}

 Equations (\ref{constraint2.1}) to (\ref{constraint2.4}) are used to ensure that $z_{ijkl}$=1 when $g_{ijkl}$, $a_{ij}$ and $a_{kl}$ all are one. Equation (\ref{constraint2.5}) is used to maintain covariates constraint such that only when the precomputed parameter $s_{ij}$ is 1 then $a_{ij}$ is allowed to take a value of 1.
 Constraints (\ref{constraint2.10})-(\ref{constraint2.12}) are the same constraints as in the sign test, to make sure we have $n$ pairs with one treatment and control observation in each pair. 
Equation (\ref{constraint2.6}) defines binary variables $a_{ij}$. Equation (\ref{constraint2.7}) is used to calculate rank $h_{ij}$ for each pair $ij$. If the pair $ij$ is not being used then $a_{ij}$ will be 0, which means $z_{ijkl}$ will be 0 by Constraint (\ref{constraint2.2}). This means the only $ij$ pairs that have positive heights are those that are being used as matched pairs. The objective will only add up heights of pairs $ij$ for which the absolute differences are positive, which was precomputed as $\delta_{ij}$.

The z-score is computed from the optimal value of $w_+$ through the following formula, and the pvalue is computed as usual.
\[
z=\frac{w_+(\mathbf{a},\mathbf{z},\mathbf{h})-n(n+1)/4-1/2}{\sqrt{n(n+1)(2n+1)/24}}.
\]

\section{Integer Linear Programming Basics}
ILP techniques have become practical for many large-scale problems over the past decade, due to a combination of increased processor speeds and better ILP solvers. Any type of logical condition can be encoded as linear constraints in an ILP formulation with binary or integer variables. Consider two binary variables $x\in\{0,1\}$ and $y\in\{0,1\}$. The logical condition ``if $y=0$ then $x=0$" can be simply encoded as
  \begin{equation*}
      x\leq y.
      \end{equation*}
Note that this condition imposes no condition on $x$ when $y=1$.
Translating if-then constraints into linear constraints can sometimes be more complicated; suppose, we would like to encode the logical condition that if a function $f(w)$ is greater than 0, then another function $g(w)$ is greater or equal to 0. We can use the following two linear equations to do this, where $\theta$ is a binary variable and $M$ is a positive number that is larger than the maximum values of both $f$ and $g$:
\begin{eqnarray*}
-g(w)&\leq & M\theta \\
f(w)&\leq & M(1-\theta).
\end{eqnarray*}
In order for $f(w)$ to be positive, then $\theta$ must be 0, in which case, $g(w)$ is then restricted to be positive. If $f(w)$ is negative, $\theta$ must be 0, in which case no restriction is placed on the sign of $g(w)$. (See for instance the textbook of \cite{winston}, for more examples of if-then constraints).

ILP can capture other types of logical statements as well. Suppose we would like to incorporate a restriction such that the integer variable $S_i$ takes a value of $K$ only if $i=t$, and 0 otherwise. The following four if-then constraints can be used to express this statement, where $\lambda_1$ and $\lambda_2$ are binary variables:
\begin{eqnarray*} 
       \lambda_1=1 & \textrm{if} &  i+1>t\\
       \lambda_2=1 & \textrm{if} & t+1>i\\
          S_i=k  & \textrm{if}  &\lambda_1+\lambda_2>1\\
          S_i=0  & \textrm{if} &  \lambda_1+\lambda_2<2.
      \end{eqnarray*}

Each of these if-then constraints (4)--(7) can be converted to a set of equivalent linear equations, similar to what we described above. (See also \cite{noor} and \cite{winston}).

There is no known polynomial-time algorithm for solving ILP problems as they are generally NP-hard, but they can be solved in practice by a number of well-known techniques (\cite{wolsey}). The LP relaxation of an ILP provides bounds on the optimal solution, where the LP relaxation of an ILP is where the integer constraints are relaxed and the variables are allowed to take non-integer (real) values. For instance, if we are solving a maximization problem, the solution of the LP relaxation can serve as an upper bound, since it solves a problem with a larger feasible region, and thus attains a value at least as high as that of the more restricted integer program. ILP solvers use branch-and-bound or cutting plane algorithms combined with other heuristics, and are useful for cases where the optimal integer optimal solution is not attained by the LP relaxation. The branch-and-bound algorithms often use LP relaxation and semi-relaxed problems as subroutines to obtain upper bounds and lower bounds, in order to determine how to traverse the branch-and-bound search tree \citep{chen,wolsey}. The most popular ILP solvers such as CPLEX, Gurobi and MINTO each have different versions of branch-and-bound techniques with cutting plane algorithms and problem-specific heuristics.



\end{document}